\documentclass[11pt,a4paper]{article}
\usepackage[pdftex]{graphics}
\usepackage{amsmath,amssymb,amsfonts}
\usepackage{tcolorbox}
\usepackage{enumitem}
\usepackage{multirow}
\usepackage{array,booktabs}
\usepackage{slashed}
\usepackage[utf8]{inputenc}
\usepackage[english]{babel}
\usepackage{amsthm}
\usepackage{graphicx}
\usepackage{float}

\usepackage{jheppub}
\usepackage{amsmath,amssymb,amsfonts}

\usepackage{tcolorbox} 
\usepackage{enumitem} 
\usepackage{multirow}
\usepackage{array,booktabs}
\usepackage{slashed}
\usepackage{wallpaper}
\usepackage[utf8]{inputenc}
\usepackage[english]{babel}
\usepackage{amsthm}

\usepackage{graphicx}
\usepackage{float}
\usepackage{cancel}
\usepackage{tabularx}
\usepackage{physics}
\usepackage{hyperref}
\usepackage{bbm}

\newcommand{\bn}{\begin{enumerate}}
\newcommand{\en}{\end{enumerate}}
\newcommand{\bl}{\begin{align}}
\newcommand{\el}{\end{align}}
\newcommand{\ie}{\begin{equation}\begin{aligned}}
\newcommand{\fe}{\end{aligned}\end{equation}}

\newcommand{\vecslashed}[1]{{\slashed{#1}}}

\newcommand{\bm}{\boldsymbol}
\newcommand{\comred}{\textcolor[rgb]{0.82,0.01,0.11}}
\newcommand{\res}{\mathop{\mathrm{Res}}}

\newcommand{\intp}[1]{\int \frac{d^3 p_{#1}}{(2\pi)^3}}
\newcommand{\zero}{\partial}
\newcommand{\bchi}{\bm{\chi}}
\newcommand{\bbarchi}{\bm{\bar\chi}}
\newcommand{\bpsi}{\bm{\psi}}
\newcommand{\bbarpsi}{\bm{\bar\psi}}
\newcommand{\disc}{\mathop{\mathrm{Disc}}}

\newcommand{\beqas}{\begin{eqnarray*}}
\newcommand{\eeqas}{\end{eqnarray*}}
\newcommand{\beqa}{\begin{eqnarray}}
\newcommand{\eeqa}{\end{eqnarray}}			
\newcommand{\beq}{\begin{equation}}
\newcommand{\eeq}{\end{equation}}

\newcommand{\braketw}[2]{\langle #1  #2 \rangle}
\newcommand{\ab}[2]{\langle #1  #2 \rangle}
\newcommand{\abBB}[2]{\langle \bar{#1}  \bar{#2} \rangle}
\newcommand{\abB}[2]{\langle {#1}  \bar{#2} \rangle}

\def\Trans{\text{T}}

\def\p{\partial}

\def\co{\mathcal{O}}

\DeclareUnicodeCharacter{2212}{-}

\preprint{}

\title{Flat space Fermionic Wave-function coefficients}
\author[1]{Bo-Ting Chen}
\author[2]{Wei-Ming Chen}
\author[3,4,5]{Yu-tin Huang}
\author[3]{Zi-Xun  Huang}
\author[3]{Yohan Liu}
\affiliation[1]{Department of Physics, Princeton University, Washington Road, Princeton, NJ, USA}
\affiliation[2]{Department of Physics, National Sun Yat-sen University, Kaohsiung 80424, Taiwan}
\affiliation[3]{Department of Physics and Center for Theoretical Physics, National Taiwan University, Taipei 10617, Taiwan}
\affiliation[4]{Physics Division, National Center for Theoretical Sciences, Taipei 10617, Taiwan}
\affiliation[5]{Max Planck{-}IAS{-}NTU Center for Particle Physics, Cosmology and Geometry, Taipei 10617, Taiwan}
\emailAdd{bc2490@princeton.edu}
\emailAdd{tainist@gmail.com}
\emailAdd{yutinyt@gmail.com}
\emailAdd{d11222011@ntu.edu.tw}
\emailAdd{youan1997@icloud.com}
\abstract{In this work we analyze the analytic structure of tree-level flat-space wavefunction coefficients (WFCs), with particular attention to fermionic operators, and derive cutting rules for internal-fermion lines. Building on these results, we set up an iterative procedure that, starting from the flat-space S-matrix, reconstructs the 3- and 4-point WFCs with the correct partial- and total-energy poles and satisfying the requisite cutting rules. Consequently, the “four-particle test” for flat-space WFCs imposes no additional constraints beyond the consistency of the flat-space S-matrix. }

\begin{document}
\maketitle
\section{Introduction}
The idea that fundamental consistency conditions can reveal the structure of physical theories has a long and rich history. Some of the well known examples are the Weinberg–Witten~\cite{Weinberg:1980kq} and Coleman–Mandula~\cite{Coleman:1967ad} theorems, which show that the very symmetries of a Lorentz-invariant S-matrix severely restrict the possibilities for consistent interactions—yielding celebrated no-go results for certain high-spin or mixed-symmetry systems.

When one focuses on tree-level S-matrices—those that best capture the behavior of Lagrangian theories accessible to perturbation theory—the power of these consistency arguments becomes even more tangible. From the requirement that four-particle scattering amplitudes behave consistently under factorization and Lorentz invariance, one can already rediscover many cornerstones of modern field theory: the emergence of gauge algebra for massless spin-1 particles~\cite{Benincasa:2007xk}, the impossibility of elementary states with spin higher than two~\cite{Arkani-Hamed:2017jhn}, the inevitability of local supersymmetry and gravity once spin-3/2 fields are present~\cite{McGady:2013sga}, and even the appearance of the Higgs mechanism and anomaly structures~\cite{Huang:2013vha}.

Scattering amplitudes are the natural observables in flat spacetime. For curved or non-trivial backgrounds, one instead considers boundary observables, defined on either time-like (AdS) or space-like (dS) slices. This raises an analogous question: how does the consistency of boundary observables constrain the dynamics of the bulk theory? In recent years, remarkable progress has been made in bootstrapping de Sitter boundary correlators, or equivalently, wave-function coefficients (WFCs)~\cite{Arkani-Hamed:2018kmz, Baumann:2019oyu, Baumann:2020dch, Baumann:2021fxj, Pajer:2020wxk, Jazayeri:2021fvk, Bonifacio:2022vwa, Mei:2024sqz}. Notably, no-go theorems for partially-massless higher-spin particles have been derived within this framework~\cite{Baumann:2025tkm, Sleight:2021iix}. 

One of the fascinating properties of boundary correlators is the emergence of flat-space amplitudes when the total energy is analytically continued to zero, appearing as the residue of the total energy pole ~\cite{Raju:2012zr, Raju:2012zs, Arkani-Hamed:2015bza}. Indeed how fundamental principles of amplitudes, such as Lorentz invariance and unitarity, emerge from correlators where none of these notions exists is a profound question that is only answered in some simple settings~\cite{Benincasa:2018ssx, Arkani-Hamed:2018bjr, Figueiredo:2025daa}. In this work, we turn the question around: if one were to perform a ``four-particle test" on WFCs, would new consistency conditions arise—constraints that go beyond those already implied by a consistent flat-space S-matrix? Put differently, given a consistent S-matrix in flat space, does a consistent WFC necessarily follow?

To address this, we focus on the three- and four-point WFCs. Starting from the flat-space S-matrix as the seed, we construct a step-by-step procedure that systematically builds the corresponding WFCs while enforcing their analytic structure in the energy variables, which are conjugate to the bulk time coordinate. If this iterative construction proceeds without obstruction, satisfying all consistency conditions automatically, then the existence of a consistent WFC follows directly from that of the S-matrix. Otherwise, the breakdown of the procedure indicates that not every consistent flat-space theory remains consistent in a non-trivial background, thereby pinpointing the tension in an explicitly on-shell manner. 

In this paper, we focus on flat-space WFCs, leaving their extension to curved spacetimes (dS/AdS) to a companion paper~\cite{CurvedPaper}. We pay particular attention to fermionic WFCs in anticipation of exploring the tension associated with spin-$3/2$ fields in de Sitter space. Fermionic boundary correlators have been studied primarily in the context of AdS/CFT, beginning with two-point functions~\cite{Corley:1998qg, Henningson:1998cd} and three-point functions~\cite{Mueck:1998iz}, followed by exchange diagrams involving massive fermions~\cite{Kawano:1999au} and, more recently, massless spin-$1/2$ fermions~\cite{Chowdhury:2024snc, Goyal:2025hdm}. Because the kinetic action is first order in derivatives, the canonical conjugate of the fermionic field is the field itself, and one cannot directly impose Dirichlet boundary conditions to define the boundary profiles. This issue is resolved by introducing a boundary term. Moreover, the first-order nature of the action causes the on-shell bulk action to vanish, so the boundary action is the sole source of tree-level WFCs.

First, we demonstrate that the perturbative expansion of the boundary action gives rise to a diagrammatic expansion analogous to its bosonic counterpart. We derive cutting rules for fermionic bulk-to-bulk propagators that determine how WFCs behave under internal-energy flips. These rules allow us to extract constraints at the partial-energy poles, whose residues themselves contain total-energy poles~\cite{Baumann:2020dch}. With these ingredients in place, we initiate the bootstrap construction of the four-point WFC. Starting from the three- and four-point S-matrices, we first construct the three-point WFC by weighting the three-point amplitude with the appropriate total-energy poles. Its polynomial terms are then determined either by power counting or by Ward–Takahashi (WT) identities. We subsequently use this result to construct the four-point WFC: we begin with the residues at the partial-energy poles and fix the total-energy pole by matching to the amplitude. The resulting four-point WFC takes the form
\ie\label{eq: Intro1}
c^T_4=\sum_{e \in s,t,u}\left(\frac{A_R^e}{E_R^e}+\frac{B_L^e}{E^e_L}\right)+\frac{C}{E_T}+D\,,
\fe
where $E_{L,R}^e$ represents left/right partial energies for channel $e=s,t,u$ and $D$ are pure polynomial terms. The numerators $(A, B, C)$ are defined in section \ref{sec:bootstrapping}. Here the super-script $T$ represents transverse WFC, which means that for conserved spinning WFCs the free vector indices are contracted transverse projectors and polarization vectors. The longitudinal pieces are given via WT identities.

We demonstrate this procedure for WFCs involving spin-$\frac{1}{2}$, $\frac{3}{2}$ as well as currents and stress-tensors. The results are given both in terms of polarization factors as well as massive spinor helicities. Interestingly, while for helicity sectors that has an amplitude limit, the amplitude appears as residue on the total energy pole singularity, for helicity configurations without amplitude limit, the result does not have total energy pole. Note that since eq.(\ref{eq: Intro1}) gives the complete solution, a consistent flat-space S-matrix automatically leads to a consistent WFCs. This will no longer be the case when non-trivial backgrounds are considered~\cite{CurvedPaper}.

This paper is organized as follows. In section \ref{sec: review of wavefunction coefficients}, we review the definition of WFC and its perturbative calculation with emphasis on fermionic and conserved WFC. Specifically, we discuss how to properly choose the fermionic Dirichlet boundary conditions as well as demonstrate the diagrammatic expansion of the boundary action. We also set up linearly-independent decompositions of the spinning WFC that would be useful in implementing constraints of WFCs involving conserved currents. In the end we discuss the WT identities and their relation to the bulk residual gauge symmetry. In section \ref{sec:Analytic properties in energy variables}, we review and discuss the analytic properties of the energy variables in the WFC from perturbative aspects, including discussion on the appearance of total energy poles, its relation to the amplitude, the cutting rules and how to extract the partial energy poles therein. Note that for the cutting rules we derive the fermion-exchange cases and found interesting universal structure for tree-level cuts regardless the spin of the exchanged particle \eqref{eqn: cutting rule}. We also derive loop-level cuts and test via inspecting the $\phi^3$ and $\phi^4$ corrections to the scalar 2-pt functions. Finally, we provide an alternative derivation of the fermionic cutting rules \emph{without} using the explicit forms of the propagators. In section \ref{sec:bootstrapping}, we systematically apply all the constraints to reconstruct fermionic 3- and 4-pt WFCs with at least one conserved current insertion. We also provide some results in the 3D helicity basis and find interesting interplay between the amplitude limit and the presence of the total energy pole. Appendix \ref{app: conventions} sets up the notation and convention in this paper. Appendix \ref{app: BoundarConstraint} elaborates on the issue of constraints on the boundary profiles from the bulk equations of motion. Appendix \ref{app:WTofWFC} provides an explicit calculation on how to realize the bulk residual gauge symmetry on the classical solution to the boundary profile. This appendix also collects all the WT identities used in this paper. Appendix \ref{App: Majorana Condition} shows why requiring consistent factorization of the 4-gravitino amplitude inevitably leads to Majorana condition on the gravitino. Appendix \ref{app:CPT} explores the implications of bulk CPT invariance on the fermionic WFCs. At last, in the appendix \ref{app:matching identity}, we provide "useful identities" needed for showing terms carrying the partial energy poles extracted from the cutting rules indeed correctly reproduce the amplitude factorization under $E_T\to 0$.

\section{Review of Wavefunction Coefficients}
\label{sec: review of wavefunction coefficients}
Consider a generic field  $\varphi({x}, t)$ evolving in flat 4-dimensional spacetime, leaving a 3-dimensional imprint on the time slice at $t = 0$. We denote $\varphi_\partial$ as the boundary profile of the field, i.e. $\varphi(t = 0) =\varphi_\partial$. The wavefunction is then the overlap between a state $\ket{\varphi_\zero}$ at $t = 0$ and the vacuum state $\ket{\Omega}$ :
\begin{equation}
    \Psi[\varphi_\zero] = \braket{\varphi_\zero}{\Omega}:
\end{equation}
The wavefunction $\Psi[\varphi_\zero]$ encodes information about the dynamics of $\varphi$ in the bulk, and gives the equal-time correlation functions in  \emph{in-in} formalism as:
\begin{equation}
\label{inindef}
\bra{\Omega} \hat \varphi(  x_1) \hat \varphi(  x_2) \dots \hat \varphi(  x_n) \ket{\Omega}
=
\int \mathcal{D}\varphi_\zero \ \varphi_\zero(  x_1) \varphi_\zero(  x_2) \dots \varphi_\zero(  x_n) |\Psi[\varphi_\zero]|^2.
\end{equation}
Here, $\hat \varphi$ represents the field operator on the boundary. It will be convenient to expand the wavefunction  in three-dimensional momentum-space eigenstates:
\begin{equation}
\label{eqn: expand wave function in wave function coefficient}
\log \Psi[\varphi_\zero]= \sum_{n=2}^\infty \int \frac{d^3\mathbf{k}_1 \dots d^3\mathbf{k}_n }{(2\pi)^{3n}} \, \delta^{(3)}\left(\sum_{i=1}^n \mathbf{k}_i\right)\, \varphi_{\zero,\mathbf{k}_1} \dots \varphi_{\zero,\mathbf{k}_n} \, c_n(\mathbf{k}_1, \dots, \mathbf{k}_n),
\end{equation}
where the functions $c_n(\mathbf{k}_1, \dots, \mathbf{k}_n)$ are referred to as \emph{wavefunction coefficients} (WFCs). For spinning fields, the WFCs carry explicit indices to be contracted with the spinning boundary profiles.
\footnote{
	We use $i,j, \dots$ to denote the 3-dimensional spatial vector indices, while $A=(\alpha,\dot{\alpha})$ to denote the four-component spinor indices. While spinors in 3 dimensions transform under SL(2,R), from the bulk point of view it will be convenient to embed it in four-component notations. For details, see section \ref{sec:bdry action}.
}
In what follows, we use the notation $c_{n,AB}^{i_1, \dots, i_n}$ to denote a n-pt generic WFC, while we use bracket notation when referring to specific boundary operators; that is,
\begin{equation}
    c_{n,AB}^{i_1, \dots, i_n}(p_1, p_2, p_3, \dots) \rightarrow \langle J^{i_1}_1\, \bar\psi^{i_2}_{2,A}\, \psi^{i_3}_{3,B}\, \phi_4 \cdots \rangle,
\end{equation}
where the subscripts on the fields denote momentum labels, and the vector indices likewise carry a subscript indicating the momentum of the field to which they belong. Note that for operators with spin, we may also employ contracted notation. For example, in the case of a vector operator,
\begin{equation}
    c_n({p}_1) := \epsilon_{1,i,\zero} c^{i}_n \rightarrow \langle \mathcal{O}_1 \cdots \rangle := \epsilon_{1,i,\zero} \langle \mathcal{O}_{1}^{i} \cdots \rangle,
\end{equation}
where $\epsilon_{1,i,\zero}$ denotes the vector boundary polarization profile.

The wavefunction is given as a path-integral over field configurations:
\begin{equation}
\Psi[\varphi_\partial] = \int_{BD} \mathcal{D} \varphi\, e^{i S[\varphi]}.
\end{equation}
where we impose the Bunch-Davies (BD) vaccuum boundary condition:
\begin{equation}
\text{BD}:\quad \lim_{t\rightarrow-\infty_{-}}\varphi(t,{x})=0\,.
\end{equation}
Here, $\infty_{-} := \infty(1 - i\epsilon)$ denotes infinity tilted slightly into the lower half of the complex plane, ensuring that positive-energy solutions are selected in the far past. The wavefunction can then be computed perturbatively by expanding around classical solutions, with the leading ``tree-level'' contribution given by evaluating the action on the classical solution:
\begin{equation}
\label{eqn: expand wave function in classical solution}
\Psi[\varphi_\zero]  \approx e^{i S[\varphi_{cl}]},
\end{equation}
where $\varphi_{cl}$ is the classical solution to the equation of motion:
\begin{equation}\label{eq: Schwinger-Dyson}
\mathcal{D} \varphi_{cl} = -\frac{1}{2} \left. \frac{\delta L_{\text{int}}}{\delta \varphi} \right|_{\varphi = \varphi_{cl}}.
\end{equation}
The operator $\mathcal{D}$ arises from the variation of the kinetic term in the Lagrangian (with $\mathcal{D} = \Box$ for scalars), and $L_{\text{int}}$ denotes the interaction part of the Lagrangian. To equate the  expansion in \eqref{eqn: expand wave function in wave function coefficient} with the classical action \eqref{eqn: expand wave function in classical solution}, we expand $\varphi_{cl}$ on free field solutions, i.e. Schwinger-Dyson equations, and identify $\varphi_{\partial,k}$ with the fourier transform of the latter. The WFCs can be obtained order by order in couplings.

The perturbative computations of WFCs can be organized into a Feynman diagram representation. Let us first use the tree-level WFCs of scalars to illustrate this point. The basic building blocks are the \emph{bulk-to-boundary} and \emph{bulk-to-bulk} propagators. The \emph{bulk-to-boundary} propagator $K({x}', {x}, t)$ is a solution to the free equation of motion, subject to the following boundary conditions:
\begin{equation}
\label{eqn: conditions on K}
	\mathcal{D}_{  x, t} K(  x',   x, t) = 0\ ;\quad K(  x',   x, t{=}0) = \delta^3(  x -   x')\ ;\quad K(  x',   x, t{=}{-}\infty_-) = 0\,.
\end{equation}
The \emph{bulk-to-bulk propagator} $G(  x,   x', t, t')$ 
the Green’s equation with appropriate boundary conditions:
\begin{equation}
\label{eqn: conditions on G}
	\mathcal{D}_{  x, t} G(  x,   x', t, t') = \delta^4(x_\mu - x'_\mu)\ ;\quad G(  x,   x', t{=}0, t') = 0\ ;\quad G(  x,   x', t{=}{-}\infty_-, t') = 0.
\end{equation}
In general it is useful to Fourier transform the boundary spatial coordinates to momentum space. For example, the scalar propagators take the form,
\ie
	\label{eq:scalarpropagtor}
	K({p}, t) &= e^{i E t}, \quad
	G({p}, t, t') &= \frac{i}{2E} \left[ e^{iE(t - t')}\theta(t' - t) + e^{-iE(t - t')}\theta(t - t') - e^{iE(t + t')} \right] \,.
\fe
The classical solution is then given as a series expansion in the coupling(s) of $L_{\rm int}$,
\begin{equation}
\varphi_{cl}=\sum_{i=0}^\infty g^i\varphi^{(i)}
\end{equation}
which is solved by substituting into both sides of (\ref{eq: Schwinger-Dyson}). The solution is given by the Schwinger-Dyson (SD) series:
\footnote{Here $\varphi=(\varphi^{(0)}, \varphi^{(1)})$ indicates one of the fields in the interaction will be $\varphi^{(1)}$, while the remaining $\varphi^{(0)}$.}
\ie
\label{eq: perturbative solution}
\varphi^{(0)}(\varphi_\partial,t,  x) &= \int d^3 x' \, K(  x,   x', t) \, \varphi_\partial(  x') \\
\varphi^{(1)}(\varphi_\partial,t,  x) &= \int d^3x' \, d^3 t' \, G(  x,   x', t, t') \left. \left( - \frac{\delta L_{\text{int}}}{2\delta \varphi(  x', t')} \right) \right|_{\varphi=\varphi^{(0)}} \\
\varphi^{(2)}(\varphi_\partial,t,  x) &= \int d^3x' \, d^3 t' \, G(  x,   x', t, t') \left. \left( - \frac{\delta L_{\text{int}}}{2\delta \varphi(  x', t')} \right) \right|_{\varphi=(\varphi^{(0)}, \varphi^{(1)})}\,.
\fe
 Substituting the solution into the classical action and expanding in $\varphi_\partial$ yields the tree-level WFCs. At zeroth order in coupling we have the two-point function. At linear order, we have ``contact diagrams" where a bulk interaction vertex is connected to the boundary via bulk to boundary propagators 
\begin{equation}
\label{contact Fey structure}
c_{n,\text{contact}} 
= \sum_{\text{perm}} \int dt\, (ig)\, V(  p_1,  p_2, \dots,  p_n, \partial_t)\, K_1(  p_1,t) K_2(  p_2,t) \dots K_n(  p_n,t),
\end{equation}
where $V$ represents the vertex factor coming from $L_{\rm int}$. Beyond linear order, we have ``exchange" diagrams where the bulk-to-bulk propagators connect two or more interaction vertices. For example starting at four-points one can have:
\begin{align}
\label{Exchagne Fey structure}
c_{4,\text{exchange}} 
= \sum_{\text{perm}} \int dt\, dt'\, (ig^2)\, &K_1(  p_1,t) K_2(  p_2,t)\, V_L(  p_1,  p_2,  p_s,\partial_t) \notag \\
&\cdot G(  p_s, E_s, t, t') \cdot V_R(  p_3,  p_4,-  p_s,\partial_{t'})\, K_3(  p_3,t') K_4(  p_4,t'),
\end{align}
where $\cdot$ denotes contractions over internal vector or spinor indices as needed. Thus each interaction vertex introduces an additional time integral. For a detailed derivation of such diagrammatic expansion for WFCs see for example appendix A of~\cite{Goodhew:2020hob}. We now discuss the new features that arise when one considers fermions.

\subsection{Fermionic Wave Function Coefficient}
\label{sec:bdry action}

As we are considering path integral with boundaries, boundary terms will contribute in the variation of the action, leading to bulk equations of motion ill-defined. For scalars and vectors, the boundary contributions can be set to zero by imposing Dirichlet boundary conditions. For gravity, due to second derivatives on the metric in the variation, Dirichlet boundary conditions are no longer sufficient to remove boundary contributions. The remedy is to introduce a boundary action whose variation cancels the unwanted boundary terms generated from the bulk action. This is the well-known Gibbons-Hawking-York (GHY) boundary term~\cite{PhysRevLett.28.1082, PhysRevD.15.2752}.

For fermionic fields, the opposite issue arises. Instead of being insufficient, Dirichlet boundary conditions are too restrictive. As was pointed out in the context of AdS/CFT~\cite{Henningson:1998cd}, since the Lagrangian is first-order in derivatives, one can only impose Dirichlet condition on half of the field, since the other half is its canonical conjugate. Thus once again one needs to introduce boundary action to remove the remaining boundary terms. For example, for a free massless spin-$\frac{1}{2}$ fermion, the combined action reads:
\begin{equation}
\label{eqn: boundary action of fermion action}
S= \int d^4x \,
\left(
-\frac{1}{2} \bm{\bar{\chi}} \slashed{\partial}^{[4]} \bm{\chi} 
+ \frac{1}{2} \bm{\bar{\chi}} \overleftarrow{\slashed{\partial}}^{\raisebox{2pt}{$\scriptstyle{[4]}$}} \bm{\chi}
\right)
+S_{b}\, , \quad
S_{b} = \frac{i}{2} \int d^3 x\, \bm{\bar{\chi}}_b \bm{\chi}_b \,.
\end{equation}
For spin-$\frac{3}{2}$ there is a similar boundary action which can be understood as the supersymmetry counterpart of GHY~\cite{Papadimitriou:2017kzw}. Through out the paper, we use the following notations:
\begin{align}
\label{cvt}
\notag \slashed{a}^{[4]}&:=a_\mu \gamma^\mu=-a_0\gamma_0+\slashed{a}\,,~~~\mu=0,1,2,3\,,\\
\slashed{a}^{[4]}_{\,-}&:=a_\mu \gamma^\mu=-a_0\gamma_0-\slashed{a}\,,\\
\notag    \slashed{a}&:=a_i \gamma^i\,,~~~~~~~~~~~~~~~\,~~~~i=1,2,3\,.
\end{align}
Here $a_\mu$  denotes a four-vector with spatial components $a_i$ and $a_0:=|\vec a|$, and $\gamma_\mu$ represents the four-dimensional gamma matrices. Our convention takes Greek indices to range from 0 to 3 and Roman indices from 1 to 3.
Note that here $\bm{\chi}_b=\bm{\chi}(t=0,  p)$ are bulk fields evaluated at the boundary. We will differentiate between $\bm{\chi}_b$ and boundary profile $\bm{\chi}_\partial$, since the latter only constitutes half of the former as we will now see.  Let's consider the variation of the combined action:
\ie
\label{eq:spinor action variation}
\delta S &= \int d^4x 
\left( 
- \delta \bm{\bar{\chi}}\, \slashed{\partial}^{[4]} \bm{\chi}
+ \bm{\bar{\chi}}\, \overleftarrow{\slashed{\partial}}^{\raisebox{2pt}{$\scriptstyle\scriptstyle[4]$}}\, \delta\bm{\chi}
\right)  \\
&\quad + \int d^3 x
\left(
-\frac{1}{2} \bm{\bar{\chi}}_b \gamma_0\, \delta \bm{\chi}_b
+ \frac{1}{2} \delta \bm{\bar{\chi}}_b \gamma_0\, \bm{\chi}_b
+\frac{i}{2} \bm{\bar{\chi}}_b \delta \bm{\chi}_b  
+ \frac{i}{2} \delta \bm{\bar{\chi}}_b \bm{\chi}_b
\right)\,.
\fe
The first two terms on the second line is the boundary contributions from the bulk action, which can be set to zero if one were to naively set both $\delta \bm{\chi}_b$ and $\delta\bm{\bar{\chi}}_b$ to zero. However, since $\bm{\chi}$ and $\bm{\bar{\chi}}$ are canonical conjugates, this would tantamount to imposing both Dirichlet and Neumann-type conditions simultaneously. Fortunately, with the boundary action included, we can show that the boundary contribution vanishes if we impose Dirichlet boundary condition on half of the degrees of freedom of $\bar\chi$ and $\chi$, and those components that are not canonical conjugates of each other.

With a space-like boundary in mind, it is natural to decompose the four-component spinor as $\bm{\chi} = \bm{\chi}_+ + \bm{\chi}_- \, $ where 
\begin{align}
\gamma_0 \bm{\chi}_{\pm} = \pm i\, \bm{\chi}_{\pm}\,,  ~~~~~\bm{\bar{\chi}}_{\pm} \gamma_0 = \pm i\, \bm{\bar{\chi}}_{\pm}\,.
\label{eq: SpinorEigen}
\end{align}
Note that under Dirac conjugation,
\begin{equation}
  \bm{\chi} _{+}\ \longleftrightarrow\ \bm{\bar{\chi}} _{+}\,,~~~
\bm{\chi} _{-}\ \longleftrightarrow\ \bm{\bar{\chi}} _{-}\,.
\end{equation}
Thus, without loss of generality, Dirichlet boundary conditions can be consistently imposed on $(\bm{\chi}_{-,b},\bm{\bar{\chi}}_{+,b})$, which will be identified as the 3D boundary profiles $(\chi_\zero,\bar\chi_\zero)$: 
\begin{equation}
\label{eq: spinor embedding}
\bm{\chi}_{-,\zero} :=\Pi_-\bchi_b= \left(\begin{array}{c}
     \sqrt{2}\chi_\zero  \\
     ~\\[-3mm]
     0 
\end{array}\right)\,,
\qquad
\bm{\bar{\chi}}_{+,\zero} := \bbarchi_b\Pi_+
=\left(\begin{array}{cc}
\,0~\,&~\,\sqrt{2}\bar\chi_\zero\,           
\end{array}\right)\,,
\end{equation}
where $\Pi_\mp=\frac{1\pm i\gamma_0}{2}$, and the last expression is the embedding of three-dimensional two-component spinors in the bulk four-component form. From now on, we will use bolded symbols ($\bm\chi_\zero$) to denote spinors in four-component notions and un-bolded ones ($\chi_\zero$) for two-components.  
With our choice of boundary profiles, i.e. the components where we impose Dirichlet boundary conditions, we immediately see that the remaining boundary terms in eq.~\eqref{eq:spinor action variation} indeed cancel.
\begin{equation}
-\frac{1}{2} \bm{\bar{\chi}}_{+,\partial} \gamma_0\, \delta \bm{\chi}_{+,\partial}
+ \frac{1}{2} \delta \bm{\bar{\chi}}_{-,\partial} \gamma_0\, \bm{\chi}_{-,\partial}
+\frac{i}{2} \bm{\bar{\chi}}_{+,\partial} \delta \bm{\chi}_{+,\partial}  
+ \frac{i}{2} \delta \bm{\bar{\chi}}_{-,\partial} \bm{\chi}_{-,\partial}
=0\,.
\end{equation}
From now on, we will suppress the subscripts $\pm$ on the boundary profile with the understanding:
\begin{equation}\label{eq: ChiBoundaryDef}
\bm{\chi}_{\partial} \rightarrow \bm{\chi}_{-,\partial}\,,~~~~~ \bm{\bar{\chi}}_{\partial} \rightarrow \bm{\bar{\chi}}_{+,\partial}\,.
\end{equation}

\noindent\textbf{Propagators}:
As discussed in the introduction, the first-order nature of the action causes the on-shell bulk action to vanish, leaving only the boundary action. Solving the Schwinger--Dyson equation perturbatively and substituting the resulting solution into the boundary action yields a Feynman-diagram expansion analogous to that of the scalar case.

To proceed, we construct the spin-$\frac{1}{2}$ version of the bulk-to-boundary and bulk-to-bulk propagators. They are given as: 
\footnote{
	The bulk-to-bulk and bulk-to-boundary propagators for the massless spinor are similar to those in de Sitter space \cite{Chowdhury:2024snc}. Two completely different methods for writing the classical solution perturbatively are presented earlier in one of the author's previous works \cite{YAflat}.}
\ie
\label{eq:fermionpropagtor}
K_{\chi}(  p,t) &= (1+i \slashed{\hat{p}})e^{iEt},\quad K_{\bar\chi}(  p,t) = e^{iEt}(1-i \slashed{\hat{p}}) \\
G_{\chi}(  p,t,t') &=G_{\bar\chi}(  p,t,t')
= \frac{1}{2E}\left(\slashed{p}^{[4]}e^{iE(t-t')}\theta(t'{-}t){-}\slashed{p}^{[4]}_-e^{-iE(t{-}t')}\theta(t{-}t'){+}i\gamma_0\slashed{p}^{[4]}_-e^{iE(t+t')}\right)
\fe
where ${p}$ is the spatial momentum, $\hat{p}={p}/E$, with $E=|{p}|$.
It is straightforward to verify the following spinor version of the boundary condition of the propagators are satisfied,
\begin{equation}
\label{eq: spinor propagator boundary condition}
    \begin{array}{rclrclrcl}
\bbarchi_{\zero} K_{\bar\chi}(p,0)\Pi_+&=&\bbarchi_\zero\,,  
        &~~ \Pi_+G_{\chi}(p,0,t')&=& G_{\chi}(p,0,t')\,,
               &~~G_{\chi}(p,t,0)\Pi_+&=&0\,,
        \\[4mm]
        \Pi_- K_{\chi}(p,0)\bchi_{\zero}&=&\bchi_\zero\,,     
       &~~ G_{\chi}(p,t,0)\Pi_-&=&G_{\chi}(p,t,0)\,, &~~ \Pi_-G_{\chi}(p,0,t')&=&0\,,
    \end{array}
\end{equation}
and the same result for $G_{\bar\chi}(p,t,0)$. We insert the propagators into the Schwinger-Dyson series \eqref{eq: perturbative solution} to yield the classical solution and the boundary behaviour of the propagators reproduces the Dirichlet boundary conditions \eqref{eq: spinor embedding} by   
\begin{equation}\label{eq: BoundaryBehaviour}
\bchi^{(0)}_{-,b}=\bchi_\zero\,, \quad~~
 \bbarchi^{(0)}_{+,b}=\bbarchi_\zero \,,\quad~~ 
 \bchi^{(n \geq 1)}_{-,b}=\bbarchi^{(n \geq 1)}_{+,b} =0\,.
\end{equation}
Note that \eqref{eq: spinor propagator boundary condition} is analogous to the scalar boundary conditions \eqref{eqn: conditions on K} and \eqref{eqn: conditions on G}, except that in the spinor case an explicit projection onto the Dirichlet boundary condition by $\Pi_{\pm}$ is required.

\noindent\textbf{Diagrammatic expansion from boundary action}:

We now substitute the Schwinger-Dyson series to the boundary action. At zeroth order in coupling we have the two-point function. To first order, due to eq.(\ref{eq: BoundaryBehaviour}), the boundary action is composed of $\bbarchi^{(0)}_{b,+}\bchi^{(1)}_{b,+}+\bbarchi^{(1)}_{b,-}\bchi^{(0)}_{b,-}$. The first half yields,
\footnote{
	The Feynman rules structure for the massive spinor, derived from the classical solution insertion into the boundary action in EAdS space, can be found in \cite{Kawano:1999au}.
}
		\ie\label{eq: 3ptWFC}
			\frac{i}{2} \int d^3 x  \bbarchi^{(0)}_{+,b} \bchi^{(1)}_{+,b}
			&=
			-\frac{1}{2}\int^0_{-\infty_-} d^4 x \;\partial^\mu\left( \bbarchi^{(0)} (-\gamma_\mu)\bchi^{(1)}\right) 
			-\frac{1}{2}
			\bbarchi^{(0)} (\gamma_\mu\overleftarrow{\partial}^\mu)\bchi^{(1)} \\
			&=
			+\frac{1}{2}
			\int^0_{-\infty_-} d^4 x \bbarchi^{(0)} (\gamma_\mu\partial^\mu)\bchi^{(1)} 
			= 
			\frac{1}{2}\int^0_{-\infty_-} d^4 x \, g\bbarchi^{(0)} \left. \left( \frac{\delta L_{\text{int}}}{\delta \bchi} \right) \right|_{\varphi^{(0)}, \bchi^{(0)}} \,,
		\fe
where we have integrated by parts. The second term in the first equality vanishes because $\bbarchi^{(0)}$ is a free solution of the equations of motion, while the last equality follows from the equation of motion for $\bchi^{(1)}$ at first order. The other half of the boundary action, $\int d^3 x  \bbarchi^{(1)}_{-,b} \bchi^{(0)}_{-,b}$, yields an analogous contribution. Expanding eq.~(\ref{eq: 3ptWFC}) in the boundary profiles then gives either a three- or four-point WFC, depending on the interaction vertex. For QED in temporal gauge, the contracted contact WFC is\footnote{With the mostly-plus metric convention, we have $g=ie$.}
		\ie
		\label{3pt Fermion Expansion}
		c_{3, \bar\chi J \chi} = i g\int^0_{-\infty_-} d x_0 \, \bbarchi^{(0)}(x_0,   p_1) \slashed{A}^{(0)}(x_0,   p_2)  \bchi^{(0)}(x_0,   p_3)\,. 
		\fe
		At the next order we have exchange diagram structure as follows:
		\ie
		\label{4pt Fermion Expansion}
		&c_{4,s,\bar\chi J \chi J}\\ 
		&=
		\frac{i g}{2} \int^0_{-\infty_-} dx_0 \,
		\left[ 
		\bbarchi^{(0)}(x_0,   p_2) \slashed{A}^{(0)}(x_0,   p_1)   \bchi^{(1)}(x_0,   p_s){+} 
		\bbarchi^{(1)}(x_0, {-}p_s) \slashed{A}^{(0)}(x_0,   p_3)  \bchi^{(0)}(x_0,   p_4)
		\right]
		\,,\\
		&c_{4,s,\bar\chi \bar\chi \chi \bar\chi} 
		=
		i g\int^0_{-\infty_-} dx_0 \,
		\left[ 
		\bbarchi^{(0)}(x_0,   p_2)  \slashed{A}^{(1)}(x_0,   p_s)  \bchi^{(0)}(x_0,   p_1)
		\right]\,,
		\fe 
		where $p_s={-}p_1{-}p_2$. Upon inserting the Schwinger-Dyson series \eqref{eq: perturbative solution} one sees that the first-order classical solutions $\bbarchi^{(1)}$, $\slashed{A}^{(1)}$, and $\bchi^{(1)}$, give rises to exchange diagrams through the bulk-to-bulk propagators.

\subsection{Spinning WFC and projectors}
As noted earlier, for spinning boundary profiles the uncontracted WFCs carry explicit spacetime (or spinor) indices. It is therefore natural to decompose both the operator and the corresponding WFCs into irreducible representations, allowing one to bootstrap the different components independently. A subtlety specific to flat space is that, in the late-time limit, the bulk equations of motion can impose “space-like” constraints on certain boundary profile components—namely, equations of motion that involve no time derivatives. This phenomenon arises for spin-2 and spin-3/2 fields. As a consequence, the WFCs must be dressed with appropriate projector factors.

\paragraph{Operator Decomposition}
For spin-1 operator, one simply decomposes into transverse and longitudinal pieces:
\begin{equation}
\begin{array}{c}
    \mathcal O^i=\left(\mathcal{O}^T\right)^i+\left(\mathcal{O}^L\right)^i\,,\\[2mm] \left(\mathcal{O}^T\right)^i:= \pi^{ij}\mathcal{O}_{j}:=
    \left(\delta^{ij}-\hat{p}^i\hat{p}^j\right)\mathcal{O}_j\,\mathcal{O}_j\,, ~~~~\left(\mathcal{O}^L\right)^i    := \hat{p}^i\hat{p}^j\,,
\end{array}
\label{eq: decomposition}
\end{equation}
For the symmetric spin-2 current $T^{ij}$, one similarly decomposes into $T^{ij}_{TT}$, $T^{ij}_{TL}$ and $T^{ij}_{LL}$. It is useful to further decompose the transverse part into the following,
\begin{equation}
    \begin{array}{rcl}
          \Big(T^{TT}\Big)^{ij}&=&\pi^{im}\pi^{jn}T_{mn}
= \Big(T^{\hat{T}\hat{T}}\Big)^{ij}+\Big(T^{\underline{TT}}\Big)^{ij}\,,\\[4mm]
\Big(T^{\hat{T}\hat{T}}\Big)^{ij}&:=&\hat{\Pi}^{ijmn}T_{mn}:=\left(\pi^{im}\pi^{jn}-\frac{1}{2}\pi^{ij}\pi^{mn}\right)T_{mn}\,, \\[2mm]
\Big(T^{\underline{TT}}\Big)^{ij}&:=&\underline{\Pi}^{ijmn}T_{mn}:=\frac{1}{2}\pi^{ij}\pi^{mn}T_{mn}\,,
\label{eq:Tdecompose}
    \end{array}
\end{equation}
~\\
~\\[-6mm]
where $\hat{\Pi}^{ijmn}$ is the projector onto the transverse-traceless part of $T_{mn}$, satisfying 
$\delta_{ij}\hat{\Pi}^{ijmn} = \delta_{mn}\hat{\Pi}^{ijmn} = 0$. 
Additionally, the projector $\underline{\Pi}^{ijmn}$ is defined such that it satisfies the trace condition 
$\delta_{ij}\underline{\Pi}^{ijmn} = \pi^{mn}$ and $\delta_{jm}\underline{\Pi}^{ijmn} = \frac{1}{2}\pi^{in}$. For WFCs involving spin-3/2 currents $\bm\psi^i$, it is also useful to further decompose the transverse projector into components that are orthogonal (parallel) to the gamma matrices,
\begin{equation}
\begin{array}{rcl}
\Big(\bpsi^{T}\Big)^i&=&\pi^{ij}\bpsi_{j}=\Big(\bpsi^{\hat{T}}\Big)^i+\Big(\bpsi^{\underline{T}}\Big)^i\,,\\[4mm]
\Big(\bpsi^{\hat{T}}\Big)^i&:=& \hat{\Pi}^{ij}\bpsi_j:=\left(\pi^{ij}-\frac{1}{2} \slashed{\pi}^{i}\slashed{\pi}^{j} \right)\bpsi_{j}\,,~~\slashed{\pi}^i:= \pi^{ij}\gamma_j\,,\\[2mm]
\Big(\bpsi^{\underline{T}}\Big)^i&:=&\underline{\Pi}^{ij}\,\bpsi_j
:=\frac{1}{2}   \slashed{\pi}^{i}\slashed{\pi}^{j} \bpsi_{j}\label{eq:gammadecompose}\,,
      \end{array}
\end{equation}
~\\
~\\[-6mm]
where the $\hat{\Pi}^{ij}$ is orthogonal to the gamma matrices ($\gamma_i\hat{\Pi}^{ij}=\hat{\Pi}^{ij}\gamma_j=0$), and we have called such component as \emph{transverse-gamma-traceless}  whereas $\underline{\Pi}^{ij}$ satisfies $\gamma_j\underline{\Pi}^{ji}=\underline{\Pi}^{ij}\gamma_{j}=\slashed{\pi}^{i}$.

\paragraph{Constrained Boundary Profile}

In flat space, the equations of motion for the trace part of spin-2 and 3/2 involve only  spatial derivatives and implies a constraint on the boundary value of the graviton $h_{ij,b}$ and the gravitino $\bm{\psi}_{i,b}$ (see appendix \ref{app: BoundarConstraint} for review):  
\begin{equation}
\label{eq:constrainedboundary}
\pi^{ij} h_{ij,b} = 0, \quad
\slashed{\pi}^{i} \bm{\psi}_{i,-,b} = 0, \quad
\bm{\bar{\psi}}_{i,+,b} \slashed{\pi}^{i}  = 0\,.
\end{equation}
The above relates the trace-part to the longitudinal components. However, the wavefunction coefficients are defined as an expansion in terms of \emph{unconstrained} boundary,  $h_{ij,\zero}$, $\bm{\psi}_{i,\zero}$, and $\bm{\bar{\psi}}_{i,\zero}$. The constrained boundary profiles are then obtained by acting with projectors:
\begin{equation}
\label{eq:constrainedboundary projector}
h^{ij}_b = P^{ijmn}_h \, h_{mn,\zero}, \quad
\bm{\psi}^{i}_{-, b} = P^{ij}_\psi \, \bm{\psi}_{j,\zero}, \quad
\bm{\bar{\psi}}^{i}_{+,b} = \bm{\bar{\psi}}_{j,\zero} \,
P^{ji}_\psi,
\end{equation}
where the projectors are defined as:
\begin{equation}
P^{ijmn}_h = \delta^{im} \delta^{jn} - \frac{1}{2} \pi^{ij} \pi^{mn}\,, \quad
P^{ij}_\psi = \delta^{ij} - \frac{1}{2} \slashed{\pi}^{i}\slashed{\pi}^{j} \,.\label{eq:constrainedP}
\end{equation}
It is straightforward to verify that Eq.~\eqref{eq:constrainedboundary projector} satisfies the constraints in Eq.~\eqref{eq:constrainedboundary}.

In summary, a spinning WFC is decomposed as \eqref{eq: decomposition}, \eqref{eq:Tdecompose}, \eqref{eq:gammadecompose} and dressed with constrained projectors \eqref{eq:constrainedP}. The general expression takes the form,
\begin{equation}
    \langle\mathcal{O}^{\{i_s\}}\dots\rangle=\left(P\cdot\sum_g\mathbb{P}_g\mathbb{A}^g\right)^{\{i_s\}}\,,
    \label{eq:generalPdecompose}
\end{equation}
where $\{i_s\}$ densely labels the spin-$s$ Lorentz indices (for spin-2 it contains 2 indices and for spin-3/2 it contains 1 since we are suppressing all spinor indices). The overall $P$ matrix denotes the constrained projector. The $g$ runs over various combinations of transverse, longitudinal and trace projectors denoted as $\mathbb{P}^g$. Thus the non-trivial information of the WFC is encoded in $\mathbb{A}^g$ which is the main subject of this paper. More explicitly, the decomposition of a single spin-2 or 3/2 operator in the WFC,
\begin{align}
\langle T_{ij,1} \dots \rangle &= P_{h,1,ij\,kl} \left[\pi^{km}_1 \pi^{ln}_1 \mathbb{A}^{TT}_{mn} 
+ \left( \pi^{km}_1 \hat{p}_1^l 
+ \pi^{lm}_1 \hat{p}_1^k 
+ \hat{p}_1^k \hat{p}_1^l \hat{p}_1^{m} \right) \mathbb{A}^L_{m} \right]\,, \nonumber\\
\langle \psi_{1,i}\dots \rangle &= \left( \pi_1^{kj} \mathbb{A}^T_j 
+ \hat{p}_1^{k} \hat{p}_1^j \mathbb{A}^L_j \right) P_{\psi,1,ki}\,, \nonumber\\
\langle \bar{\psi}_{1,i}\dots \rangle &= P_{\psi,1,i} \left( \pi_1^{kj} \mathbb{A}^T_j 
+ \hat{p}_1^{k} \hat{p}_1^j \mathbb{A}^L_j \right)\,.
\label{eq: graviton and gravitino decomp}
\end{align}
The matrices contracted in front of each $\mathbb{A}^g$ are the various $\mathbb{P}_g$ defined in \eqref{eq:generalPdecompose}, except that $\underline{\Pi}^{ij}$ and $\underline{\Pi}^{ijmn}$ vanish when contracted with $P_h$ and $P_\psi$, respectively. Similarly, $\Pi^{ijmn}_{\widehat{TT}}$ and $\Pi^{ij}_{\widehat{T}}$ reduce to the projectors $\pi$'s. One can check that the trace and $\gamma$-trace components of the WFCs are fully determined by the longitudinal components which is completely fixed by the Ward-Takahashi identity we discuss later in Section \ref{sec: WT identity}:
\begin{equation}
\label{eq: trace and gamma trace}
\langle T^{i}_{\ i} \rangle = \langle T^{LL} \rangle, \quad
\sigma^i \langle \psi_i \rangle = \slashed{\hat{p}} \langle \psi^L \rangle\,, \quad
\langle \bar{\psi}_i \rangle \sigma^i = \langle \bar{\psi}^L \rangle \slashed{\hat{p}}\,.
\end{equation}

For contracted WFCs, it will be useful to factor out the $P_h,P_\psi$ projectors and boundary profile and replace with auxiliary tensors as place holders. For simplicity we will use factorized form for these auxiliary polarization tensors:
\ie
\label{eq:reference bi-vector}
P^{iji'j'}_h h_{i'j',\zero}
\to
\epsilon^i \epsilon^j\,,~~~~ P^{ii'}_\psi \, \bpsi_{i',\zero}
\to
\epsilon^{i} \bchi\,,~~~~
 \bbarpsi_{i',\zero} {P}^{i'i}_\psi
\to
\bm{\bar\chi} \epsilon^{i} \,.
\fe
in which $\epsilon$ and $\bar\chi, \chi$ should satisfy the constraint \eqref{eq:constrainedboundary},
\ie
\label{eq:boundary reference polarization constraint}
\pi_{ij} \epsilon^i \epsilon^j  = 0\,, \quad~~
\epsilon^i \slashed{\pi}_i\, \bm \chi=0\,, \quad~~
\bm{\bar\chi} \,\slashed{\pi}_{i} \,\epsilon^i=0\,.
\fe
These constraints can be solved for $\epsilon, \bbarchi, \bchi$, yielding two linearly independent solutions for each equation. We remind the reader that $\epsilon$ and $\bchi$ are merely place holders and differ from $\epsilon_\partial$ and $\bchi_\partial$ which are true unconstrained boundary profiles for vectors and fermions.

\subsection{Ward-Takahashi Identity}
\label{sec: WT identity}

For systems with space-like boundaries, it is convenient to adopt the temporal gauge for the bulk gauge field. The residual gauge symmetry—characterized by transformation parameters that are independent of time—can then be identified with the boundary limit of the bulk gauge parameter. Furthermore, for spacetime gauge symmetries such as diffeomorphisms and local supersymmetry, bulk gauge transformations generate boundary contributions that are precisely canceled by the variation of the boundary action.\footnote{A discussion of this cancellation in Lorentizian AdS can be found in \cite{Papadimitriou:2017kzw}.} To derive the consequence of residual gauge symmetry on WFCs, one simply notes that the residual gauge symmetry of the boundary field can be inherited by a transformation of the boundary profile, 
\begin{equation}
\delta_\xi \varphi_{cl}(\varphi_{\partial})=\varphi_{cl}(\delta_\xi\varphi_{\partial})\,.
\end{equation}
Let us take scalar QED as an example, where the boundary profile transforms as:
\ie
\delta_\alpha \phi _{\zero}({x}) = -ie\alpha ({x}) \phi _{\zero}({x})\,, \quad \delta_\alpha \phi _{\zero}^{*} = ie\alpha ({x}) \phi _{\zero}({x})\,, \quad \delta_\alpha \epsilon _{i,\zero}({x}) = \partial _{i} \alpha ({x})\,.
\fe
Substitute into the vector version of the Schwinger Series \eqref{eq: perturbative solution},
\footnote{Here we use the notation:
	\ie
	\frac{\delta \mathcal{L}_{\text{int}}}{\delta \phi^{(0)}}
		:= 
		\left.\frac{\delta \mathcal{L}_{\text{int}}}{\delta \phi}\right|_{\phi^{(0)}, A^{(0)}_i}\,, \quad
	\frac{\delta \mathcal{L}_{\text{int}}}{\delta A_\mu^{(0)}}
		:=
		\left.
			\frac{\delta \mathcal{L}_{\text{int}}}{\delta A_\mu}
		\right|_{\phi^{(0)}, A^{(0)}_i}\,.
	\fe

}
\ie
[\delta A_{i,cl}]^{(0)} (  x, x_0)
&=
\int d^3 x' K_{A,i i'} (  x-  x', x_0)\p^{i'}_{x'} \alpha(x')\,, \\
 [\delta A_{i,cl}]^{(n+1)} (  x, x_0)
&=
e \int d^4 x' G_{A,i \mu}(  x-  x', x_0, x_0') \delta\left[ \frac{\delta \mathcal{L}_{\text{int}}}{\delta A^{(n)}_\mu}(  x',x_0') \right], \quad n>0\,.
\fe
For QED, owing to the abelian nature of the gauge transformation,  
$
\delta_\alpha \!\left[\frac{\delta \mathcal{L}_{\text{int}}}{\delta A^{(n)}_\mu}(x',x_0') \right] = 0 .
$
Consequently, the boundary variation does not contribute to higher orders in the vector Schwinger series. 
The remaining contribution therefore reduces to the zeroth-order term. 
By explicitly inserting the bulk-to-boundary propagator of the vector field, we obtain
\ie
\delta A_{i,cl}=[\delta A_{i,cl}]^{(0)} (  x, x_0)
=
i\int \frac{d^3 p}{(2\pi)^3}e^{i   p \cdot   x} (\pi_{ij,p}e^{i E_p x_0} + \hat{p}_i \hat {p}_{i'}) p^{i'} \alpha(  p)
=
\partial_i \alpha(  x) \,.
\fe
The resulting transformation is indeed a residual gauge transformation on the bulk classical solution for the temporal gauge $A_{cl,0}=0$, where the gauge parameter is constrained by $\p_0 \alpha=0$, making it a spatial function $\alpha=\alpha({x})$. We leave the discussion for the scalar field to appendix \ref{app:WTofWFC}.

The invariance of the classical action now becomes, after fourier transform, 
\begin{equation}\label{eq: sclvariation}
\delta_\xi S[\varphi_c]=\sum_{n=2}^\infty \int \frac{\prod_i d^3k_i }{(2\pi)^{3n}} \, \delta^{(3)}\left(\sum_{i} k_i\right)\,\left(\sum_j \varphi_{\zero,1} \cdots \delta_\xi \varphi_{\zero,j}\cdots\varphi_{\zero,n} \right)\, c_n=0\,. 
\end{equation}
Now note that the transformation of the boundary gauge field is given by $p_i \xi$, whereas the transformation of the gauged fields should begin from first order in the coupling constant and contributes only at the next orders. Thus, at each order in the coupling-constant expansion of $\delta_\xi S[\varphi_c]$, the longitudinal component of the WFC is related to lower-point WFCs. This relation is precisely what we refer to as the Ward--Takahashi identity.
\footnote{This is, of course, the standard Ward--Takahashi identity for correlation functions. The reasoning here is slightly different, however, since we are working directly with WFCs, without assuming the existence of a boundary theory whose correlators are identified with them.}

If the symmetry is abelian, the WT identity directly relates the longitudinal part of a conserved current in $c_n$ with $c_{n-1}$. For non-abelian transformations, however, the identity also involves WFCs of even lower multiplicity. For instance, consider the case of diffeomorphisms acting on the graviton and scalar fields,
\begin{equation}
    \delta h_{ij,b}=2\ \partial _{( i} \xi _{j)} -2\kappa\xi^{m} \partial_{( i} h_{j) m,b} +\kappa\xi ^{m} \partial _{m} h_{ij,b} \,,\quad\delta\phi_{\zero}=-\kappa \xi^i\partial_i\phi_{\zero}-\kappa^2 h^{ij}_{b}\xi_i\partial_j\phi_{\zero}\,,
\end{equation}
in which the $\kappa$ is the Einstein gravitational constant.
The WT identity for $\langle TOO\rangle$ involves $\langle OO\rangle$ whereas $\langle TOTO\rangle$ iinvolves $\langle TOO\rangle$ as well as $\langle OO\rangle$,
\ie
&p_{1i}\xi_{1j}\langle T^{ij}\co T\co\rangle
\\
=&-\frac{\kappa}{2}\Big(p_{2}\cdot\xi_{1}\langle\co_{1+2}T_3\co_4\rangle+p_4\cdot\xi_1\langle\co_{1+4}T_3\co_2\rangle+p_3\cdot\xi_1\langle\co_{1+3}T_3\co_2\rangle\Big)
+
\kappa\left(\xi_1\cdot\epsilon_3\right)p_{3i}\langle T^i_{1+3} O_2O_4\rangle
\\
&+\frac{\kappa^2}{2}\left((p_{2}\cdot\epsilon_{3})(\xi_1\cdot\epsilon_3)\langle\co_{1+2+3}\co_4\rangle+(p_{4}\cdot\epsilon_{3})(\xi_1\cdot\epsilon_3)\langle\co_{1+3+4}\co_2\rangle\right),
\fe
where the first line arises from the variation of $h_\partial,\phi_\zero$ at first order in $\kappa$, while the second line originates from the variation of $\phi_\zero$ at second order in $\kappa$. For convenience, we shall henceforth set $\kappa=1$ in the remainder of the discussion.



\paragraph{Locality}

The left-hand side of the Ward-Takahashi (WT) identity depends only on momentum and WFCs, and therefore must remain free of any \emph{pure} singularities in the external energies. This imposes a locality constraint on the explicit form of the RHS of the WT identity:
\begin{equation}
\lim_{p_1 \to 0,\, E_{\text{phys}} \neq 0}  \left( \frac{1}{E_1} \cdot p_1^i \langle \mathcal{J}_{i} \, O O \dots \rangle \right) < \infty \label{eq:WT locality}\,.
\end{equation}
Here, the term \emph{pure} indicates that the limit $p_1 \to 0$ should be taken while keeping all physical poles, denoted $E_{\text{phys}}$ (such as total or partial energy poles), \emph{nonzero}. This condition will guarantee that the longitudinal component $\mathbb{A}_L$ of the WFC, as determined by the WT identity (see Section~\ref{sec:bootstrapping}), remains regular in the external energy. Nevertheless, because the energies of different legs aren't independent variable, they're realted by momentum conservation, we need to take a special care to see the locality indeed satisfied for explicit WT identities.

For example, let us consider the massless QED contact example (the explicit WT identity is given in Appendix \ref{app:WTofWFC}):
\ie
\lim_{p_1 \to 0} \frac{1}{E_1} \cdot p_1^i\langle J_{i} \, \bar\chi \chi \rangle
&= \left( \frac{(E_2 - E_3)\bar\chi_{2,\zero}\vecslashed{p}_2\chi_{3,\zero}}{E_1 E_2 E_3} \right)\,.
\label{WTSpurious}
\fe
At first glance, this expression appears to be singular when $p_1 = 0$. However, this apparent singularity is resolved kinematically: in the limit $p_1 \to 0$, momentum conservation enforces ${p}_{2} \to -{p}_{3}$, so $E_2^2 - E_3^2 \to 0$. This implies that either $E_2 - E_3 \to 0$ or $E_2 + E_3 \to 0$. If $E_2 - E_3 \to 0$, the limit is manifestly finite. In the case $E_2 + E_3 \to 0$, the limit appears divergent; however, in this situation, as $E_1 \to 0$ together with $E_2 + E_3 \to 0$, the total energy $E_T \to 0$, which is precisely the situation we exclude in the locality statement.

\section{Analytic properties in energy variables}
\label{sec:Analytic properties in energy variables}
In this section, we review the analytic properties of tree-level WFCs that reveal their origin in a local bulk theory. These properties manifest as singularities and discontinuities in variables conjugate to the bulk time coordinate, namely the energy variables. Most of the features discussed here rely on the existence of a perturbative bulk description, although certain aspects—such as singularities at total energy—are expected to persist even at the non-perturbative level.

\subsection{Total energy pole}
\label{sec:total-energy-pole}
WFCs are rational functions of energy and momentum subject to spatial momentum conservation. Upon analytic continuation to the configuration where the total energy vanishes, these functions acquire support on full four-dimensional momentum conservation. In this regime, energy conservation is  restored, time-translation invariance emerges, and boundary contributions effectively disappear. It is thus natural to expect a direct correspondence between WFCs and the flat-space S-matrix. To establish this relation concretely, we work in perturbation theory and study the Feynman diagram representation of the WFCs.

\paragraph{Contact Diagrams}

We begin with contact diagrams by focusing on the time integral of the scalar Feynman rule \eqref{contact Fey structure}. For scalars, bulk-to-boundary propagator is simply $e^{iEx_0}$ and hence the contact diagrams yield:
\ie
c_{n,\phi,\text{contact}}=
\sum_{\text{perm}} \int^0_{-\infty_-} dx_0\ (ig)\, V({p}_1,{p}_2,\dots,{p}_n)\, e^{iE_T x_0}
=
\frac{gV}{E_T}\,.
\fe
in which we define the total energy $E_T:=\sum_{i=1}^n\,E_{i}$. 
This is that well known total energy pole with the flat-space amplitude $gV$ as the residue. 

The same argument applies to spinning fields, as the bulk-to-boundary propagators retain the same $e^{iEx_0}$ factor. An important subtlety in making the connection between WFC to flat-space S-matrix is how the boundary profiles are mapped into spinor/polarization wavefunctions. Let's first consider spinors. The bulk-to-boundary propagators \eqref{eq:fermionpropagtor} take the form of exponentials of the energy multiplied by $(1\pm i\slashed{p})$. When a pair of spinors interacts with scalars, the contact diagram yields
\ie
c_{n,\chi,\text{contact}}=
\sum_{\text{perm}} \int^0_{-\infty_-} dx_0\ (ig)\,
\bbarchi_\zero (1+i\slashed{\hat{p}}_1) 
V({p}_1,{p}_2,\dots,{p}_n)
(1-i\slashed{\hat{p}}_2)
\bchi_\zero
\, e^{iE_T x_0}
=
\frac{g\bar u V u}{E_T}\,.
\fe
Note that in the above, the boundary profile is combined  with $(1\pm i\slashed{p})$ into the polarization spinors $u,\bar u$:
\ie
\label{eq: fermion polarization}
u =
(1 - i\vecslashed{\hat{p}})\bm{\chi}_{\zero}\,, \quad
\bar u =
\bm{\bar\chi}_{\zero}(1 + i\vecslashed{\hat{p}})\,.
\fe
Indeed one can check that $u,\bar u$ solve the positive-energy massless Dirac equation. For conserved currents, since the WT identity relates the longitudinal modes to lower-point functions, it cannot depend on the $n$ energies independently and thus there are no $E_T$ singularities. Let us consider the transverse polarizations $\epsilon^T\equiv \epsilon \cdot \pi$.  To see that the residue of $E_T$ can directly yield contact contributions to amplitude we simply note that 
\ie
\label{eq:ampdecompose}
{\epsilon}^T \cdot {V}
= (\epsilon\cdot V) -(\epsilon\cdot \hat{p})(\hat{p}\cdot {V}) 
=\epsilon_{\mu} V^\mu +\frac{p_\mu V^\mu}{E}  \,,
\fe
where in the last equality we've used that the polarization vectors for the amplitude satisfies $\epsilon_\mu p^\mu=0$. The second term in the last equality will cancel against the exchange part of the amplitude and hence can be dropped.

	The correspondence can be straightforwardly applied to massive field. For vectors, the WT identity no longer removes the longitudinal mode at the total energy pole. The temporal gauge cannot be imposed on the classical solution. Instead, the condition $p^\mu \epsilon_\mu = 0$ gives $\epsilon_{0} = \frac{p^i \epsilon_{i}}{\sqrt{p^2 + m^2}}$. This temporal component appears both in the total energy pole residue and in the amplitude, enabling a direct mapping of polarization structures:
	\ie
	\epsilon_{\mu}
	= \left(\frac{p^i\epsilon_{i,\zero}}{\sqrt{p^2+m^2}} ,\epsilon_{i,\zero}\right), \quad
	u=
	\left(1 - \frac{i\vecslashed{p}}{E-m}\right)\bm{\chi}_{\zero}, \quad
	\bar u =
	\bm{\bar\chi}_{\zero}\left(1 + \frac{i\vecslashed{p}}{E-m}\right).
	\fe

As we will be interested in scenarios where the amplitude limit involves Majorana fermions, we will need boundary profiles that reflects this fact. From \eqref{eq: fermion polarization}, we find:
\begin{equation}
	\label{eq:Majorana Condition on placeholder spinor}
	\bbarchi_{\zero}(p)=\bchi^{\text{T}}_{\zero}(p) C_{-}\,,
\end{equation}
where we use $\text{T}$ to denote the transpose and the charge conjugation operator is defined as 
$C_{-} = \gamma_2 \gamma_{0}$. Note that CPT invariance imposes non-trivial constraint on fermionic WFCs. As we will not use these constraints for our bootsrap program, we refer  interested readers to appendix~\ref{app:CPT} for details.

\paragraph{Exchange Diagrams}
In each exchange channel, the left and right total energies are denoted by 
$E^e_{L/R}$, where $e \in \{s, t, u\}$ labels the specific channel. 
The internal 3-momentum sums on the right side vertex for each channel are defined as\,,
\begin{equation}
p_{s}^i=p_3^i+p_4^i\,,~~~~p_t^i=p_1^i+p_4^i\,,~~~~p_u^i=p_2^i+p_4^i\,,
\end{equation}
with the corresponding internal energies given by\,,
\begin{equation}
E_s=|p_s|=|p_3+p_4|\,,~~E_t=|p_t|=|p_1+p_4|\,,~~E_u=|p_u|=|p_2+p_4|\,.    
\end{equation}
Taking the $s$-channel as an example, the right and left total energies correspond to  $E_{R}^s=E_{34s}$ and $E_{L}^s=E_{12s}$, respectively.
We now extend our analysis to exchange diagrams using four-points as the primary example. Firstly, for scalar exchanges it is straightforward to integrate out the two time integrals in the Feynman rule in a given channel ($s$) in \eqref{Exchagne Fey structure} with the scalar propagators \eqref{eq:scalarpropagtor}, 
\ie
& c_{4,s,\phi}=
\frac{g^2 V_L V_R}{E_{L}^s E_{R}^sE_T}\,,
\fe
where we define $V_L:=V({p}_1,{p}_2,{p}_s),V_R:=V({p}_3,{p}_4,-{p}_s)$ and the partial energy pole whose residue we'll discuss in the next next section.  It is easy to see that
\ie
 E_{L}^sE_{R}^s|_{E_T\to 0}&=(-E_{34}^2+E_s^2)=: S\,,\\   E_{L}^tE_{R}^t|_{E_T\to 0}&=(-E_{14}^2+E_t^2)=: T\,,\\
 E_{L}^u E_{R}^u|_{E_T\to 0}&=(-E_{24}^2+E_u^2)=: U\,.
\fe
Combining all channels, along with the contact terms, we have
\ie
\label{eq:total energy pole scalar exchange diagram}
c_4|_{E_T \to 0} 
&= \left.\left(
	\sum_{e \in s,t,u} \frac{g^2 V_{L,e} V_{R,e}}{E_{L}^e E_{R}^eE_T}
	+\frac{g^2 V_c}{E_T}
\right)\right|_{E_T\to0}\\
&= \frac{g^2}{E_T}\left(
	\frac{V_{L,s} V_{R,s}}{S}
	+\frac{V_{L,t} V_{R,t}}{T}
	+\frac{V_{L,u} V_{R,u}}{U}
	+V_c
\right)\,,
\fe
thus confirming that the total-energy pole residue exactly reproduces the amplitude. For external fields with spin, the promotion of the boundary profile into four-dimensional external line factors are identical to the contact diagram.

For internal spinning fields the time dependence of the bulk-to-bulk propagator is more complicated. For example for vectors: 
\ie
\label{eq: vector bulk to bulk}
G_{A,i}^\mu(E_s,t,t') &= \pi_{s,ij} \eta^{j\mu} G_{\phi}(E_s,t,t') + \frac{p_i}{p_s} \eta^{0\mu}  \theta(t'-t)\,.
\fe
The first term on the right, being proportional to the scalar propagator, will yield a total energy pole upon integration, whose residue will be proportional to $1/S$. However the $\pi_{s,ij}$ prefator differs from the standard numerator of Feynman propagators. Furthermore the second term also yields none-trivial total energy singularity as well. To demonstrate the presence of total energy poles and how Feynman propagators emerge, let us study the time integrals in detail.

To start, redefinning $x_0 = \frac{1}{2}(\tau + \delta)$ and $x_0' = \frac{1}{2}(\tau - \delta)$ the general time integral takes the form
\ie
\label{eq:exchange diagram integral in tau delta}
c_{4,s} = \frac{1}{2} \int^{0}_{-\infty_-} d\tau\, e^{i E_T \tau/2} \int^{\infty_-}_{-\infty_-}  d\delta\, (A(\delta) + B(\delta)e^{iE_s \tau})\,,
\fe
where $A(\delta)$ and $B(\delta)$ are determined by the vertices and the propagators, with $\tau$-dependence appearing only in $e^{iE_s\tau}$. We're interested in the limit $E_T\rightarrow 0$, where the integral behaves as 
\ie
\label{eq: total energy pole int prop }
\lim_{E_T \rightarrow 0} \int^a_{-\infty_-} e^{i E_T \tau} (A(\delta) + B(\delta)e^{iE_s \tau}) \; d \tau = \frac{A(\delta)}{i E_T}{+}\mathcal{O}(E^0_T)\,.
\fe
We've kept the upper bound of the time integral unfixed to demonstrate that the leading $E_T$ behaviour is insensitive to the boundary. Thus, to extract the residue of the total-energy pole, we can simply take the  $a \rightarrow -\infty_- +\epsilon$ as a upper bound. That is, the residue of the total energy pole is controlled by the physics of the far-past.  

Let us now consider the boundary conditions of bulk-to-bulk propagators, i.e. \eqref{eqn: conditions on G}, in the far-past region.  In $(\tau,\delta)$ variables they read
\ie
\label{eq: bulk to bulk in tau delta}
\mathcal{D}_{\delta}G(p_s,\tau,\delta) = \delta(\delta)\,, \qquad
G(p_s,\tau,\delta=-\tau)=0\,, \qquad
G(p_s,\tau,\delta=\tau-\infty_-)=0\,.
\fe
By comparison, the Feynman propagator, which is translation invariant and depends only on $\delta$, satisfies
\ie
\label{eq: Feynman propagator in tau delta}
\mathcal{D}_{\delta}G_{\text{Fey}}(p_s,\delta)=\delta(\delta)\,, \qquad
G_{\text{Fey}}(p_s,\delta=\infty_-)=0\,, \qquad
G_{\text{Fey}}(p_s,\delta=-\infty_-)=0\,.
\fe
One finds that \eqref{eq: Feynman propagator in tau delta} is precisely the $\tau \to -\infty_-$ limit of \eqref{eq: bulk to bulk in tau delta}. Hence,
\ie
\label{eq: bulk to bulk limit}
\lim_{\tau \to -\infty_-} G(p_s,\tau,\delta) = G_{\text{Fey}}(p_s,\delta)\,.
\fe
It is straightforward to check that the scalar bulk-to-bulk propagator in \eqref{eq:scalarpropagtor} indeed satisfies \eqref{eq: bulk to bulk limit}. Thus we see that for the residue of total energy pole, the bulk-to-bulk propagator becomes the Feynman propagator. This of course applies to the scalar exchange which we began with.


\subsection{Cutting Rules and Partial Energy Poles}
\label{sec:cutting rules}

As functions of energy and spatial momentum \((E, p)\), the WFCs naturally inherit branch cuts originating from the dispersion relation \(E = \sqrt{p^2 + m^2}\). The associated discontinuity is obtained by taking the difference under the exchange \(E \leftrightarrow -E\), which we will use the shorthand notation:
\begin{equation}
\disc_{E}  f(E)\equiv f(E)-f(-E)\,.
\end{equation}
Since the internal energies (denoted as $E_I$)  appear exclusively through the bulk-to-bulk propagator, taking the discontinuity in $E_I$ allows one to exploit analytic properties of the propagator which are agnostic to the details of the interaction, and reflects the nature of time evolution in the bulk. These universal features of the WFCs generally referred to as \emph{cutting rules}. 

Indeed originally it was shown that unitarity of time evolution operator $U$, i.e. \( UU^\dagger = 1 \),  yields relations amongst complex-conjugated and (external) energy-flipped de-Sitter WFCs~\cite{Goodhew:2020hob}. Such ``Cosomological Optical Theorem (COT)" for four-scalars are written as,
\ie\label{eq: COT}
c^s_4+c^{s,*}_4(-E_{e},E_{s},p)=(c_{3,L}-c_{3,L}(-E_s,p)) \frac{1}{2c_2(E_s)} (c_{3,R}-c_{3,R}(-E_s,p))\,,
\fe
where $C_2(E_s)$ is the two-point function and depends on the spin of the exchanged state, 
\ie
C_{2,\phi}(E_s) = 1,\quad 
C_{2,J}^{i_1 j_1}(E_s)= \pi_{I}^{i_1 j_1}\,, \quad 
C_{2,T}^{i_1 i_2 j_1 j_2}(E_s)= \Pi_{(2,2),I}^{i_1 i_2 j_1 j_2}\,.
\fe
These results can be derived more directly from the analytic properties and relations of the propagators. In particular, considering the discontinuity in particular internal energies, which allows one to zoom in on factorization in particular channels, the discontinuity of bulk-to-bulk propagators factorize into the product of that of the bulk-to-boundary propagators~\cite{Meltzer:2021zin}.
\ie
\label{eq:disc of boson G : Meltzer}
\disc_{E_s}  G(E_s,t,t') = \disc_{E_s} K(E_s,t) \cdot \left(-\frac{i}{2c_2(E_s)}\right) \cdot \disc_{E_s} K(E_s,t')\,.
\fe
For correlators, this implies
\ie
\disc_{E_s} c_4 
=\disc_{E_s} c_{3,L } \cdot \frac{1}{2 c_2(E_s)} \cdot \disc_{E_s} c_{3, R}\,.
\fe
Similar relations can be extended to external conserved higher-spins~\cite{Baumann:2021fxj} and to multi-cuts at tree and loop-level~\cite{Melville:2021lst}. 
Note that the cutting rules above are slightly different than the original relations derived from COT in eq.(\ref{eq: COT}). Their equivalence is a consequence of CPT invariance of the WFC, which at tree-level takes the form 
\cite{Goodhew:2024eup} (7.62),
\ie
c_{n} = (-1)^{4L-1} c_{n}^{*}(\{-E_e\},\{-E_I\},\{-p\})\,.
\fe
We will derive the fermionic version of the CPT theorem in the App. \ref{app:CPT}.

In this subsection, we derive the cutting rules for fermion WFCs and extract their implications. In particular, the residues of partial energy poles.

\noindent\textbf{Tree-level Cutting Rules:}
Let us begin by the discontinuity of bulk-to-bulk propagator:
\ie
\label{eq:disc of boson G}
\disc_{E_I}  G(E_I,t,t') = \disc_{E_I} K(E_I,t) \cdot \left(-\frac{i\,C_2(  p_I)}{2E_I}\right) \cdot \disc_{E_I} K(E_I,t')\,.
\fe
Surprisingly, even though the massless spinor bulk-to-boundary and bulk-to-bulk propagators listed in \eqref{eq:fermionpropagtor} are not proportional to the scalar one, the equation \eqref{eq:disc of boson G} remains valid with the factor 
\ie
\label{C of the spinor}
{C}_{2,\chi}(p_I) &= \Pi_- \cdot i\vecslashed{p}_{I} \cdot \Pi_+\,.
\fe
Similarly, for the gravitino, as in the case of spinning bosons, the factor is simply the spinor one dressed with the transverse gamma-traceless projector:
\ie
{C}_{2,\psi}^{ij}(p_I)  &= \Pi_- \cdot i\hat{\Pi}^{ij}_{I}\vecslashed{p}_{I} \cdot \Pi_+\,.
\fe

Substituting the discontinuity of the propagator into the exchange diagram, we arrive at the cutting rules for the generic fields: \cite{Baumann:2021fxj}
\ie
	\label{eqn: cutting rule}
	\disc_{E_s} c_4(E_s,p_{1\sim 4}) 
	=
 \disc_{E_s} c_{3, i_1 \dots }(p_{1},p_{2},p_{s}) \cdot \frac{C_{2}^{i_1 \dots j_1 \dots}(  p_s)}{2E_s} \cdot \disc_{E_s} c_{3, j_1 \dots}(-p_{s},p_{3},p_{4})\,,
\fe
Note that $c_{3,i_1 \dots}$ denotes the WFC with all internal boundary polarizations stripped off. The result above immediately implies that the longitudinal pieces, along with (gamma) trace parts under \eqref{eq: trace and gamma trace}, of conserved currents do not contribute to the cutting rules. Once again this is because they are given in terms of lower-point WFCs that do not depend on $E_I$. 

The cutting rule can also be expressed in the spinor-helicity formalism introduced in App.~\ref{app: conventions}. Writing the spinning field $C_{2}$ in terms of the three-dimensional spinors $\lambda_s$ and $\bar\lambda_s$, with vector indices converted into spinor indices using sigma matrices, is equivalent to decomposing it into products of opposite-helicity polarizations. For the conserved vector and stress tensor, for example, one finds \cite{Baumann:2021fxj},
\ie
&C_{2,J,aa'}^{bb'} = \epsilon^{(+)}_{s,aa'} \epsilon^{(-),bb'}_s+ \epsilon^{(-)}_{s,aa'} \epsilon^{(+),bb'}_s, \\
&C_{2,T,a a'cc'}^{bb'dd'} = \epsilon^{(+)}_{s,aa'} \epsilon^{(+)}_{s,bb'}\epsilon^{(-),cc'}_s \epsilon^{(-),dd'}_s+ \epsilon^{(-)}_{s,aa'} \epsilon^{(-)}_{s,bb'} \epsilon^{(+),cc'}_s \epsilon^{(+),dd'}_s
\fe
For fermions, analogous decompositions can be written directly:
\ie
&C_{2,\chi,a}^{b} = \chi^{(+)}_{s,a} \bar\chi^{(-),b}_s+ \chi^{(-)}_{s,a} \bar\chi^{(+),b}_s, \\
&C_{2,\psi,acc'}^{bdd'} = \chi^{(+)}_{s,a} \epsilon^{(+)}_{s,cc'} \bar\chi^{(-),b}_s \epsilon^{(-),dd'}_s+ \chi^{(-)}_{s,a} \epsilon^{(-)}_{s,cc'} \bar\chi^{(+),b}_s \epsilon^{(+),dd'}_s
\fe
Here the helicity spinor polarizations are defined by $\bar \chi^{(+),a}=\frac{\bar\lambda^a}{\sqrt{2}}$, $\bar\chi^{(-),a}=\frac{\lambda^a}{\sqrt{2}}$, $\chi^{(+)}_a=\frac{\bar\lambda_a}{\sqrt{2}}$, and $\chi^{(-)}_a=\frac{\lambda_a}{\sqrt{2}}$. Contracting these decompositions with the stripped WFC coefficients $c_{i_1,\dots}$, which project onto the different internal-helicity sectors $c^{(\pm)}$, then gives the simpler helicity-space relation
\ie
	\disc_{E_s} c_4(E_s,p_{1\sim 4}) 
	=
	 \frac{1}{2E_s} 
 \sum_{h \in \pm} \disc_{E_s} c_{3}^{(h)}(p_{1},p_{2},p_{s})  \disc_{E_s} c_{3}^{(-h)}(-p_{s},p_{3},p_{4})\,.
\fe
where the external helicity labels have been suppressed. Importantly, the helicity projection is performed only after taking the discontinuity:
\footnote{
	The interplay between the discontinuity and the helicity decomposition is made explicit by writing
	\ie
	\disc_{E_s} c^{(h)}:= \left. c^{(h)}(E_s)-\left(c^{(-h)}(E_s) \right|_{\lambda_s \leftrightarrow \bar\lambda_s, E_s \leftrightarrow -E_s} \right) 
	\fe
	Thus, the discontinuity in a fixed helicity sector is in fact a mixed-helicity combination. Consequently, the vanishing of a helicity component does not by itself imply that its discontinuity vanishes. For example, although $\langle J^+ \bar \chi^+ \chi^+ \rangle=0$, the component $\langle J^+ \bar \chi^+ \chi^- \rangle$ entering the corresponding mixed-helicity combination need not vanish.}
\ie
\disc_{E_s} c^{(h)}:=(\disc_{E_s} c)|_{\epsilon \to \epsilon^{(h)}, \bar\chi \to \bar\chi^{(h)},  \chi \to \chi^{(h)}}.
\fe

\noindent\textbf{Loop Diagram Cutting Rules:}

\begin{figure}[t]
	\centering	\vspace{-1cm}\includegraphics[width=0.8\textwidth]{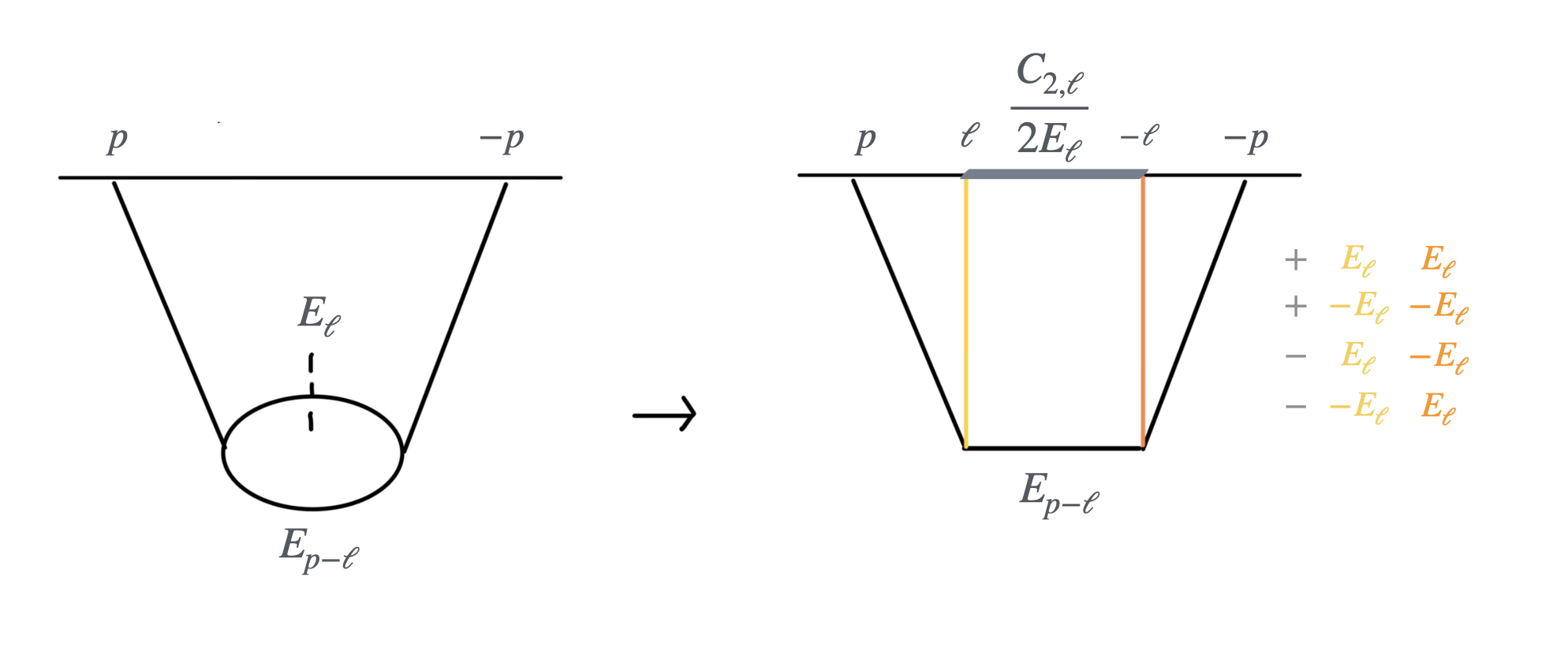}
	\vspace{-1cm}\caption{The cutting rule for the one-loop two-propagator WFC with two external legs. The internal leg is cut into two external legs with energies color-coded to match \eqref{eq: LoopCutting1}.}
	\label{fig:1L2G}
\end{figure}

The cutting rules derived above can be straightforwardly applied to loop-WFCs.  As an example, let us consider the one-loop diagram with two bulk-to-bulk propagators with two external legs. The corresponding Feynman rule is given by: 
\ie
&\int d^3 \ell  \; {c}^{\rm 1{-}loop}_{2}(\ell)
= 
\sum_{perm}\int d^3 \ell \int dt \int dt' \ g^2  K(  p,t) V_L(  p,  p{-}\ell,   \ell,\partial_t) \\
&\cdot G(  p{-}\ell,E_{p{-}\ell},t,t') \cdot G(  \ell,E_{\ell},t,t') \cdot V_R({-}p,{-}  p{+}\ell, {-}\ell,\partial_t') K({-}p,t')\,.
\fe
If we consider the discontinuity in $E_\ell$, only one of the bulk-to-bulk propagators are cut, and we have:
\footnote{
	Here we focus on a single-cut relation for the loop WFC, rather than the multiple-cut relations discussed in \cite{Melville:2021lst}. This choice allows for a direct derivation for internal fermions, whereas extending the derivation to double and higher cuts is less straightforward.
}
\ie\label{eq: LoopCutting1}
&\disc_{E_\ell} {c}^{\rm 1{-}loop}_2 (E_\ell,E_{\ell{-}s})=
\left[
\begin{aligned}
	&c^{\rm tree}_{ i_1 j_1 \dots }(E_{\ell},E_{\ell}) + c^{\rm tree}_{ i_1 j_1 \dots }({-}E_\ell,{-}E_\ell)\\
	& -c^{\rm tree}_{i_1 j_1 \dots }(E_\ell,{-}E_\ell) -c^{\rm tree}_{ i_1 j_1 \dots }({-}E_\ell,E_\ell)
\end{aligned}
\right]\cdot \frac{C_{2}^{ i_1 \dots j_1 \dots}(  p_\ell)}{2E_\ell}\,,
\fe
where $c^{\rm tree}$ is the tree-level WFC obtained from cutting open the loop,  in which the internal leg with momentum $p_\ell$ is cut into two external legs with energies $E_{R,\ell}$ and $E_{L,\ell}$ on the right/left side denoted as $c^{\rm tree}(E_{L,\ell},E_{R,\ell})$. The cutting rule could be diagrammatically shown as Fig. \ref{fig:1L2G}.

As a test of the cutting rule above, let us consider $\phi^3$ theory.  Although we use this theory as a simple illustration, the cutting rule is expected to apply to loop WFCs more generally. The one-loop two-point function is given by \cite{Anninos:2017eib}
\ie
{c}^{\rm 1-loop}_2= -\frac{1}{4E_p\cdot(E_p+E_\ell+E_{p-\ell})^2}\left( \frac{1}{E_p+E_\ell} + \frac{1}{E_p+E_{p-\ell}} \right)
\,.\fe 
The cut in $E_\ell$ is given as:
\ie
 -\frac{E_\ell \left(5E_p^2{-}E_\ell^2 {+} 4E_p E_{p-\ell} {+} E_{p-\ell}^2\right)}{2E_p(E_p {-}E_\ell)(E_p {+}E_\ell)(E_p{-} E_\ell {+} E_{p-\ell})^2(E_p{+} E_\ell {+} E_{p-\ell})^2}\,.\\
\fe
This matches eq.(\ref{eq: LoopCutting1}) if one identifies, $C_{2,\phi}= 1$ and 
\begin{equation}
c^{\rm tree}_4(E_{R,\ell},E_{L,\ell})=\frac{g^2}{(2E_p{+}E_{R,\ell}{+}E_{L,\ell})(E_p{+}E_{p-\ell}{+}E_{L,\ell})(E_p{+}E_{p-\ell}{+}E_{R,\ell})}\,.
\end{equation}

\begin{figure}[t]
	\centering    \vspace{-1cm}\includegraphics[width=0.8\textwidth]{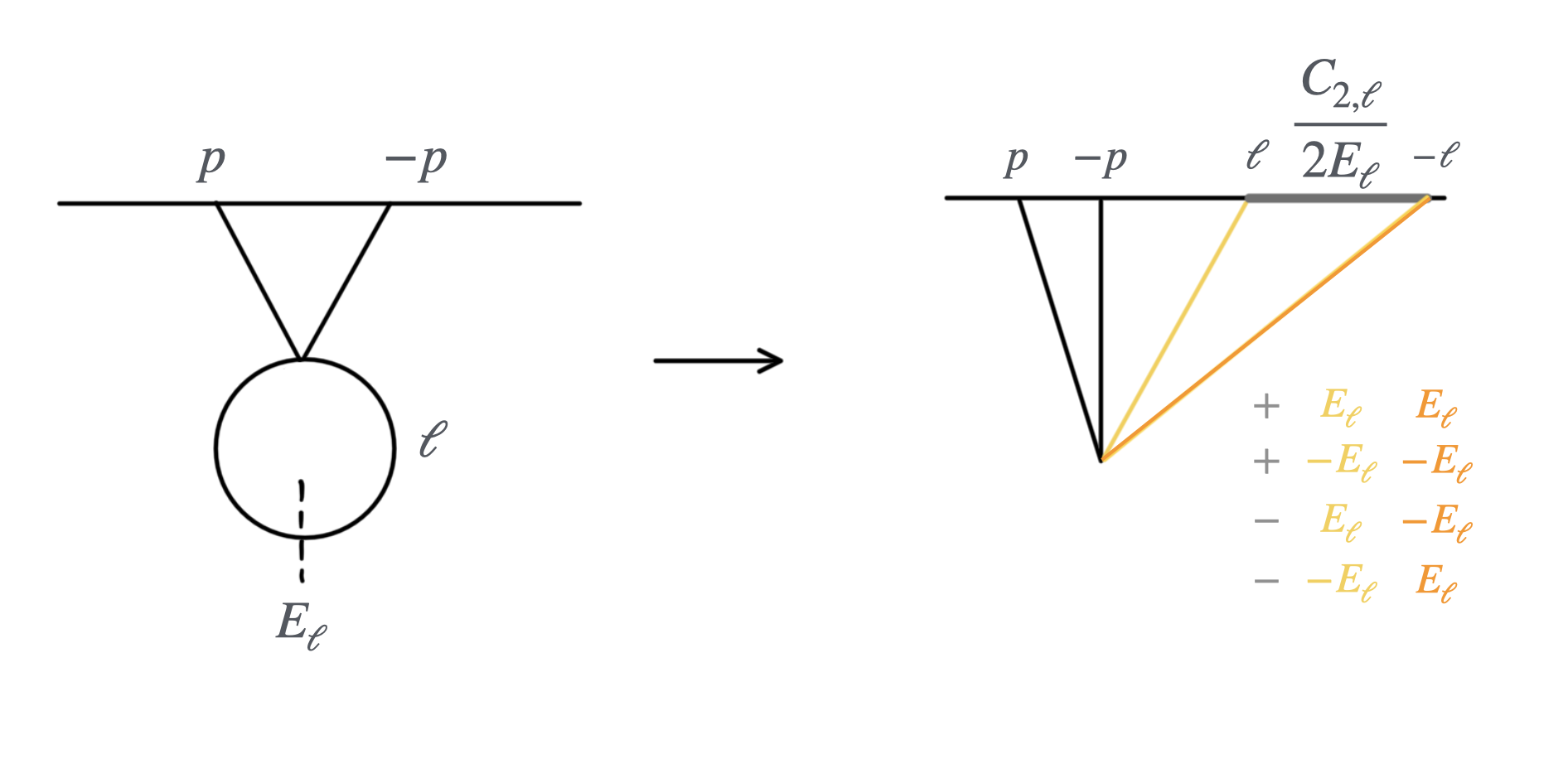}
	\vspace{-1cm}\caption{The cutting rule of the 1-loop 1-propagator WFC with two external legs. The internal leg is cut into two external legs with energies color-coded to match \eqref{eq: LoopCutting1}.}
	\label{fig:1L1G}
\end{figure}

As another example, consider the two-point tadpole diagram in Fig.~\ref{fig:1L1G}. Its Feynman rule is
\ie
\label{eq: Loop WFC 2}
\int d^3 \ell \; {c}^{\rm 1-loop}_{2}(\ell)
= 
\sum_{perm}\int d^3 \ell \int dt \ g  K(  p,t) K(-  p,t) V(  p,  \ell,\partial_t) \cdot G(  \ell ,E_{\ell},t,t)\,,
\fe
where we define $\theta(0)=\tfrac{1}{2}$. For example, the scalar case takes the form
\ie
G_\phi(p_\ell,E_{\ell},t,t) 
= \frac{i}{2E_\ell}\Big(K_\phi(E_\ell,t)K_\phi(-E_\ell,t)-K_\phi^2(E_\ell,t)\Big)\,.
\fe
One may check that the discontinuity of equal-time propagators still obeys the relation \eqref{eq:disc of boson G} at $t'=t$. For $\phi^4$ theory, we have
\ie
{c}^{\rm 1-loop}_{2}(\ell)=\frac{g}{4 E_p (E_\ell+E_p)}\,.
\fe 
The discontinuity in the internal energy $E_\ell$ of the tadpole WFC integrand $c_{2}$ reproduces \eqref{eq: LoopCutting1}, where now $c^{\rm tree}$ is given by the contact diagrams as illustrated in the right of fig.~\ref{fig:1L1G}. Indeed the cut in $E_\ell$ is given as:
\ie
\frac{g E_{\ell}}{2 E_p (E^2_\ell-E_p^2)}\,.
\fe
This matches eq.(\ref{eq: LoopCutting1}) if one identifies, $C_{2,\phi}= 1$ and $$ c^{\rm tree}_4(E_{L,\ell},E_{R,\ell})=\frac{g}{2 E_p + E_{L,\ell} + E_{R,\ell}}\,.$$

\noindent\textbf{Partial Energy Poles}

From the cutting rules, we see that the discontinuity in each internal energy of the WFC produces the product of two lower-point shifted WFCs. Each will carry their own total energy pole singularities where one sums over the energy of subgraphs. This means that the parent WFC, i.e. the LHS of eq.(\ref{eqn: cutting rule}) must contain partial energy poles~\cite{Baumann:2020dch}. That is, the presence of partial energy poles and it's residue can be viewed as a corollary of the total energy pole constraint and cutting rules. Let us use  the exchange diagram of four-point WFC as a primary example.

 Taking the limit $E_{L}^s\rightarrow0$ in  \eqref{eqn: cutting rule},  only the unflipped WFC containd the partial energy pole. Thus, we obtain:
\begin{equation}
	\label{eq: bosonic partial energy pole}
    c_4|_{E_{L}^s\to 0}=\frac{M_{3,L}}{E_{L}^s}\cdot \frac{C_2}{2E_s}\cdot \disc_{E_s} c_{3,R}\,.
\end{equation}
where we have utilized the fact that the $E_{L}^s$ pole corresponds to the total energy pole of $c_{3,L}$ with residue $M_{3,L}$. This analysis directly extends to fermion exchanges,
\ie
\label{eq: Spinor Partial Pole}
	{c}_{4}|_{E_{12s} \to 0}
	 =\frac{M_{3,L}^{-}}{E_{12s}} \cdot \frac{C_{2}}{2E_s}
	 \cdot
	 \disc_{E_s}c_{3,R}\,,
	 \quad
	 {c}_{4}|_{E_{34s} \to 0}
	 =
	\disc_{E_s} {c}_{3,L}
	 \cdot \frac{C_{2}}{2E_s}
	 \cdot
	 \frac{M_{3,R}^{+}}{E_{34s}}\,,
\fe
where the $\pm$ superscript marks the amplitude where the appropriate boundary profiles are stripped:	
        \ie
        M_{3,L}|_{u_s=(1-i \vecslashed{\hat{p}_s})\bchi_{s,\zero}}
		& =: M_{3,L}^{-} \cdot \bchi_{s,\zero}\,, 
		\quad
		M_{3,R}|_{\bar u_{-s}=\bbarchi_{-s,\zero}(1-i\vecslashed{\hat{p}_s})} =:\bbarchi_{-s,\zero} M_{3,R}^{+}\,.
        \fe

\subsection{Alternative Derivation of (Massive) Fermionic Cutting Rules}
\label{sec: alternative derivation of fermionic cutting rules}

In the previous section, we observed that the discontinuity of the massless fermion bulk-to-
bulk propagator exhibits a factorized structure analogous to the scalar
case, thereby leading to the cutting rules. This was derived using the explicit analytic property of bulk-to-bulk propagators. However, this might leave one with the impression that this is a special property of flat space. In this section we take an alternative route that only uses the differential equation and boundary conditions of the fermionic propagators. This approach will be more useful in curved backgrounds. For readers only interested in flat-space, this section is optional.

The key ingredient is a reorganization of the  Schwinger-Dyson series. We begin by first decomposing the EOM in terms of $\bchi_{+}$  $\bchi_{-}$, and derive a second order differential equation.  To illustrate, let us use the QED example. The EOM can be expressed in terms of $\bchi_+$ and $\bchi_-$ as
\begin{equation}
	\label{linearchiEOM}
	\left(\pm i \p_t+m\right)\bchi_\pm=i\slashed{p}\bchi_\mp-g \Pi_{\pm}\slashed{A}\bchi\,.
\end{equation}
Substituting the EOM with $\bchi_-$ on the LHS into that with $\bchi_+$ on the LHS yields a second-order equation for $\bchi_-$,
\ie
	\label{chiposEOMint}
  \left(\p_t^2+E^2\right)\bchi_-=g\Pi_-\left(i\slashed{p}+i\p_t+m\right)\slashed{A}\bchi\,.
\fe
Note that this is compatible with our boundary conditions in \eqref{eq: spinor embedding} was set for $\bchi_-$. Thus, $\bchi_-$ can be solved as a scalar with a dressed interaction term and the boundary condition, which makes it straightforward to construct the Schwinger-Dyson series for $\bchi_-$ by analogy with the scalar case. At zeroth order,
\ie
\label{eq:chiminusInt_0}
	\bchi^{(0)}_-(p,t)
		&=K_{\phi}\left(p,t\right)\bchi^-_\zero\,.
\fe  
Substituting into \eqref{linearchiEOM} gives $\bchi_+$ at the same order,
\ie
	\label{eq:chiplusInt_0}
	\bchi_+^{(0)}(p,t)
	  	&=\left(\frac{-i\slashed{p}}{E^2-m^2}\right)\left(-i \p_t+m\right)\bchi^{(0)}_{-}\,.
\fe
Iterating to the next order, one finds:
\ie
\bchi^{(1)}_-(p,t)
	&=\int dt' G_{\phi}(p;t,t')\cdot (-g)\,\Pi_-\left(i\slashed{p}+i\p_{t'}+m\right)\slashed{A}^{(0)}\bchi^{(0)}\,,\\
\bchi_+^{(1)}(p,t)
	&=\left(\frac{-i\slashed{p}}{E^2-m^2}\right)\cdot g\Pi_-\left(i\slashed{p}+i\p_{t'}+m\right)\slashed{A}^{(0)}\bchi^{(0)}.
\fe
This yields the complete $\bchi^{(1)}$. In the same manner, higher-order terms in the SD series, as well as the conjugate field series, can be constructed sequentially.

Substituting into the action  gives an alternative represenation of  the WFCs as a bulk integral using the scalar propagators,
\ie
\label{eq:fermion new WFC}
c_{3, J_1 \bar\chi_2 \chi_3}
	&= ig \int dt \, \bbarchi^{(0)}(  p_1,t) \cdot \slashed{A}^{(0)}(p_2,t) \cdot \bchi^{(0)}(  p_3,t) \,,\\
c_{4, J_1 \bar\chi_2 J_3 \chi_4}
	&= g^2 \int dt \int dt' 
	\bbarchi^{(0)}(  p_1,t) \cdot \slashed{A}^{(0)}(p_2,t) \cdot 
	\left[
	1-\left(\frac{i\slashed{\mathbf{p_s}}}{E_s^2-m^2}\right)\left(-i\p_t+m\right)
	\right]\\
	& \left[
		\Pi_- \cdot \vecslashed{p_s} \cdot \Pi_+ \cdot G_\phi(  p_s, t, t')
	\right]\cdot \left[
		1-\left(\frac{i\slashed{\mathbf{p_s}}}{E_s^2-m^2}\right)\left(-i\overleftarrow{\p}_t'+m\right)
		\right]
	\left[\slashed{A}^{(0)}(  p_3,t')\bchi^{(0)}(  p_4,t')  \right]\\
&-
ig^2 \int dt  \bbarchi^{(0)}(  p_1,t) \cdot \slashed{A}^{(0)}(p_2,t) \cdot \Pi_- \left(\frac{i\slashed{\mathbf{p_s}}}{E_s^2-m^2}\right)\Pi_+ \cdot\slashed{A}^{(0)}(  p_3,t) \bchi^{(0)}(  p_4,t)\,,
\fe
in which we already do some integration by parts and the first order in the SD series reads,
\ie
\label{eq:chi0}
\bchi^{(0)}(  p,t) 
=
e^{i E t} \left(1-\frac{i\vecslashed{p}}{E-m}\right)\bchi_{\zero}(  p)\,,~~~ \bbarchi^{(0)}(  p,t) 
=
e^{i E t} \cdot \bbarchi_{\zero}( p)\left(1+\frac{i\vecslashed{p}}{E-m}\right).
\fe

Now, let us consider the discontinuity of the internal energy. We consider the discontinuity of the $s$-channel internal energy in Compton scattering as an example. Note that $\disc_{E_s}$ extracts only the term with $e^{i E_s t}$ dependence, which appears solely in $G_\phi$. Thus, using the discontinuity identity of the scalar bulk-to-bulk propagator \eqref{eq:disc of boson G}, we obtain
\ie
\label{app: disc of ferm derivation}
\disc_{z_s}{c}_{J\bar\chi J \chi}
&= 
g^2
\int dt \left[\bbarchi^{(0)}(  p_2,t) \slashed{A}^{(0)}(  p_1,t)\right] \left( \disc_{E_s} K_\phi(E_s, t)  \left(1-\frac{i\vecslashed{p_s}}{E_s-m}\right)  \right) \\
&\cdot \Pi_- \cdot \frac{-i\vecslashed{p_s}}{2 E_s} \cdot \Pi_+\cdot \int dt'  \left( \disc_{E_s}  K_\phi(E_s, t') \left( 1 - \frac{i\vecslashed{p_s}}{E_s-m} \right) \right) \left[ \slashed{A}^{(0)}(  p_3,t') \bchi^{(0)}(  p_4,t')\right].
\fe
The factor $\left( 1 \pm \frac{i\vecslashed{p_s}}{E_s-m} \right)$ is precisely the one appearing in $\bbarchi^{(0)}(-p_s)$ and $\bchi^{(0)}(p_s)$ written in \eqref{eq:chi0}. Comparing with the $c_{3, J \bar\chi \chi}$ in \eqref{eq:fermion new WFC}, we see that the left and right terms correspond to ${c}_{3, J \bar\chi \chi}$ with the internal boundary profiles extracted, denoted ${c}_{3, J \bar\chi \chi,A}$. The discontinuity version of the COT is then
\ie
\label{fermion discCOT}
    \begin{aligned}
        \disc_{E_s}{c}_{J\bar\chi J \chi}
        &=
        \disc_{E_s}{{c}}_{J\bar\chi\chi,A} (p_1,p_2,p_s) 
        \cdot \left[\Pi_- \cdot \frac{i\vecslashed{p_s}}{2 E_s} \cdot \Pi_+\right]^{A}_B
       \disc_{E_s}{{c}}_{J\bar\chi\chi,B} (p_3,p_4, p_s).
    \end{aligned}
\fe
which reproduces the cutting rule \eqref{eqn: cutting rule} with the spinor factor \eqref{C of the spinor}. We find the same rule also applied to the WFC exchaging \emph{massive} spinors.

\section{Bootstrapping (Fermionic) WF coefficients}
\label{sec:bootstrapping}
Equipped with the analytic constraints in energy variables, in this section we demonstrate that starting with a consistent flat-space amplitude, we can feed the amplitude through a sequence of operations whose result gives the WFCs. We will consider the scenario where at least one  \emph{conserved} operator are involved. Since longitudinal part of conserved currents are determined by lower point WFCs via WT identities 
\footnote{
	One can readily verify that if the three-point and four-point WT identities are generated by the same gauge transformation, then the cutting rules in Eq.~\eqref{eqn: cutting rule}, with the longitudinal components on both sides fixed by the WT identities, are satisfied.
}, we will only focus on transverse components, which we denote as $c^T$.

\paragraph{3-point  WFCs}

Based on the discussion in  Sec. \ref{sec:total-energy-pole} and \ref{sec:cutting rules}, the transverse component should include the total energy pole, whose residue is the amplitude. As there are no partial or internal energy poles, the only remaining unfixed terms must be polynomial. \footnote{The unit vector like $\hat p_i$ for the external leg could only appear in the projector in the decomposition \eqref{eq:generalPdecompose}.} We will demonstrate that unfixed polynomial terms that are consistent with dimensional analysis will be removable via field redefinition.

\paragraph{4-point  WFCs}
The 4-point WFC now involve total and partial energy poles. We will start our ansatz with the partial energy pole residues, and gradually build our answer by enforcing the correct total energy pole residue. In particular, our ansatz takes the form :
\ie
\label{eqn: 4-point WFC ansatz}
c^T_4=\sum_{e \in s,t,u}\left(\frac{A_R^e}{E_R^e}+\frac{B_L^e}{E^e_L}\right)+\frac{C}{E_T}+D\,.
\fe
This can be determined through the following three steps:
\begin{enumerate}

    \item \emph{Matching Partial Energy Residues} ($A^e_{R},B_L^e$)

	We begin by matching the partial energy pole $E_R^e$ in each channel. The resulting residue 
    will be the product of the  amplitude and the discontinuity of WFCs on the other side. Importantly on the support of $E_R^e=0$, 
    \begin{equation}
E_L^e\bigg|_{E_e\rightarrow -E_e}=E_T\,.
    \end{equation}
 Thus the discontinuity of the WFC will introduce total energy pole. For example, using the result in Sec.~\ref{sec:cutting rules}, the residue of the partial energy poles can be reorganized as:
	\ie
		\res_{E^s_R \rightarrow 0} c_4^T &=
	M_L^s\cdot \frac{1}{2E_s}\left(\frac{O^s_{P}}{E^s_L}-\frac{O^s_{R}}{E_T}\right)\cdot M_R^s\,,\\
	\res_{E^s_L \rightarrow 0} c_4^T &=
		M_L^s\cdot \frac{1}{2E_s}\left(\frac{O^s_{P}}{E^s_R}-\frac{O^s_{L}}{E_T}\right)\cdot M_R^s\,, 
	\fe
where the $M_{L/R}^s$ are the amplitude of the left and right diagrams in the $s$ channel, and we have reorganized $\disc_{E_s} c_{3,R}$ using the explicit 3-pt WFCs derived in the previous step. $O$'s arise from the polarization sums of the specific exchanged fields:
    \begin{equation}
   \notag  \begin{array}{rcl}
         	\label{eqn: ORTLT}
	O_{P,\phi}^s
&=&O_{R,\phi}^s=O_{L,\phi}^s=1\,, \\
	O_{P,J}^{s,ij}
&=&O_{R,J}^{s,ij}=O_{2,L,J}^{s,ij}=\pi^{ij}_{s}\,, \\
	O_{P,T}^{s.ijkl}&=&
O_{R,T}^{s,ijkl}=O^{s,ijkl}_{L,T}=\hat\Pi^{ijkl}_{s}\,,
  \end{array}
    \end{equation}
~\\[-0.9cm]
    \ie
\label{PoS}        O^s_{P,\chi} 
	=
	-\gamma_0 \slashed{p}_{s\,-}^{[4]}\,,~~~~~O^{s}_{L,\chi}&=& -i\slashed{p}_s^{[4]}\,,~~~~~~O^{s}_{R,\chi}=i\slashed{p}^{[4]}_{s\,-}\,, 
    \fe
    ~\\[-1cm]
    \ie
	\notag O^{s,ij}_{P,\psi}
			&=-\gamma_0\pi^{ij}_{s}\slashed{p}_{s\,-}^{[4]} +
			\frac{i}{2} (1+i\slashed{\hat{p}}_s)(\slashed{\pi}_s^i \slashed{p}_s \slashed{\pi}_s^j)\left(\frac{1-i\gamma_0}{2}\right)(1-i\slashed{\hat{p}}_s),\\
\notag	O^{s,ij}_{L,\psi}
			&=-i\pi^{ij}_{s}\slashed{p}_s^{[4]} +
			\frac{i}{2} (1-i\slashed{\hat{p}}_s)(\slashed{\pi}_s^i \slashed{p}_s \slashed{\pi}_s^j)\left(\frac{1-i\gamma_0}{2}\right)(1-i\slashed{\hat{p}}_s),\\
	\notag O^{s,ij}_{R,\psi}
&=i\pi^{ij}_{s}\slashed{p}_{s\,-}^{[4]} +
			\frac{i}{2} (1+i\slashed{\hat{p}}_s)(\slashed{\pi}_s^i \slashed{p}_s \slashed{\pi}_s^j)\left(\frac{1-i\gamma_0}{2}\right)(1+i\slashed{\hat{p}}_s)\,,
	\fe
    where $\phi$, $J$, $T$, $\chi$, $\psi$ label the exchanged fields with spin-$0$, spin-$1$, spin-$2$, spin-$1/2$ and spin-$3/2$, respectively. 
	 Here, all $M$s  are written with the external polarizations already replaced by the transverse-traceless ones, and with the internal polarizations removed, as detailed in Sec.~\ref{sec:total-energy-pole}.

    We can readily determine $A_R^s$ by matching the first line:
	\begin{equation}
		A_{R}^s=M_L^s\cdot \frac{1}{2E_s}\left(\frac{O^s_{P}}{E^s_L}-\frac{O^s_{R}}{E_T}\right)\cdot M_R^s\,.
	\end{equation}
	Then, it is straightforward to write $B_L^s$ based on $A_R^s$ to match the other $E_L^s$ partial energy pole:
	\ie
	B_{L}^s=M_L^s\cdot \frac{1}{2E_s}\left(-\frac{O^s_{L}}{E_T}\right)\cdot M_R^s\,.
	\fe
    
    \item \emph{Matching Amplitude Limit} ($C$)
    
	Next, we examine the behavior of the WFCs constructed from $A_R^e$ and $B_L^e$ as $E_T \rightarrow 0$:
	\footnote{Steps 2 and 3 could in principle be interchanged. However, imposing the cutting rules before matching the amplitude limit would require a more general ansatz for $D$, including terms of the form $D=D'/E_T$.}
	\ie
	\label{eqn: total pole of A B}
	\res_{E_T \rightarrow 0} 
	\sum_{e \in s,t,u} \left(\frac{A_R^e}{E_R^e}+\frac{B_L^e}{E^e_L}\right) 
	&=
	\sum_{e \in s,t,u} M_L^e\cdot \left(-\frac{O^e_{R}}{2E_e E_R^e}-\frac{O^e_{L}}{2E_e E_L^e}\right)\cdot M_R^e 
	=:
	M_{fact}\,.
	\fe
	This expression already aligns with the amplitude at the factorization pole, which we will demonstrate later. Then the discrepancy between the total energy pole residue and the amplitude is a contact term, which we address by including $\frac{C}{E_T}$ in the ansatz \eqref{eqn: 4-point WFC ansatz} and reads
    \ie 
    \label{ctamp}
	C =
		M_4
	- 
		\lim_{E_T \rightarrow 0} M_{\text{fact}}\,.
	\fe

	To clarify why \eqref{eqn: total pole of A B} correctly represents the residue on the factorization pole, we could focus on the $s$ channel without loss of generality. The limit $S \rightarrow 0$ can be approached via two paths: the partial energy pole limits $E_R^s \rightarrow 0$ or $E_L^s \rightarrow 0$. These paths cause the two distinct terms in \eqref{eqn: total pole of A B} to converge to the polarization sum in the amplitude limit, $O^s_{\rm A}$ , in different manners:
	\ie
	\label{eqn: total pole of A B limit}
	\lim_{S \rightarrow 0} \left(\frac{A_R^s}{E_R^s}+\frac{B_L^s}{E^s_L} \right)
	=
	\begin{cases}
	\displaystyle	\frac{M_L^s O^s_{L} M_R^s}{2E_s E_R^s}
			\rightarrow~
			\frac{M_L^s O_{\rm A}^s M_R^s}{S}\,,
			\text{ when } E_R^s = E_{34}-E_s \rightarrow 0\,.
			\\
	\displaystyle	\frac{M_L^s O^s_{R} M_R^s}{2E_s E_L^s}
			\rightarrow~
			\frac{M_L^s O_{\rm A}^s M_R^s}{S}\,,
		\text{ when } E_L^s  = E_{12}-E_s \rightarrow 0\,.
	\end{cases}
	\fe
	For specific fields, the polarization sum reads:
	\begin{equation}	    
    \begin{array}{cll}
 	&O_{{\rm A},\phi}^s=1\,,
			&~~~~O_{{\rm A},J}^{s\,,\mu\nu}=\eta^{\mu\nu}\,,
		~~~~~~~~~~~	O_{{\rm A},T}^{s\,,\mu\nu\rho\sigma}=\eta^{\mu\nu}\eta^{\rho\sigma}\,,\\[2mm]
			&O_{{\rm A},\chi}^s= -i\left(\slashed{p}_{3}^{[4]}+\slashed{p}_4^{[4]}\right)\,,
			&~~~~O_{{\rm A},\psi}^{s\,,\mu\nu}= -i \eta^{\mu\nu} \left(\slashed{p}_{3}^{[4]}+\slashed{p}_4^{[4]}\right) \label{eqn: ORTLT to polarization sum}\,.
	\end{array}
    	\end{equation}
	This demonstrates that the ansatz $(A^e_R, B^e_L)$ in \eqref{eqn: 4-point WFC ansatz} fixed in the previous step indeed leads to amplitude factorization under the total energy pole. What remains is the algebraic step ensuring that \eqref{eqn: total pole of A B limit} holds, thereby establishing the consistency between the cutting rule and the total-energy-pole residue. We present the detailed calculations and useful identities to get the $C$ term for specific theories in Appendix~\ref{app:matching identity}.

    \item \emph{Back to the Cutting Rules} ($D$)

    After addressing all singularity constraints, we return to the cutting rules \eqref{eqn: cutting rule} to verify their validity. If they are not satisfied, we add terms $D$ without partial or total energy poles in order to restore consistency.
\end{enumerate}

\subsection{3-pt WFC}

\paragraph{$\underline{\langle J \bar{\chi}\chi\rangle}$}~\\[-3mm]
\par We begin with the current fermion fermion WFC. The decomposition reads,
\begin{equation}
    \langle J\bar\chi\chi\rangle
=\epsilon_{1,i,\zero}\left(\pi_1^{ij}\mathbb{A}^{T}_j+\hat{p}^i\mathbb{A}^L\right)
	\equiv 
    \langle J^T \bar\chi \chi \rangle
	+
   	\langle J^L \bar\chi \chi \rangle\,.
\end{equation}
Consider the QED and use its flat space amplitudes as an input,
the result is,
\footnote{
	The total energy pole in the longitudinal part is spurious, one can show that
	\ie
 	\langle J^L \bar\chi \chi \rangle 
	=
	-\frac{i \epsilon_{1, \partial}\cdot  \hat{p}_{1}}{E_T}  \bar{u}_2 \gamma_0 u_3 = \frac{i \epsilon_{1, \partial}\cdot \hat{p}_{1}}{E_1} \cdot \bm{\bar{\chi}}_{2, \partial}(\slashed{p}_2 + \slashed{p}_3) \bchi_{3, \partial}.
\fe
It's what we discussed in Sec. \ref{sec: WT identity}. The longitudinal part is fixed by the WT identity which is the combination of the lower point functions.}
\begin{equation}
    \langle J^L\bar\chi \chi \rangle
		= 
		-\frac{i (\epsilon_{1,\zero}\cdot \hat p_{1})\,(\bar u_2 \gamma_0 u_3)}{E_T}  \,,~~
        \langle J^T \bar\chi \chi \rangle
        =
		\frac{i \bar u_2 \vecslashed{\epsilon}_{1,\zero}^T u_3}{E_T}  \,.
\end{equation}
	A straightforward dimensional analysis shows that there is no room to introduce any polynomial term. Therefore, the residue at the total-energy pole fully fixes the transverse component. 
   	The polarization spinors $\bar u, u$ are related to the boundary spinors $\bbarchi_\zero, \bchi_\zero$ by equation \eqref{eq: fermion polarization}. These results can be directly matched to Feynman rules~\cite{YAflat}.
We can project the transverse part onto the three-dimensional massive spinor-helicity basis defined in App.~\ref{app: conventions}. The demonstration is implemented in \texttt{src/3pt SPH.nb}. The result is
 \begin{equation}
     \langle J^+\bar\chi^+\chi^-\rangle
     =-2i\frac{\langle \bar{1}\bar{2} \rangle^2}{\langle\bar{2}\bar{3}\rangle}\left(\frac{E_{23}{-}E_1}{E_TE_1}\right),\quad  
      \langle J^+\bar\chi^-\chi^-\rangle
    = \langle J^+\bar\chi^+\chi^+\rangle
    =0.
 \end{equation} 
Only the helicity component with a non-vanishing amplitude limit is non-zero. This feature persists for the other operators considered below and is characteristic of flat-space correlators; it does not generally hold in curved space~\cite{CurvedPaper}.
The $1/E_1$ does not represent a true singularity, as it appears in the combination $\bar\lambda_1 \bar\lambda_1/E_1$. 

\paragraph{$\underline{\langle \phi \bar\chi \chi \rangle}$}~\\[2mm]
For the Yukawa theory, we take the corresponding flat-space amplitude as input:
\ie
 \langle \phi \bar\chi \bar \chi \rangle 
    &=  \frac{\bar u_2 u_3}{E_T}
\fe
A straightforward dimensional analysis shows that no additional polynomial term can be present. Projecting the external spinors $\bar u_2$ and $u_3$ onto the three-dimensional massive spinor-helicity basis defined in App.~\ref{app: conventions} then gives
\ie
    \langle \phi \bar\chi^+ \bar \chi^+ \rangle 
    &=  \frac{4i\langle \bar 2 \bar 3 \rangle}{E_T} ,\, \quad 
	\langle \phi \bar\chi^+ \bar \chi^- \rangle 
    &=  0.
\fe
Thus, only the helicity component with a non-vanishing amplitude limit is non-zero.

\paragraph{$\underline{\langle T \bar{\chi}\chi\rangle}$}~\\[-4mm]
\par The decomposition reads,
\ie
\langle T \bar\chi \chi \rangle &= \epsilon_{1,k} \epsilon_{1,l} \left[ \pi^{ki'}_1 \pi^{lj'}_1 \mathbb{A}^{TT}_{i'j'} 
+ \left( \pi^{ki'}_1 \hat{p}_1^l 
+ \pi^{li'}_1 \hat{p}_1^k 
+ \hat{p}_1^k \hat{p}_1^l \hat{p}_1^{i'} \right) \mathbb{A}^L_{i'} \right]
\\
&\equiv 
\langle T^{TT} \bar\chi \chi \rangle
+
\underbrace{\langle T^{TL} \bar\chi \chi \rangle
+
\langle T^{LT} \bar\chi \chi \rangle
+
\langle T^{LL} \bar\chi \chi \rangle}_{\langle T^L_1 \bar\chi_2 \chi_3 \rangle}\,.
\fe
~\\[-4mm]
Once again, following the same procedure, we find\,,
\ie
\langle T^L \bar\chi\chi \rangle 
&:= (\epsilon_{1} \cdot \hat p_1)\hat p_{1,i} \epsilon_{1,j} \langle T^{ij} \bar\chi \chi \rangle \\
&=
\frac{i\epsilon_{1} \cdot \hat p_1 }{16E_1} 
\bm{\bar\chi}_{2,\zero}\left\{\vecslashed{\hat p}_2\left[[\vecslashed{p}_1,\,\vecslashed{\epsilon}_1]+8 (p_3\cdot \epsilon_1)\right]+
	\left[  [\vecslashed{p}_1,\,\vecslashed{\epsilon}_1]-8 (p_2 \cdot \epsilon_1)\right]\vecslashed{\hat p}_3  \right\}
\bchi_{3,\zero}\,, \\
\langle T^{TT} \bar{\chi} \chi \rangle
&=
\frac{i}{E_T} \epsilon^T_{1}\cdot (p_2-p_3) (\bar u_2 \vecslashed{\epsilon}_{1}^T u_3)\,.
\fe
 These results can also be directly matched to Feynman rules~\cite{YAflat}. Below we project the transverse component onto various helicity configurations, the results written in the kinematic variables defined in App. \ref{app: conventions} read,
\begin{equation}
    \langle T^+\bar\chi^+\chi^-\rangle=i\frac{\langle \bar{1}\bar{2} \rangle^3\langle \bar{3}\bar{1} \rangle}{\langle \bar{2}\bar{3} \rangle^2}\left(\frac{(E_1-E_{23})^2}{E_TE_1^2}\right),\quad\langle T^+\bar\chi^-\chi^-\rangle=\langle T^+\bar\chi^+\chi^+\rangle=0\,.
\end{equation}   
\paragraph{$\underline{\langle T \bar{\psi}\psi\rangle}$}~\\
\par As a further application, we consider the gravitino-graviton contact WFC. The same procedure gives\,,
\ie
\langle T^{L}\bar{\psi } \psi \rangle  
&= i\frac{(\epsilon_{1} \cdot \hat p_{1})}{2E_1}\bbarchi_{2}
	\bigg[  \cancel{\hat p}_{2} \left(\epsilon_{2} \cdot \pi_2 \cdot p_{1}\right)(\epsilon_{3} \cdot \epsilon_{1}) -\cancel{\hat p}_{3} \left(\epsilon_{3} \cdot \pi_3 \cdot p_{1}\right)(\epsilon_{2} \cdot \epsilon_{1})\\
&	~~+ \left( \epsilon_{2}\cdot \pi_2 \cdot \epsilon_{3}\right)\cancel{\hat p}_{2} \left( p_{3} \cdot \epsilon_{1}+\frac{1}{8}\left[\cancel{p}_{1} ,\cancel{\epsilon}_{1}\right]\right)+ \left(\epsilon_{2} \cdot \pi_3 \cdot \epsilon_{3}\right)\left(\frac{1}{8}\left[\cancel{p}_{1} ,\cancel{\epsilon}_{1}\right]-p_{2} \cdot \epsilon_{1}\right)\cancel{\hat p}_{3}\bigg]\bchi_{3}\,,\\
~\\[-5mm]
\langle T\bar{\psi }^{L} \psi \rangle  
&= -i\frac{\epsilon_{2} \cdot \hat p_{2}}{E_2}  
	\bbarchi_{2}\big[
	\left(\epsilon_{1} \cdot \pi_1 \cdot \epsilon_{3}\right)\left(\cancel{\epsilon}_{1} \right) E_{1} \ +\frac{1}{8}\left(\epsilon_{1} \cdot \pi_3 \cdot \epsilon_{3}\right)\left({\left[\cancel{p}_{1} ,\cancel{\epsilon}_{1}\right]\cancel{\hat p}}_{3} \right)\big]\bchi_{3}\,, \\
    ~\\[-5mm]
\langle T\bar{\psi } \psi^{L} \rangle  
&=	
	i\frac{\epsilon_{3} \cdot \hat p_{3}}{E_3}
	\bbarchi_{2}\big[
	\left(\epsilon_{1} \cdot \pi_1 \cdot \epsilon_{2}\right)\left(\cancel{\epsilon}_{1}\right) E_{1} \ -\frac{1}{8}\left(\epsilon_{1} \cdot \pi_2 \cdot \epsilon_{2}\right)\left({\cancel{\hat p}}_{2}\left[\cancel{p}_{1} ,\cancel{\epsilon}_{1}\right] \right)
	\big]\bchi_{3}\,,\\
    ~\\[-5mm]
\langle T^{TT} \bar{\psi}^{T} \psi^{T}
\rangle
&=
	\frac{i}{E_T} (\bar u_2 \vecslashed{\epsilon}_{1}^T u_3)
	\left[(2 \epsilon^T_{3} \cdot p_{1}) (\epsilon^T_{2} \bm \cdot \epsilon^T_{1})
		+ (\epsilon^T_{2} \cdot \epsilon^\Trans_3) (p_2- p_3) \cdot \epsilon^T_{1} 
	- (2\epsilon_{2}^T \cdot p_1) (\epsilon_{3}^T \cdot \epsilon_1^T) \right]\,.  
\fe

The first three lines represent the single longitudinal components for each operator \footnote{
	There's a freedom we could add unfix term written as,
	\ie
	\langle T^{TT}\bar\psi^{T} \psi^{T} \rangle'
		&= \langle T^{TT}\bar\psi^{T}\psi^{T} \rangle + a \; (\epsilon^T_{1}\cdot\epsilon^{T}_{2})(\epsilon^T_{1}\cdot\epsilon^{T}_{3}) \left[ \overline{\chi }\cancel{\epsilon }_{2}^{T}\left( \slashed{\hat{p}}_{2}-\slashed{\hat{p}}_{3}\right)\cancel{\epsilon }_{3}^{T} \chi \right]\,.
	\fe
	It corresponds to the non-dynamic field redefinition freedom on the boundary profile $\bm{\psi}_{i,\zero} \rightarrow \bm{\psi}_{i,\zero}+\kappa h_{i,\zero}^j\bm{\psi}_{j,\zero}$ and $\bm{\bar\psi}_{i,\zero} \rightarrow \bm{\bar\psi}_{i,\zero}+\kappa h_{i,\zero}^j\bm{\bar\psi}_{j,\zero}$. And there will be also introduce an additional unfix term by $h_{ij,\zero} \rightarrow h_{ij,\zero}+\kappa h_{ik,\zero}h^k_{j,\zero}$ in,
	\ie
	\langle T^{TT}T^{TT}T^{TT} \rangle'
		&= \langle T^{TT}T^{TT}T^{TT} \rangle + a \; E_T (\epsilon^T_{1}\cdot\epsilon^{T}_{2})(\epsilon^T_{1}\cdot\epsilon^{T}_{3})(\epsilon^T_{2}\cdot\epsilon^{T}_{3})\,.
	\fe
	In the paper, we fix the field redefinition freedom by setting $a=0$.
}. The exact components in the decomposition, such as $\langle T^{TL} \bar\psi^{L} \psi^{T} \rangle$, can be obtained by further decomposing either $\langle T^{L} \bar\psi \psi \rangle$ or $\langle T \bar\psi^{L} \psi \rangle$,
\ie
\langle T^{TL} \bar{\psi}^{L} \psi^{T}
\rangle 
\,&=\,
\langle T^{L} \bar\psi \psi \rangle\,
\Big|
\resizebox{2.1cm}{!}{
\hspace{-1mm}
$\begin{array}{rcl}
      \epsilon_{1} &\to& \epsilon_{1}^T\\ 
      \epsilon_2 &\to& (\epsilon_2 \cdot \hat p_2) \hat p_2
      \\ \epsilon_{3} &\to& \epsilon_{3}^T
\end{array}$}
\,=\,
\langle T \bar\psi^L \psi \rangle\Big|
   \resizebox{3cm}{!}{$\begin{array}{rcl}
   &&\\[-8mm]
         \epsilon_1^i \epsilon_1^j&\to & (\epsilon_1^{T})^{\raisebox{0.2ex}{\footnotesize $(i$}} \hat p_1^{\,j)}(\epsilon_1 \cdot \hat p_1)\\
         &&\\[-3mm]
        \epsilon_3 &\to &\epsilon_3^T
    \end{array}$}\\
&=
-\frac{\left( \epsilon _{1} \cdot \hat p_1)( \epsilon _{2} \cdot \hat p_2\right)(p_{1} \cdot \epsilon_{3}^{T})}{16 E_{1} E_{2}}
	\bbarchi_{2}{\left[\cancel{p}_{1} ,\cancel{{\epsilon }}^T_{1}\right]\cancel{\hat p}}_{3} \bchi _{3}\,.
\fe
Notice in the \S App. \ref{app:WTofWFC}, it's direct to check that the longitudinal parts given by different WT identities are the same. Following the previous example, we express the result in spinor helicity form as well\,,
\ie
&\langle T^+\bar\psi^+\psi^-\rangle=i\frac{\langle \bar{1}\bar{2} \rangle^5}{\langle \bar{2}\bar{3} \rangle^2\langle \bar{3}\bar{1} \rangle}\left(\frac{(E_T-2E_1)^2(E_T-2E_3)(E_T-2E_2)}{8E_TE_1^2E_2E_3}\right)
\\
&\langle T^+\bar\psi^-\psi^-\rangle=\langle T^+\bar\psi^+\psi^+\rangle=0
\fe
~\\
\subsection{4-pt WFC}

\paragraph{$\underline{\langle J \bar{\chi } J \chi \rangle}$}
~\\~\\
The longitudinal mode is completely fixed by the WT identity, and therefore we focus on the pure transverse part of the WFC,
\begin{equation}
    \langle J^T \bar{\chi} J^T \chi \rangle 
	=
	\sum_{s,t}\left(\frac{A_{R,J\bar\chi J\chi}^e}{E_R^e}+\frac{B_{L,J\bar\chi J\chi}^e}{E^e_L}\right)+\frac{C_{J\bar\chi J\chi}}{E_T}+D_{J\bar\chi J\chi}\,.
\end{equation}
First we write down the $A_R^s$ for the s-channel exchanged, to match the residue of $E_R^s=E_{34s}$ pole in \eqref{eq: Spinor Partial Pole},
\ie
A_{R,J\bar\chi J\chi}^s
=
\frac{i}{2E_s}  \bar u_2 \vecslashed{\epsilon}_1^T 
	  \left( \frac{i\gamma_0 \slashed{p}_{s\,-}^{[4]}}{ E_{12s}} - \frac{\slashed{p}_{s\,-}^{[4]}}{E_T} \right) 
\vecslashed{\epsilon}_3^T u_4\,.
\fe 
Then to match $E^s_L=E_{12s}$ pole in \eqref{eq: Spinor Partial Pole}, we need to add a term $B_{R,J\bar\chi J\chi}^s$,
\ie
B_{L,J\bar\chi J\chi}^s 
=
\frac{i}{2E_sE_T}  \bar u_2 \vecslashed{\epsilon}_1^T 
	   \slashed{p}_s^{[4]}
\vecslashed{\epsilon}_3^T u_4\,.
\fe    
Similarly, we could write down $A_{R,J\bar\chi J\chi}^t$ and $B_{L,J\bar\chi J\chi}^t$ for the $t$-channel exchanged. If we combine them we could see we already generate the correct amplitude limit of the total energy pole.
\begin{tcolorbox}[text width= 417pt, top=-3pt, boxsep=0pt]
 \ie
\label{eq:JchiJchitrans}
\langle J^T \bar{\chi} J^T\chi \rangle
&=
-i
\bar u_2 \vecslashed{\epsilon}_1^T 
	 \left[ \frac{ \slashed{p}_{3}^{[4]}+\slashed{p}_{4}^{[4]} }{E_T E_{12s} E_{34s} } - \frac{1-i\gamma_0}{2}\frac{\slashed{p}^{[4]}_{s\,-}}{E_s E_{12s} E_{34s}} \right] 
\vecslashed{\epsilon}_3^T u_4 +(1\leftrightarrow 3)\,.
\fe   
\end{tcolorbox}
 We set $C_{J\bar\chi J\chi}$ to zero because the amplitude limit of the total energy pole is already matched. Similarly, $D_{J\bar\chi J\chi}$ is set to zero since, based on dimensional analysis, there is no viable contact term ansatz to construct.
Alternatively, we can rewrite the expression in terms of spinor helicity variables. The corresponding calculation is implemented in \texttt{src/4pt SPH.nb}, and gives
\begin{tcolorbox}[text width= 417pt, top=-4pt, boxsep=0pt]
\begin{align}
\langle J^+ \bar{\chi}^+ J^- \chi^- \rangle
&= -\frac{i \langle 3 4 \rangle \langle \bar{1} \bar{2} \rangle}{2E_1 E_3 E_{12s} E_{34s}}
\left[
  \left(\frac{2}{E_T}+\frac{1}{E_s}\right)\langle 3 4\rangle \langle \bar{1} \bar{4}\rangle
  -\left(1-\frac{4E_{34}}{E_T}-\frac{E_{34}}{E_s}\right)\langle 3 \bar{1}\rangle
\right] \nonumber \\
&\quad
+\frac{i \langle 3 \bar{2}\rangle \langle 4 \bar{1}\rangle}{2E_1 E_3 E_{23t} E_{14t}}
\left[
  \left(\frac{2}{E_T}+\frac{1}{E_t}\right)\langle 3 4\rangle \langle \bar{1} \bar{4}\rangle
  +\frac{E_{14t}}{E_t}\langle 3 \bar{1}\rangle
\right], \\[4mm]
\langle J^+ \bar{\chi}^+ J^+ \chi^- \rangle
&= -\frac{i \langle \bar{1} \bar{2}\rangle \langle 4 \bar{3}\rangle}{2E_1 E_3 E_{12s} E_{34s}}
\left[
  \left(\frac{2}{E_T}+\frac{1}{E_s}\right)\langle 4 \bar{1}\rangle \langle \bar{3} \bar{4}\rangle
  -\frac{E_{34s}}{E_s}\langle \bar{1} \bar{3}\rangle
\right] \nonumber 
+(\,1\leftrightarrow 3\,).
\end{align}
\end{tcolorbox}

The leading total energy pole which appears in the WFC $\langle J^+ \mathcal \chi^+ J^+ \chi^- \rangle$ is, in fact, spurious, and the order of the total energy pole starts from $O(E_T^0)$. The cancellation of the spurious pole can be seen by arranging the expression into independent kinematic variables via momentum conservation. This matches consistently the amplitude limit where the flat space amplitudes is zero for the given helicity configuration. 

Similarly, we can build 4-point WFCs with four fermionic operators,
\begin{tcolorbox}
	\ie
	\langle \bar\chi \chi \bar\chi \chi \rangle
	&=
	\frac{\bar u_1 \gamma^\mu u_2 \cdot \bar u_3 (\gamma_\mu) u_4}{E_T E_{12s} E_{34s}} -\frac{\bar u_1 \gamma_0 u_2 \cdot \bar u_3 \gamma_0  u_4}{E_	{12s} E_{34s} E_s} + (1\leftrightarrow 3).
	\fe
\end{tcolorbox}

We can rewrite the expression by spinor helicity variables and it gives
\begin{tcolorbox}
\begin{align}
\langle\bar\chi^-_1\chi^+_2\bar\chi^-_3\chi^+_4\rangle
&={}\frac{16}{E_T E_{12s}E_{34s}}
\bigg[
-\langle 1 3\rangle\langle\bar 2\bar 4\rangle
+\langle1\bar 4\rangle\langle3\bar 2\rangle
-\langle1\bar2\rangle\langle3\bar4\rangle
\bigg]\nonumber\\
&\quad-\frac{16\langle1\bar2\rangle\langle3\bar4\rangle}
{E_s E_{12s}E_{34s}} + (1 \leftrightarrow 3),\\ 
\langle\bar\chi^-_1\chi^+_2\bar\chi^+_3\chi^-_4\rangle &= \left.\langle\bar\chi^-_1\chi^+_2\bar\chi^-_3\chi^+_4\rangle\right|_{3\leftrightarrow 4}
\end{align}
\end{tcolorbox}
All other helicity components, except for those which are opposite to the ones above, vanish.

\paragraph{\underline{$\langle \phi \bar\chi \phi \chi \rangle$}}~\\~\\
For the WFC $\langle \phi \bar\chi \phi \chi \rangle$ in Yukawa theory, the only change relative to QED is the interaction vertex. Thus, replacing $\vecslashed{\epsilon}$ by the identity in \eqref{eq:JchiJchitrans} gives
\begin{tcolorbox}
\ie
\langle \phi \bar{\chi} \phi \chi \rangle
        &= 
        \bar u_2 \cdot \left[ \frac{\slashed{p}^{[4]}_{3} + \slashed{p}^{[4]}_{4}}{E_T E_{12s} E_{34s} } - \frac{1-i\gamma_0}{2}\frac{\slashed{p}^{[4]}_{s,-}}{E_s E_{12s} E_{34s}} \right] \cdot u_4 + (1 \leftrightarrow 3)
\fe
\end{tcolorbox}
For $\langle \bar\chi \chi \bar\chi \chi \rangle$, both the vertex and the cutting-rule factors are modified. The two $\gamma_\mu$ matrices in the vertex are replaced by the identity, and the polarization tensor $\pi_{ij}$ in the cutting rule is replaced accordingly. In this case, no contact term $C$ is required to match amplitude factorization. The result is
\begin{tcolorbox}
\ie
\langle \bar\chi \chi \bar\chi \chi \rangle
        &=
        \frac{\bar u_1 u_2 \cdot \bar u_3 u_4}{E_T E_{12s} E_{34s}} + (1 \leftrightarrow 3)
\fe
\end{tcolorbox}
In the spinor-helicity basis, the non-vanishing components are
\begin{tcolorbox}
\ie
    \langle \bar\chi^- \chi^- \bar\chi^+ \chi^+ \rangle
        &=
        \frac{16\langle 1 2 \rangle \langle \bar 3 \bar 4 \rangle}{E_T E_{12s} E_{34s}}
        \,\,;
	\langle \bar\chi^+ \chi^- \bar\chi^- \chi^+\rangle = \langle \bar\chi^- \chi^- \bar\chi^+ \chi^+\rangle|_{1\leftrightarrow 3}	
	,\\
	\langle \bar\chi^+ \chi^+ \bar\chi^+ \chi^+ \rangle
		&=
		-\frac{16\,\langle \bar{1}\bar{2}\rangle
		\langle \bar{3}\bar{4}\rangle}
		{E_T E_{12s} E_{34s}},
		\\
	\langle \phi \bar{\chi}^+ \phi \chi^+ \rangle
        &=
		\frac{
		2(E_s+E_4)\langle\bar{2}\bar{4}\rangle
		+\langle 3\bar{2}\rangle\langle\bar{3}\bar{4}\rangle
		-\langle\bar{2}\bar{3}\rangle\langle 3\bar{4}\rangle
		}{
		E_s E_{12s} E_{34s}
		}+ (1 \leftrightarrow 3),
		\\
    \langle \phi \bar{\chi}^- \phi \chi^+ \rangle
        &=
        \frac{
            4 \langle 2 3 \rangle \langle \bar 3 \bar 4 \rangle
        }{E_T E_{12s} E_{34s}}
        -
        \frac{
            2E_{34S}\langle 2 \bar 4 \rangle    
        -
            2\langle 2 3 \rangle \langle \bar 3 \bar 4 \rangle
        }{E_s E_{12s} E_{34s}}+ (1 \leftrightarrow 3).
\fe
\end{tcolorbox}
All other helicity components, except for those which are opposite to the ones above, vanish.

\paragraph{$\underline{\langle T \bar{\chi} T \chi \rangle}$}
~\\~\\
The longitudinal mode is completely determined by the WT identity, so we focus on the pure transverse part of the WFC:
\begin{equation}
    \langle T^{TT} \bar{\chi}\, T^{TT} \chi \rangle 
	=
	\sum_{s,t,u}\left(\frac{A_{R,T\bar\chi T\chi}^e}{E_R^e}+\frac{B_{L,T\bar\chi T\chi}^e}{E^e_L}\right)+\frac{C_{T\bar\chi T\chi}}{E_T}+D_{T\bar\chi T\chi}\,.
\end{equation}
We define the $s$ and $t$-channels for fermion exchange and the $u$-channel for graviton exchange. Initially, the $s$-channel partial energy pole residue matches $\langle J^{T}_1 \bar{\chi_2}^{-} J^{T}_3 \chi^{+}_4 \rangle$, except for the factor $(2\epsilon_1^T \cdot p_2) \cdot (-2\epsilon_3^T \cdot p_4)$. The $t$-channel partial energy pole residue is similar, with an additional factor. Thus, we can extend the bootstrapped result in \eqref{eq:JchiJchitrans} to:
\begin{tcolorbox}[text width= 417pt, top=0pt, boxsep=0pt]
\ie
&\sum_{s,t} \left(\frac{A_{R,T\bar\chi T\chi}^e}{E_R^e}+\frac{B_{L,T\bar\chi T\chi}^e}{E^e_L}\right)\\
&=
4i(\epsilon_1^T \cdot p_2)  (\epsilon_3^T \cdot p_4) 
\,
\bar u_2 \vecslashed{\epsilon}_1^T 
	 \left[ \frac{ (\slashed{p}_{3}^{[4]}+\slashed{p}_{4}^{[3]}) }{E_T E_{12s} E_{34s} } - \frac{1-i\gamma_0}{2}\frac{\slashed{p}^{[4]}_{s,-}}{E_s E_{12s} E_{34s}} \right] 
\vecslashed{\epsilon}_3^T u_4 +(1\leftrightarrow 3)\,.
\fe
\end{tcolorbox}
Next, we focus on the $u$-channel. We derive $A_R^u$ and $B_L^u$ for the $u$-channel exchange to match the residue of $E_R^u=E_{24u}$,  $E_L^u=E_{13u}$ poles in \eqref{eq: bosonic partial energy pole}, following the procedure outlined at the beginning of the section:
\ie
&A_{R,T\bar\chi T\chi}^u
=
\frac{1}{2E_u}\left(\frac{1}{E_{13u}}-\frac{1}{E_T}\right) N^u_{T\bar\chi T\chi}\,,~~~~B_{L,T\bar\chi T\chi}^u 
=
-\frac{i}{4E_u E_T}
N^u_{T\bar\chi T\chi}\,,
\fe
where the shorthand terms are defined as:
\begin{align}
\notag N^u_{T\bar\chi T\chi}
&:=	2\left\{ 
\begin{array}{cc} 
		& 4\left(p_{3} \cdot \epsilon_{1}^{ T}\right)\left(p_{2} \cdot \pi_u \cdot \epsilon_{3}^{ T}\right) -4\left(p_{1} \cdot \pi_u \cdot \epsilon_{3}^{ T}\right)\left(p_{2} \cdot \epsilon_{1}^{ T}\right)\\
        &+(\epsilon_{3}^T \cdot \epsilon_{1}^T)\left[( p_{2} -p_{4})\cdot\pi_u\cdot ( p_{1} -p_{3}) \right]
	\end{array}\right\} \left[L_u \cdot(\bar u_2\slashed{\pi}_{u} u_4)\right]
\\[2mm]
\notag	&~~~~-
	 (
    L_u\cdot \pi_u \cdot L_u
    ) 
	\left[(p_2-p_4)\cdot (\bar u_2\slashed{\pi}_{u} u_4)\right]\,,
	\\
\label{eqn: definition of J3L} 
(L_u)_i
&:= 
	(\epsilon_1^T \cdot \epsilon_3^T)( p_1 -p_3)_i
	 +2(\epsilon_1^T \cdot p_3) (\epsilon_3^T)_i
	 -2(\epsilon_3^T \cdot p_1) (\epsilon_1^T)_i\,.
\end{align}
Finally, we can write down the $C_{T\bar\chi T\chi}$ term in \eqref{eqn: 4-point WFC ansatz} by \eqref{ctamp}. 
For convenience, we incorporate the $C_{T\bar\chi T\chi}/E_T$ term into the $u$-channel $A_{R,T\bar\chi T\chi}^u/E_R^u+B_{L,T\bar\chi T\chi}^u/E_L^u$.
We can reorganize the result as follows:
\begin{tcolorbox}[text width= 417pt, top=6pt, boxsep=0pt]
\ie
\left(\frac{A_{R,T\bar\chi T\chi}^u}{E_R^u}+\frac{B_{L,T\bar\chi T\chi}^u}{E^u_L}\right)+\frac{C_{T\bar\chi T\chi}}{E_T}
=
	i\frac{4\mathcal{N}^u_{T\bar\chi T\chi}
	- 2E_T T^c_{T\bar\chi T \chi}
	}{E_TE_{13u}E_{24u}}
		+
	i\frac{ \mathcal{N}^c_{T\bar\chi T\chi}}{E_T}\,,\\[5pt]
\fe    
\end{tcolorbox}
where we define:
\begin{equation}
\begin{array}{rcl}
      \mathcal{N}^u_{T\bar\chi T\chi}&:=&\left\{ 
		\begin{array}{c}
		\vphantom{\frac{1}{2}} \left(p_{3} \cdot \epsilon_{1}^{T}\right)\left(p_{2} \cdot \epsilon_{3}^{T}\right) 
		-
		\left(p_{1} \cdot \epsilon_{3}^{T}\right)\left(p_{2} \cdot \epsilon_{1}^{T}\right) \\
		+\frac{(\epsilon_{3}^T \cdot \epsilon_{1}^T)}{4}\left[( p_{2} -p_{4})^{\mu }( p_{1} -p_{3})_{\mu }- E_T \frac{(E_1-E_3)(E_2-E_4)}{E_u} \right]
		\end{array}
	\right\} \\[6mm]
       && \times \bar u_2
	\left[\slashed{L}_u-(\epsilon^T_1\cdot \epsilon^T_3)(E_1-E_3)\left(1+\frac{E_T}{E_u}\right)\gamma_0 \right]
	u_4\,,\\[6mm]
         	T^c_{T \bar \chi T \chi}&:=&\displaystyle\left(\frac{E_2-E_4}{4E_u^3}\right)\left\{\begin{array}{c}2(E_T E_u +E_u^2 + E_{13} E_{24} ) (E_1-E_3)^2\\
     -E_u E_{13u} 
			\left(
				E_u^2-E_{24}^2
			\right)
\end{array}\right\}
		 (\epsilon_1^T \cdot \epsilon_3^T)^2  (\bar u_2 \gamma_0 u_4)\,,\\[8mm]
	\mathcal{N}^c_{T\bar\chi T\chi}&:=&\bar u_2 
		\left\{
			\begin{aligned}
		&(\epsilon_1^T \cdot \epsilon_3^T)
		\left[
			2(p_3\cdot \epsilon_1^T)\vecslashed{\epsilon_3}^T
			-2(p_1\cdot \epsilon_3^T)\vecslashed{\epsilon_1}^T
			+(\epsilon_1^T \cdot \epsilon_3^T) (\slashed{p}^{[4]}_1-\slashed{p}^{[4]}_3)
		\right]\\
		&+
		2(\epsilon_1^T \cdot \epsilon_3^T)
		\vecslashed{\epsilon_1}^T(\slashed{p}_1^{[4]}+\slashed{p}_4^{[4]}) \vecslashed{\epsilon_3}^T
		\end{aligned}
		\right\}
		u_4\,.\label{eqn: Tc definition}
\end{array}    
\end{equation}

Furthermore, to ensure the WFC satisfies the full cutting rule, we find the mismatch occurs only in the u-channel. By comparing the RHS of the cutting rule in \eqref{eqn: cutting rule}, we can add the $D$ term to the WFC to satisfy the full cutting rule:
\begin{tcolorbox}[text width= 417pt, top=4pt, boxsep=0pt]
\ie
&D_{T \bar\chi T\chi}
=
i
	\frac{(E_1-E_3)^2(E_2-E_4)}{E_u^3}
\,(\epsilon^T_1\cdot \epsilon^T_3)^2\,(\bar u_2
\gamma_0 u_4). 
\fe    
\end{tcolorbox}

\paragraph{$\underline{\langle T \bar{\psi } T \psi \rangle}$}~\\~\\
The longitudinal mode is completely determined by the WT identity, so we focus on the pure transverse part of the WFC:
\begin{equation}
    \langle T^{TT} \bar{\psi}^T T^{TT} \psi^T \rangle 
	=
	\sum_{s,t,u}\left(\frac{A_{R,T\bar\psi T\psi}^e}{E_R^e}+\frac{B_{L,T\bar\psi T\psi}^e}{E^e_L}\right)+\frac{C_{T\bar\psi T\psi}}{E_T}+D_{T\bar\psi T\psi}\,.
\end{equation}
We define the $s$ and $t$-channels for gravitino exchange and the $u$-channel for graviton exchange.

First, we focus on the $s$-channel. We derive $A_R^s$ and $B_L^s$ for the $s$-channel exchange to match the residue of $E_R^s=E_{34s}, E_L^s=E_{12s}$ poles in \eqref{eq: Spinor Partial Pole}, following the procedure outlined at the beginning of the section:
\ie
\label{eqn: ABLs definition for TpsiTpsi}
&A_{R,T\bar\psi T\psi}^s
=
\frac{(L^s \cdot \pi_s \cdot R^s )}{2E_s}\left(\frac{N^s_{T\bar\psi T\psi}}{E_{12s}}-\frac{N^s_{R,T\bar\psi T\psi}}{E_T}\right)\,,~~~~B_{L,T\bar\psi T\psi}^s 
=
-(L^s \cdot \pi_s \cdot R^s )\frac{N^s_{L,T\bar\psi T\psi}}{2E_s E_T}\,,
\fe
where the shorthand terms are defined as:
\ie
\label{eqn: N of TpsiTpsi s-channel definition}
N^s_{T\bar\psi T\psi}
&:=
i\bar u_2 
\vecslashed{\epsilon}_1^T
	\gamma _{0}\slashed{p}_{s,-}^{[4]}  
	\vecslashed{\epsilon}_3^T u_4,\, 
N^s_{R,T\bar\psi T\psi}
:=
i\bar u_2 
\vecslashed{\epsilon}_1^T
	\slashed{p}_{s,-}^{[4]}  
	\vecslashed{\epsilon}_3^T u_4,\, 
N^s_{L,T\bar\psi T\psi}
:=-i
\bar u_2 
\vecslashed{\epsilon}_1^T
	\slashed{p}_{s}^{[4]}  
	\vecslashed{\epsilon}_3^T u_4\,.\\
(L_s)_i&:= (L_u)_i|_{3 \to 2}, \quad (R_s)_i:= (L_s)_i|_{1 \to 3 \atop 2 \to 4}\,,
\fe
and the $L_u$ is already defined in \eqref{eqn: definition of J3L}.
The $t$-channel terms, $A^t_{T\bar\psi T\psi}$ and $B^t_{T\bar\psi T\psi}$, are derived by exchanging the momentum labels $1$ and $3$ in the $s$-channel results. To determine the $u$-channel terms, $A^u_{T\bar\psi T\psi}$ and $B^u_{T\bar\psi T\psi}$, which involve graviton exchange, we align them with the residues of the poles $E_R^u=E_{24u}$ and $E_L^u=E_{13u}$ as specified in \eqref{eq: bosonic partial energy pole}:
\ie
&A_{R,T\bar\psi T\psi}^u
=
\frac{1}{4E_u}\left(\frac{1}{E_{13u}}-\frac{1}{E_T}\right) N^u_{T\bar\psi T\psi}\,,~~~~B_{L,T\bar\psi T\psi}^u 
=
-\frac{N^u_{T\bar\psi T\psi}}{4E_u E_T}
\,,
\fe
where the shorthand terms are defined as:
\begin{equation}
\hspace{-0.7cm}\begin{array}{rcl}
      \label{eqn: N of TpsiTpsi uchnl definition}
N^u_{T\bar\psi T\psi}
&:=&i
	\left(
        L_u\cdot \pi_u \cdot L_u
	\right)
	\left[
        R_u
        \cdot 
		\left(
			\bar u_2
			\slashed{\pi}_u
			u_4
		\right)
	\right]+2i\left(        L_u\cdot \pi_{u}\cdot R_u
	\right) 
\left[     L_u\cdot (\bar u_2 \slashed{\pi}_u u_4)\right]\,,\\[4mm]
(R_u)_i&:=&(L_u)_i\big|_{1\rightarrow 2\atop 3\rightarrow 4} =
	(\epsilon_2^T \cdot \epsilon_4^T)( p_2 -p_4)_i
	 +2(\epsilon_2^T \cdot p_4) (\epsilon_4^T)_i
	 -2(\epsilon_4^T \cdot p_2) (\epsilon_2^T)_i\,,
\end{array}
\end{equation}
in which the $L_u$ already defined in \eqref{eqn: definition of J3L}. Next, we can extract the contact term contribution $C_{T\bar\psi T\psi}$ via \eqref{ctamp}. Combining all the contributions, we have 
\begin{tcolorbox}[text width= 417pt, top=4pt, boxsep=0pt]
\ie
&\sum_{s,t,u}\left(\frac{A_{R,T\bar\psi T\psi}^e}{E_R^e}+\frac{B_{L,T\bar\psi T\psi}^e}{E^e_L}\right)+\frac{C_{T\bar\psi T\psi}}{E_T}\\
&=
-\mathcal{N}^s_{JJJJ}
\bar u_2 \vecslashed{\epsilon}_1^T 
	 \left[ \frac{ (\slashed{p}_{3}^{[4]}+\slashed{p}_{4}^{[3]}) }{E_T E_{12s} E_{34s} } - \frac{1-i\gamma_0}{2}\frac{\slashed{p}_{s,-}^{[4]}}{E_s E_{12s} E_{34s}} \right] 
\vecslashed{\epsilon}_3^T u_4 + (1\leftrightarrow 3)\\
&~~~~~+
\frac{1}{E_TE_{13u}E_{24u}} 
	\left[
	 \mathcal{N}^u_{T\bar\psi T\psi}
	+ 
	2
	(\epsilon^T_4 \cdot \epsilon^T_2) E_T T^c_{T \bar \chi T \chi}
	\right]
+
\frac{i\mathcal{N}^c_{T\bar\psi T\psi}}{2E_T}\,,
\fe    	
\end{tcolorbox}
~\\
~\\[-8mm]
in which we define the shorthand notation:
\ie
\mathcal{N}^s_{JJJJ} &:=
		(L_s\cdot   \pi_{s}\cdot  R_s )
		-
		(\epsilon_1^T \cdot \epsilon_2^T) (\epsilon_3^T \cdot \epsilon_4^T) E_{12s} E_{34s} \frac{(E_1-E_2)(E_3-E_4)}{E^2_s} \,,\\
	\mathcal{N}^u_{T\bar\psi T\psi} &:=
	i\left[
		(L_u\cdot   {\pi_{u}} \cdot R_u )
		-
		(\epsilon_1^T \cdot \epsilon_3^T) (\epsilon_2^T \cdot \epsilon_4^T) E_{13u} E_{24u} \frac{(E_1-E_3)(E_2-E_4)}{E^2_u} 
	\right] \\
 &  ~~~~~~ \times \bar u_2
	\left[\slashed{L}_u-(\epsilon^T_1\cdot \epsilon^T_3)(E_1-E_3)\left(1+\frac{E_T}{E_u}\right)\gamma_0 \right]
	u_4\,,\\
	\mathcal{N}^c_{T\bar\psi T\psi} &:=
	(U_t-U_u)\left[\slashed{L}_u-(E_1-E_3)(\epsilon_1^T\cdot\epsilon_3^T)\gamma_0\right]
    -
    	U_s\left[
		\slashed{\epsilon_1}^T (\slashed{p}^{[4]}_1+\slashed{p}^{[4]}_3+2\slashed{p}^{[4]}_4)\slashed{\epsilon_3}^T 
			\right]\,,\\[2mm]
    U_s &:= 
	2(\epsilon_1^T \cdot \epsilon_3^T)(\epsilon_2^T \cdot \epsilon_4^T)
	-(\epsilon_1^T \cdot \epsilon_2^T)(\epsilon_3^T \cdot \epsilon_4^T)
	-(\epsilon_1^T \cdot \epsilon_4^T)(\epsilon_2^T \cdot \epsilon_3^T)
	\,,\\[2mm]
    L_s&:=L_u\big|_{3\rightarrow 2}\,,~~~R_s:=R_u\big|_{2\rightarrow 3}\,,~~~
    U_t:=U_s|_{1\leftrightarrow 2}\,,~~~U_u:=U_s|_{1\leftrightarrow 4}\,.
\fe
Furthermore, to ensure the WFC satisfies the full cutting rule, we compare the RHS of the cutting rule relevant to the gravitino exchange in \eqref{eqn: cutting rule}, and add the $D$ term to the WFC to satisfy the full cutting rule:
\footnote{Here, we use the useful identity \eqref{useful identity for gravitino matching}}
\begin{tcolorbox}[text width= 417pt, top=-3pt, boxsep=0pt]
\ie
D_{s+t,T\bar\psi T\psi}
&= -\frac{i}{2} E_{12s}\left(1-\frac{E_1-E_2}{E_s}\right)(\epsilon_1^T \cdot \epsilon_2^T)(\epsilon_3^T \cdot  \epsilon_4^T) \\
&~~~~~\times
	\bar u_2  
	\vecslashed{\epsilon}^T_1 
     (1 -i\vecslashed{\hat p}_s)
	\left(\frac{1-i\gamma_0}{2}\right)
	\vecslashed{\hat p}_s
	\left(\frac{1+i\gamma_0}{2}\right) 
	\left(\frac{E_3-E_4}{E_s}+i\vecslashed{\hat p}_s\right)
	\vecslashed{\epsilon_3}^T u_4+ (t)
\fe    
\end{tcolorbox}
We also compare the RHS of the  cutting rule relevant to the graviton exchange \eqref{eqn: cutting rule}, and add a $D$ term to the WFC to satisfy the full cutting rule:
\begin{tcolorbox}[text width= 417pt, top=4pt, boxsep=0pt]
\ie
D_{u,T\bar\psi T\psi}
&=
i \frac{(E_1-E_3)^2(E_2-E_4)}{E_u^3}
	(\epsilon_{1}^{T} \cdot \epsilon_{3}^{T})^2 (\epsilon_{2}^{T} \cdot \epsilon_{4}^{T}) (\epsilon_1^T \cdot p_2)\, \bar u_2 \gamma_0 u_4.
\fe    
\end{tcolorbox}

The total $D$ term is the sum of the above two $D$ terms:
\ie
D_{T\bar\psi T\psi}
&=
D_{s+t,T\bar\psi T\psi}
+
D_{u,T\bar\psi T\psi}.
\fe

\section{Conclusions}
In this paper, we bootstrap flat-space Wave Function Coefficients (WFCs) using the S-matrix as input data. From the boundary perspective, the flatness of the bulk manifests in the analytic structure of the WFCs with respect to the energy variables—the total energy $E_T$ and the partial energies $E^e_{L,R}$. The residues of the partial-energy poles are fixed by cutting rules, which we have derived for fermionic exchanges for the first time. With these ingredients, we show that the four-point WFC can be constructed systematically and uniquely, without any additional ansatz. This demonstrates that the consistency of flat-space WFCs imposes no further constraints on the underlying theory beyond those already required by a consistent S-matrix. This conclusion can be understood more directly in the helicity basis: the WFCs vanish for helicity configurations that do not admit a flat-space amplitude limit. 

In a sense the result is expected. Given a consistent flat-space theory, introducing a boundary merely introduces a need for appropriate boundary conditions, which only pertains to the quadratic part of the action, and is therefore insensitive to the interactions (see~\cite{Witten:2018lgb} for a comprehensive discussion).  This is ofcourse no longer true in curved spacetime, as the interactions must be consistent with the isometries of the background. We will explore this in more detail in~\cite{CurvedPaper}. For color ordered amplitudes, by now there are many successful examples where the amplitude is identified as a geometric object~\cite{Arkani-Hamed:2012zlh,   Arkani-Hamed:2013jha, Arkani-Hamed:2017mur, Arkani-Hamed:2023jry}. In these constructions, there is a separation between the kinematics and the dynamics: the dynamics is encoded in the geometry defined in kinematic space. The current discussion highly suggests that the WFCs for these theories in flat space share similar geometry, defined in a kinematic space where four-dimensional Poincare invariance is broken down to three-dimensional one.  

\section*{Acknowledgments}
We thank Daniel Baumann, Harry Goodhew and Jiajie Mei for enlightening discussions. Y-t H  thanks Riken iTHEMS and the Yukawa Institute for
Theoretical Physics at Kyoto University. Discussions during
“Progress of Theoretical Bootstrap” were useful in completing this work.
Y-t H Z-X H and Y L are supported by the Taiwan National Science and Technology Council grant 112-2628-M-002-003-MY3 and 114-2923-M-002-011-MY5. W.-M. C. is supported by the NSTC through Grant NSTC 112-2112-M-110-013-MY3 and NSTC 114-2811-M-110-009.

\appendix
\section{Conventions}
\label{app: conventions}
In our convention, all the field in position space is expanded in the momentum space under $\varphi_{\zero}(x) = \intp{} \varphi_\zero(p) e^{- i p \cdot x}$.

For the signature in the paper, we use the metric $\eta_{\mu\nu}=(-1,1,1,1)$ and the following gamma matrix conventions:
\begin{align*}
    &\gamma_{a} = (\gamma_0,\gamma_1,\gamma_2,\gamma_3), \quad \gamma_0^2 = -1,  \quad \gamma_i^2 = 1, \quad \gamma_5= -i \gamma_0 \gamma_1 \gamma_2 \gamma_3 
\end{align*}
Dirac conjugation is defined by $\bar \chi = \chi^\dagger (\gamma_0),\gamma_0^2=1$. However, the reader could use the $\bar \chi = \chi^\dagger (i\gamma_0\gamma_5),\gamma_5^2=1$ instead, under that convention, the one should choose $(\bar \chi_+,\chi_+)$ pair to impose the Dirichlet boundary condition.

In our convention, with the gamma matrices given by
\begin{equation}
    \gamma_0 =\begin{pmatrix}
         ~-i& ~~0~~~  \\
         ~~0& ~~i~~~
    \end{pmatrix},\quad~~~
    \gamma_i=\begin{pmatrix}
         ~0&~~\sigma_i~~  \\
         ~\sigma_i&~~ 0 ~~
    \end{pmatrix}, \quad
    \gamma_5=\begin{pmatrix}
         ~0&~~-i~~  \\
         ~i&~~ 0 ~~
    \end{pmatrix}.
\end{equation}
in which Pauli matrices read,
\beqa
(\sigma_1)^a_{~\,b}=\left(\begin{array}{cc}
0&~~1\\
1&~~0
\end{array}\right)\,\,,~~(\sigma_2)^a_{~\,b}=\left(\begin{array}{cc}
0&~-i\\
i&~~0
\end{array}\right)\,\,,~~(\sigma_3)^a_{~\,b}=\left(\begin{array}{cc}
1&~~0\\
0&~-1
\end{array}\right)\,,
\eeqa
where they satisfy the familiar equation for Pauli matrices\,,
\beqa
(\sigma_\mu)^a_{~\,b} (\sigma_\nu)^b_{~\,c} =\delta^a_c \delta_{\mu\nu}+i\epsilon_{\mu\nu\rho} (\sigma^\rho)^a_{~\,c}\,.
\eeqa
The $SU(2)$ spinor indices can be raised and lowered by $\epsilon_{ab}$ and $\epsilon^{ab}$ as 
\beqa
\epsilon^{cb} T^{ a}_{\,~ b }= T^{ a c}\,,~~~~\epsilon_{ca} T^{a}_{\,~b}= T_{cb}\,,
\eeqa
and moreover, we define
$
\epsilon_{12}=-\epsilon^{12}=1\,.
$
The 3D spatial momentum could be expressed in the spinor basis as
\beqa
p_i (\sigma^i)_{ab}\equiv p_{ab}=\frac{1}{2}\left(\lambda_a\bar \lambda_b+\lambda_b\bar \lambda_a\right)=\lambda_{(a}\bar \lambda_{b)}\,.
\eeqa
We also define the inner product of two spinors\,,
\beqa
\braketw{i}{j}\equiv \epsilon_{ab}\lambda^a_i  \lambda^b_j=\lambda^a_i  \lambda_{j,a}\,.
\eeqa
Now the on-shell condition could be written as
\beqa
E^2=-\frac{1}{2} p_{ab}p^{ab}=\frac{\braketw{\lambda}{\bar\lambda}^2}{4}\,,
\eeqa
and a consistent choice is
\beqa
E\equiv -\frac{\braketw{\lambda}{\bar \lambda}}{2}\,.
\eeqa
 
Transverse polarization vectors in the helicity basis are\,,
\ie
\label{eq: SPH polarization}
(\epsilon^{(-)})^{ab}=\frac{\lambda^{a}\lambda^{b}}{\braketw{ \lambda}{\bar \lambda}}\,\,,~~~({\epsilon}^{(+)})^{ab}=\frac{\bar \lambda^{a}\bar \lambda^{b}}{\braketw{\lambda}{\bar \lambda}},
\fe
which satisfy
\begin{equation}
(\epsilon^{(+)})_{ab}(\epsilon^{(-)})^{ab}=1,\quad (\epsilon^{(\pm)})_{ab}(\epsilon^{(\pm)})^{ab}=0 ,\quad k_{ab}(\epsilon^{(\pm)})^{ab}=0.    
\end{equation}
And the 4D spinor helicity form of $\bar u$ and $u$ could be obtained by the following procedure. First, we could insert the 3D helicity spinors $ \bar\lambda_a, \lambda_a$ as $\bar{\chi}_{\zero,a},\chi_{\zero,a}$ into \eqref{eq: spinor embedding} to get their 4D embeddings
\ie
    \bbarchi^{(+)}&=\Big(0,~\sqrt{2}\bar{\lambda}^a\Big),\quad\bbarchi^{(-)}=\Big(0,~\sqrt{2}\lambda^a\Big),\\
	\bchi^{(-)}&=\left(\begin{array}{c}
    \sqrt{2}\lambda_a\\
    0
	\end{array}\right),\quad \bchi^{(+)}= \left(\begin{array}{c}
	\sqrt{2}\bar \lambda_a\\
    0
	\end{array}\right).
\fe 
Then we could use the \eqref{eq: fermion polarization} to get the corresponding spinor helicity forms of $\bar u$ and $u$,
\ie
    \bar{u}^{(+)}&=\sqrt{2}\Big(i\bar{\lambda}^a,~\bar{\lambda}^a\Big),\quad\bar{u}^{(-)}=\sqrt{2}\Big(-i\lambda^a,~\lambda^a\Big),\\
	u^{(-)}&=\sqrt{2}\left(\begin{array}{c}
	\lambda_a
	\\ -i \lambda_a
	\end{array}\right),\quad u^{(+)}= \sqrt{2}\left(\begin{array}{c}
	\bar \lambda_a
	\\ i \bar \lambda_a
	\end{array}\right).
\fe
As a consistency check, we can find that they're also the eigenbases of $\gamma_5$, with $\gamma_5 u^{(\pm)} = \pm u^{(\pm)}, \, \bar u^{(\pm)} \gamma_5 = \pm \bar u^{(\pm)}$. And satisfy Dirac equation by construction.
\paragraph{Momentum Dependence and Energy Variables}
Throughout this paper, the momentum dependence of the operators (particles) in the WFCs (amplitudes) follows their position in the bracket from left to right unless otherwise stated. For example,
\begin{equation}
\langle\mathcal{O}\mathcal{O}\mathcal{O}\rangle=\langle\mathcal{O}_1\mathcal{O}_2\mathcal{O}_3\rangle,
\end{equation}
and similarly for the amplitudes.
Energy variables with multiple lower indices are defined as sum of the individual energies. For example,
\begin{equation}
    E_{13u}\equiv E_1+E_3+E_u.
\end{equation}
\section{Constraint on Boundary Profiles from Bulk EOM}
\label{app: BoundarConstraint}

We begin by analyzing the Einstein equations $G_{\mu\nu} = 0$ under linear perturbations of the metric:
\begin{equation}
g_{\mu\nu}(  x, x_0) = \eta_{\mu\nu} + \kappa h_{\mu\nu}(  x, x_0).
\end{equation}
At zeroth order in $\kappa$, this yields the free equation of motion for the graviton. In particular, focusing on the $00$-component,
\begin{equation}
G_{00} := \partial^{i} \partial^{j} h_{ij}(  x, x_0) - \partial_i^2 h(  x, x_0) + \mathcal{O}(\kappa) = 0, \label{eq:G00flatposition}
\end{equation}
where $h \equiv \eta^{ij} h_{ij}$. Since the higher-order terms $\mathcal{O}(\kappa)$ in the coupling constants contribute only at the next order of wavefunction coefficients (WFCs) when contracted with those in the wavefunction expansion, we can focus on the zeroth-order term in \eqref{eq:G00flatposition} and omit higher-order terms in each equation of motion discussed below. Notably, the zeroth-order term in \eqref{eq:G00flatposition} contains no time derivatives, and thus represents a purely spatial constraint. Setting $x_0 = 0$, this becomes a boundary value constraint involving only the degrees of freedom fixed by the Dirichlet boundary condition. In momentum space, this becomes
\begin{equation} \label{eq: Bpconstraint}
\pi^{ij} h_{ij, b}(  p) = 0.
\end{equation}
Now, we use the subscript $b$ to denote the boundary value of the field, $h_{ij, b}(  p):=h_{ij}(  p, x_0=0)$. In de Sitter (dS) space, where the background metric is given by $\eta_{\mu\nu,\text{dS}} = \frac{1}{H^2 x_0^2} \eta_{\mu\nu}$, the zeroth-order perturbative contribution to $G_{00}$ includes an additional term:
\begin{equation}\label{eq: D4}
G_{00} = \partial^{i} \partial^{j} h_{ij}(  x,x_0) - \partial_i^2 h(  x,x_0) + \frac{2}{x_0} \partial_0 h(  x,x_0) = 0.
\end{equation}
A similar expression holds in Euclidean AdS (EAdS). Therefore, in (EA)dS spacetimes, there is no purely spatial constraint like in flat space.

Let us see how such constrained equations are consistent with properties of polarizations tensors of gravitons. Indeed this is necessary as when we take the total energy pole residue we must recover the flat-space amplitude. For amplitudes the polarization tensor for the graviton is given by $h_{\mu\nu,p} = \epsilon_\mu \epsilon_\nu$, which satisfies $p_\mu \epsilon^\mu = 0$ and $\epsilon_\mu \epsilon^\mu = 0$, we can write
\begin{equation}
h_{00,p} = \epsilon_0^2 = (\epsilon_i \hat{p}^i)^2 = \epsilon_i^2 \quad \rightarrow \quad \pi^{ij} \epsilon_i \epsilon_j = \pi^{ij} h_{ij,p} = 0.
\end{equation}
Thus this is consistent with eq.(\ref{eq: Bpconstraint}). For de-Sitter space, we see that the constraint equations emerge  in the asymptotic past $x_0 \to -\infty$, where the $1/x_0$ in eq.(\ref{eq: D4}) vanishes. 

On the other hand, a similar purely spatial constraint applies to the gravitino under Dirichlet boundary conditions. To derive it, we linearly combine the equations of motion for the free gravitino:
\begin{equation}
\left( \frac{i - \gamma^0}{2} \right) \gamma^j \gamma^0 \left( \partial_0 \frac{\delta \mathcal{L}}{ \delta (\partial_0 \bm{\psi}_j)} - \frac{\delta \mathcal{L}}{ \delta \bm{\psi}_j} \right) 
= \partial_i \bm{\psi}^i_-(  x,x_0) - (\gamma^j \partial_j)(\gamma^k \bm{\psi}_{k,-}(  x,x_0)) = 0.\label{eq:psiconstraintflat}
\end{equation}
Setting $x_0=0$ in momentum space and , this becomes
\begin{equation}
\label{eq:flat boundary gravitino EOM}
\pi^{ij} \bm{\psi}_{i,-,b} = 0.
\end{equation}
in which we also use the subscript $b$ to denote the boundary value of the field, $\bm{\psi}_{i,-,b}(  p) := \bm{\psi}_{i,-}(  p, x_0=0)$. In EAdS space, the combination of EOMs is more involved. Following the analysis in~\cite{Corley:1998qg}, we obtain the boundary constraint
\begin{equation}
\label{eq: boundary gravitino EOM}
\pi^{ij} \bm{\psi}_{i,-,p} = \frac{ 2\vecslashed{\hat{p}}}{E^2 x^2_0} \left( x_0 \partial_0 - \frac{1}{2} \right) (\gamma^i \bm{\psi}_{i,-,p}),
\end{equation}
which includes a time derivative of the boundary value and reflects the Dirichlet boundary condition. A similar structure appears in dS space. Thus, in (EA)dS, there is no purely spatial constraint analogous to the flat case.

Finally, the polarization of a massless spin-$\tfrac{3}{2}$ particle in 4D can be expressed as a tensor product of a massless spin-1 polarization $\epsilon_\mu$ and a massless spin-$\tfrac{1}{2}$ spinor $u$ satisfying $(p^\mu \gamma_\mu) u = 0$, subject to an additional 4D \emph{gamma-traceless} condition:
\begin{equation}
	\gamma^\mu(\epsilon_\mu u) = 0.
\end{equation}
This condition implies that the transverse part of the 3D polarization is also gamma-traceless:
\begin{equation}
	\gamma^\mu(\epsilon_\mu u) = 0 
	\ \rightarrow \
	({\epsilon} \cdot \hat{p})  \gamma^0 u = (\epsilon_i^T + \epsilon_i^L)\gamma^i  u 
	\ \rightarrow \
({\epsilon} \cdot \hat{p}) (\gamma^0 - \hat{\slashed{p}}) u = {\slashed{\epsilon}}_T = 0,
\end{equation}
where we have used the condition $(p_\mu \gamma^\mu) u = 0 \rightarrow (-\gamma^0 + \hat{\slashed{p}}) u = 0$. If we define $\bm{\psi}_{i,p} = \epsilon_i u$, the above equation coincides with~\eqref{eq:flat boundary gravitino EOM} under the projection operator $\frac{i - \gamma^0}{2}$.

Therefore, it is also straightforward to see that, whether in flat space or in the asymptotic past $x_0 \to -\infty$ of (EA)dS, where the additional term in~\eqref{eq: boundary gravitino EOM} vanishes, the gravitino field satisfying the equations of motion also satisfies the constraint implied by its amplitude polarization structure.

\section{WT identities}
\label{app:WTofWFC}

\subsection{Gauge Transformations: From Boundary Profiles to Classical Solutions}
We will use straightforward examples to illustrate how variations in the boundary profile lead to corresponding residual gauge transformations in the bulk classical solution, under which the action remains invariant. We demonstrate this in scalar QED. In section \ref{sec: WT identity}, we have already shown that this holds for variations of the vector field. Now, we extend the discussion to the scalar field. Let's first examine the linear order of the variation: \footnote{
	As in Section~\ref{sec: WT identity}, we use $[\delta_\alpha \phi_{\mathrm{cl}}]^{(n)}$ to denote the $n$-th order expansion of the classical solution $\phi_{\mathrm{cl}}$ after inserting the boundary variation $\delta_\alpha \phi_{\zero}$. On the other hand, we use $\delta \phi_{\mathrm{cl}}^{(n)}$ to denote the $n$-th order expansion resulting from the bulk variation $\delta \phi$ evaluated on the classical solution $\phi_{\mathrm{cl}}$. By definition, these two expansions coincide on the boundary; that is,
	$
		[\delta_\alpha \phi_{\mathrm{cl}}]\big|_{x_0=0} = \delta \phi_{\mathrm{cl}}\big|_{x_0=0} = \delta_\alpha \phi_{\zero}.
	$
}
\ie
[\delta_{\alpha} \phi_{cl}]^{(1)}(  x, x_0)
&= 
e \int d^3 x' K_{\phi} (  x-  x', x_0) \, [-i \alpha(  x') \phi_{\zero}(  x')] 
\\&+
e \int d^4 x' G_{\phi}(  x-  x', x_0, x_0') \, [2 i \p_{i',x'} \alpha(  x') \p^{i'}_{x'} \phi^{*,(0)}_{cl}(x_0',   x')]
\fe
At first glance, this does not appear to be the bulk transformation,
\begin{equation}
    \delta \phi_{\mathrm{cl}}^{(1)} = (-i e) \alpha({x})\, \phi^{(0)}_{\mathrm{cl}}(x_0, {x}).
\end{equation}
However, we will show that by using the EOM, we can rewrite the RHS to achieve our goal and generalize the result to arbitrary order. First, observe that
\ie
[\delta_\alpha \phi_{\mathrm{cl}}]^{(0)}({x}, x_0)
&=
\delta \phi^{(0)}_{\mathrm{cl}}({x}, x_0) = 0.  
\fe
We can proceed to n-th order by mathematical induction. Suppose that we have already shown
\[
    [\delta_\alpha \phi_{\mathrm{cl}}]^{(n-1)}({x}, x_0)
    =
    \delta_\alpha \phi^{(n-1)}_{\mathrm{cl}}({x}, x_0) = 0.
\]
Using the covariance of the equations of motion (EOM) under gauge transformations, we expand both sides to demonstrate the equivalence,
\footnote{
    We define $\delta \mathcal{L}_{\text{int}}/\delta \phi_{\mathrm{cl}} = [\delta \mathcal{L}_{\text{int}}/\delta \phi]|_{\phi = \phi_{\mathrm{cl}}, A = A_{\mathrm{cl}}, \ldots}$, which is the source term evaluated at the classical solution. The superscript denotes the order in the coupling constant expansion.
}
\ie
\Box( \delta \phi^{(n)}_{\mathrm{cl}} ) = \delta_\alpha \left[\frac{\delta \mathcal{L}_{\text{int}}}{\delta \phi_{\mathrm{cl}}}\right]^{(n)}
\fe
where we have used that $\delta \left[\frac{\delta \mathcal{L}_{\text{int}}}{\delta \phi^{(n)}_{\mathrm{cl}}}\right]= \delta_\alpha\left[\frac{\delta \mathcal{L}_{\text{int}}}{\delta \phi_{\mathrm{cl}}}\right]^{(n)}$, since it consists of $[\delta_\alpha \phi_{\mathrm{cl}}]^{(n-1)}$ and $[\delta_\alpha A_{\mathrm{cl}}]^{(n-1)} = \delta A^{n-1}_{\mathrm{cl}}$, as established in Section~\ref{sec: WT identity}. We can therefore rewrite the above equation and generalize it to any massless scalar theory:
\small{
\ie
\label{eq:WT from bulk:phi}
[\delta_{\alpha} \phi_{\mathrm{cl}}]^{(n)}({x}, x_0)
&= 
\int d^3 x'\, K_{\phi} ({x} - {x}', x_0)\, [\delta_{\alpha} \phi_{\zero}]^{(n)}({x}')
+
\int d^4 x'\, G_{\phi}({x} - {x}', x_0, x_0')\, \delta_\alpha \left[\frac{\delta \mathcal{L}_{\text{int}}}{\delta \phi_{\mathrm{cl}}} \right]^{(n)}({x}', x_0')
\\
&= 
\int d^3 x'\, K_{\phi} ({x} - {x}', x_0)\, [\delta_\alpha \phi_{\zero}]^{(n)}({x}')
+
\int d^4 x'\, G_{\phi}({x} - {x}', x_0, x_0')\, \Box'(\delta \phi^{(n)}_{\mathrm{cl}}({x}', x_0'))
\\
&=
\delta \phi^{(n)}_{\mathrm{cl}}({x}, x_0) 
+
\int d^3 x' 
\left[ 
    K_{\phi} ({x} - {x}', x_0)
    -
    (\partial'_0 G_{\phi})({x} - {x}', x_0, x_0' = 0)
 \right] 
[\delta_\alpha \phi_{\zero}]^{(n)}({x}')
\fe}
where we have used integration by parts twice, the vanishing of the bulk-to-bulk propagator on the boundary, $G_\phi({x} - {x}', x_0, x_0')|_{x'_0=0}=0$, and the Green's function property $\Box' G_\phi({x} - {x}', x_0, x_0') = \delta({x} - {x}') \delta(x_0 - x_0')$. Finally, by inserting the explicit forms of the bulk-to-bulk and bulk-to-boundary propagators in \eqref{eq:scalarpropagtor}, we find
\ie
(\partial'_0 G_{\phi})({x} - {x}', x_0, x_0' = 0)= K_{\phi} ({x} - {x}', x_0).
\fe
Thus, by induction, we conclude that the corresponding transformation of the classical scalar field solution is given by
\ie
[\delta_\alpha \phi_{\mathrm{cl}}]^{(n)}({x}, x_0)
&=
\delta \phi^{(n)}_{\mathrm{cl}}({x}, x_0).  
\fe
as expected. Summing over all orders, we see that the complete transformation of the classical scalar solution is indeed the bulk transformation. This procedure remains unchanged for other theories: by employing the equations of motion and the relation between bulk-to-bulk and bulk-to-boundary propagators, one can see that the corresponding transformation of the varied boundary profiles also matches the bulk gauge transformation. 

Then we could demonstrate our derivation of WT identity in the momentum space. First, under the Fourier transform, the momentum space boundary profile will be transformed as
\ie
\delta \phi_\zero(  p) &= -ie\, \int \frac{d^3 q}{(2\pi)^3} \alpha({q})\, \phi_\zero({p}-{q}), \,
\delta \phi_\zero^*(  p) = ie\, \int \frac{d^3 q}{(2\pi)^3} \alpha({q})\, \phi_\zero^*({p}-{q}), \,
\delta \epsilon_{i,\zero}(  p) = i\, p_{i}\, \alpha({p}).
\fe
Then the WT-identity will be directly from the invariance of wavefunction under the boundary profile's decomposition will be like
\ie
	0=\delta \Psi(A_{i,0},\phi_{0}) 
	&= 
		\prod_a^3 \intp{a} \delta^3 \left(\sum_a^3{p_a}\right) \\ 
	&\big\{\langle O^*_2 O_{1+3} \rangle \phi^*_\zero({p_2})\delta\phi_0({p_1}+{p_3})
		+ \langle O^*_{2+1} O_{3} \rangle \phi^*_\zero({p_2}+{p_1})\delta\phi_0({p_3})\\
		&+ \langle J_{1,i} O_2^* O_3 \rangle \delta A^i_0({p_1}) \phi^*_\zero({p_2})\phi_0({p_3})
		\big\}
 \label{WTproof}
\fe
Then we'll have the WT identity like $p_{1}^{i} \langle J_{1,i} O_{2}^{*} O_{3} \rangle =-e\ \langle O_{1+2}^{*} \ O_{3} \rangle \ +e\ \langle O_{2}^{*} \ O_{1+3} \rangle \ =\ e\ ( E_{2} -E_{3}).$ For U(1)-charged fermions the derivation is similar. Below we list for completeness all the symmetry transformation of the boundary profiles that would be used in this paper,
\ie
\label{eq:boundary_profile_transformation}
\delta \epsilon_{i,\zero} &= \partial_{i} \alpha\\
\delta h_{b,ij} \  & =2\ \partial _{( i} \xi _{j)} -2\xi _{( i}^{m} \partial h_{j) m,b} +\xi ^{m} \partial _{m} h_{ij,b}
+i \bar{\boldsymbol{\epsilon}}_{+} \gamma _{( i} \boldsymbol{\psi} _{b,j,-)} 
+
\frac{i}{2}\bar{\boldsymbol{\epsilon}}_{+} \gamma ^{a} h_{b,a( i} \boldsymbol{\psi} _{b,j,-)} +O\left( h^{2}\right) \\
\delta \bm{\chi} _{\zero,-} 
&=
-ie\alpha \bchi_{-,\zero}
+
\xi ^{m} \partial _{m} \bm{\chi} _{\zero,-} +\frac{1}{8} \partial _{a} \xi _{b}\left[ \gamma ^{a} ,\gamma ^{b}\right] \bm{\chi} _{\zero,-}\\
 & -\frac{1}{2} \xi _{a} h_{b}^{ab} \partial _{b} \bm{\chi} _{\zero,-} -\frac{1}{16} h_{b,ca} \partial ^{c} \xi _{b}\left[ \gamma ^{a} ,\gamma ^{b}\right] \bm{\chi} _{\zero,-} +O\left( h^{2}\right)\\
\delta \bm{\bar{\chi}} _{\zero,+} 
&=
ie\alpha \bbarchi_{+,\zero}
+
\bm{\bar{\chi}} _{\zero,+}\overleftarrow{\partial }_{m} \xi ^{m} -\frac{1}{8}\bm{\bar{\chi}} _{\zero,+}\left[ \gamma ^{a} ,\gamma ^{b}\right] \partial _{a} \xi _{b}\\
 & -\frac{1}{2} \xi _{a} h_{b}^{ab} \partial _{b}\bm{\bar{\chi}} _{\zero,+} +\frac{1}{16}\bm{\bar{\chi}} _{\zero,+}\left[ \gamma ^{a} ,\gamma ^{b}\right] h_{b,ca} \partial ^{c} \xi _{b} +O\left( h^{2}\right)\\
\delta \bm{\psi}_{b,-}^{i} 
&= 
\xi^{m} \partial_{m} \bm{\psi}_{b,-}^{i} 
+ 
(\partial_{i} \xi_{m}) \bm{\psi}_{b,-}^{m} 
+ 
\frac{1}{8} \partial_{a} \xi_{b} \left[\gamma^{a}, \gamma^{b}\right] \bm{\psi}_{b,-}^{i} \\
&+ 
\xi^{a} h_{b,ab} \partial^{b} \bm{\psi}_{b,-}^{i} 
+ 
(\partial_{i} \xi^{a}) h_{b,ab} \bm{\psi}_{b,-}^{b} 
+ 
\frac{1}{16} h_{b,ca} \partial^{c} \xi_{b} \left(\left[\gamma^{a}, \gamma^{b}\right] \bm{\psi}_{b,-}^{i}\right) \\
&+
\partial _{i} \ \boldsymbol{\epsilon}_{-} 
+
\frac{1}{8} \ \partial _{a} h_{b,bi}\left[ \gamma ^{a} ,\gamma ^{b}\right] \boldsymbol{\epsilon}_{-}\\
 & 
 -
 \frac{1}{16} h_{b}^{aj} \partial ^{b} h_{b,ij}[ \gamma _{a} ,\gamma _{b}] \boldsymbol{\epsilon}_{-} 
 -
 \frac{1}{32} h_{b}^{ja} \partial _{j} h_{b,i}^{b}[ \gamma _{a} ,\gamma _{b}] \boldsymbol{\epsilon}_{-} 
 + 
O(h^{3}) \\
\delta \bm{\bar{\psi}}_{b,+}^{i} 
&= 
\bm{\bar{\psi}}_{b,+}^{i} \overleftarrow{\partial}_{m} \xi^{m} 
+ 
(\partial_{i} \xi_{m}) \bm{\bar{\psi}}_{b,+}^{m} 
- 
\frac{1}{8} \bm{\bar{\psi}}_{b,+}^{i} \left[\gamma^{a}, \gamma^{b}\right] \partial_{a} \xi_{b} 
+ 
O(h_{b}) \\
&+ 
\xi^{a} h_{b,ab} \partial^{b} \bm{\bar{\psi}}_{b,+}^{i} 
+ 
(\partial_{i} \xi^{a}) h_{b,ab} \bm{\bar{\psi}}_{b,+}^{b} 
- 
\frac{1}{16} h_{b,ca} \partial^{c} \xi_{b} \left(\bm{\bar{\psi}}_{b,+}^{i} \left[\gamma^{a}, \gamma^{b}\right]\right) \\
&+
\partial _{i} \ \bar{\boldsymbol{\epsilon}}_{+} 
-
\frac{1}{8}\bar{\boldsymbol{\epsilon}}_{+} \ \ \partial _{a} h_{b,bi}\left[ \gamma ^{a} ,\gamma ^{b}\right]\\
& +
\frac{1}{16} h_{b}^{aj} \partial ^{b} h_{b,ij} \ \left(\bar{\boldsymbol{\epsilon}}_{+}[ \gamma _{a} ,\gamma _{b}]\right) 
+
\frac{1}{32} h_{b}^{ja} \partial _{j} h_{b,i}^{b}\left(\bar{\boldsymbol{\epsilon}}_{+}[ \gamma _{a} ,\gamma _{b}]\right) +O\left( h^{3}\right)
\fe
where $\alpha$ parametrizes the U(1) transform, $\xi^i$ parametrizes the diffeomorphism, $\bm{\epsilon}_-$ parametrizes the SUSY which obeys the Majorana condition $\bar{\boldsymbol{\epsilon}}_{+} =\boldsymbol{\epsilon}_{-}^{T} \ C_{-}=\boldsymbol{\epsilon}_{-}^{T}(\gamma_2\gamma_0).$ From these transformations, one can derive the WT identities for $2,3,4$ point WFCs as shown in the next subsection.	

\subsection{2-point WT identity and 2-point WFCs}
\label{app:sub:2ptWT}
The two-point WT identity based on \eqref{eq:boundary_profile_transformation} is straightforward. Following the derivation shown in \eqref{WTproof}, we can write:
\begin{align}
	0 &= p^i\langle J_i(-p) J_j(p)\rangle \\
	0 &= p^i\langle T_{ij}(-p) T_{kl}(p) \rangle\\
	0 &= p^i\langle \bar\psi_{i}(-p) \psi_{j}(p) \rangle = p^j\langle \bar\psi_{i}(-p) \psi_{j}(p) \rangle.
\end{align}
Using the WT identity and dimensional counting in the WFC expansion, we find that the only forms we can write for the two-point WFCs are:
\ie
	\langle O_{-p} O_{p}\rangle &= E \\
	\langle \bar{\chi }_{-p} \chi _{p} \rangle &=
	i \bbarchi_{\zero,-p} \vecslashed{\hat p}\bchi_{\zero,p} \\
	\langle J_i(-p) J_j(p)\rangle &= E \pi_{ij,p}\\
	\langle \bar\psi(-p) \psi(p) \rangle
	&=	
		i\bbarpsi^i_{\zero,-p} 
		P_{\psi,i i',\psi} 
		(\pi^{i'j',p} \cdot \vecslashed{\hat{p}})
		P_{\psi,j'j}
		\bpsi^j_{\zero,p} 
	= 
		i\bbarpsi^i_{\zero,-p}	
		(\hat\Pi_{ij,p}\vecslashed{\hat{p}})
		\bpsi^j_{\zero,p}\\
	\langle T_{ij}(-p) T_{kl}(p) \rangle &= E P_{h,iji'j',p} P_{kl k'l',p} (\pi^{i'j',p} \pi^{k'l',p}) = E \cdot \hat\Pi_{ijkl,p}.
\fe
Here, we set the overall normalization to 1 for bosonic fields and $i$ for fermionic WFCs. The factor of $i$ for the spinor ensures consistency with results obtained from Lagrangian calculations. Note that graviton/gravitino WFCs must be dressed with $P_h/P_{\psi}$ projectors due to the constrained boundary values of the bulk fields.
\footnote{
	For massive spinors, the two-point function should be written as $
	\langle \bar{\chi }_{-p} \chi _{p} \rangle 
	=
	\bbarchi_{\zero,-p} \cdot (i\frac{\vecslashed{p}}{E-m}) \cdot \bchi_{\zero,p}$, because now $\vecslashed{p}/E$ would have a $1/E$ pole. Notice that when $E \rightarrow 0$, we have $p^2=m^2$ instead of $p^2=0$. However, if we use $\vecslashed{p}/(E-m)$ instead, when $E \rightarrow m$ we have $p^2=E^2-m^2=0$, so there is no $1/(E-m)$ pole for real momentum. This two-point function can also be obtained from Lagrangian calculations.
}

\subsection{3pt WT identity} 
\ie
\label{WT: 3pt}
p_{1,i} {\xi}_{1,j} \langle T_{1}^{ij} T_{2} T_{3} \rangle &= 
	({\xi}_{1} \cdot {\epsilon}_{2}) \, p_{2,k} \epsilon_{2,l} \langle T_{1+2}^{kl} T_{3} \rangle 
	-\frac{1}{2} ({\xi}_{1} \cdot p_{2}) \epsilon_{2,k} \epsilon_{2,l} \langle T_{1+2}^{kl} T_{3} \rangle \\
	&\quad + ({\xi}_{1} \cdot {\epsilon}_{3}) \, p_{3,k} \epsilon_{3,l} \langle T_{2} T_{3+1}^{kl} \rangle 
	-\frac{1}{2} ({\xi}_{1} \cdot p_{3}) \epsilon_{3,k} \epsilon_{3,l} \langle T_{2} T_{3+1}^{kl} \rangle \\
p_{1}^{i} \langle J_{1,i}{\bar{\chi}}_{2} {\chi}_{3} \rangle &=
	-e\ \langle {\bar{\chi}}_{1+2} \ {\chi}_{3} \rangle \ +e\ \langle {\bar{\chi}}_{2} \ {\chi}_{1+3} \rangle \ 
	=
	e \bm{\bar{\chi}}_{2,\zero} \left(i\vecslashed{\hat p}_{2} +i\vecslashed{\hat p}_{3}\right) \bm{\chi}_{3,\zero}\\
p_{1,i} \xi _{1,j} \langle T_{1}^{ij}{\bar{\chi}}_{2} {\chi}_{3} \rangle  &=
	-
	\frac{1}{2}(p_{2} \cdot \xi _{1}) \langle {\bar{\chi}}_{1+2} \ {\chi}_{3} \rangle 
	-
	\frac{1}{2}(p_{3} \cdot \xi _{1}) \langle {\bar{\chi}}_{2} {\chi}_{1+3} \rangle \\
	&-
	\frac{1}{16}\left[\cancel{p}_{1} ,\cancel{\xi_1}\right] \langle {\bar{\chi}}_{1+2} {\chi}_{3} \rangle 
	+
	\frac{1}{16} \langle {\bar{\chi}}_{2} {\chi}_{1+3} \rangle \left[\cancel{p}_{1} ,\cancel{\xi_1}\right] \\
p_{2,k} \langle T_{1} \bar{\psi }_{2}^{k} \psi _{3} \rangle &= 
	-i\langle T_{1} T_{2+3}^{kl} \rangle \ \epsilon _{l,3} \ (\bar{\chi }_{2} \gamma _{k} \chi _{3}) 
	+ \frac{1}{8} \left[ \cancel{p}_{1}, \cancel{\epsilon}_{1} \right] \left( \epsilon _{1,k} \langle \bar{\psi }_{1+2}^{k} \psi _{3} \rangle \right) \\
p_{3,k} \langle T_{1}\bar{\psi }_{2} \psi _{3}^{k} \rangle &= 
	i\langle T_{1} T_{2+3}^{kl} \rangle \ \epsilon _{l,2} \ (\bm{\bar{\chi}}_{2} \gamma _{k} \bm{\chi}_{3}) - \frac{1}{8}\left( \epsilon _{1,k} \langle \bar{\psi }_{2}^{k} \psi _{1+3} \rangle \right)\left[\cancel{p}_{1} ,\cancel{\epsilon}_{1}\right] \\
p_{1,k} \xi_{1,l} \langle T_{1}^{kl}\bar{\psi }_{2} \bm{\psi} _{3} \rangle 
	&= 
	-\frac{1}{2} \langle \bar{{\psi}}_{2+1}  {\psi}_{3} \rangle (p_{2} \cdot \xi_{1}) 
	- \frac{1}{2} \bar{\bm{\chi}}_{2,A} \left( p_{1,k} \langle \bar{{\psi}}_{2+1}^{k,A} {\psi}_{3} \rangle \right) ({\epsilon}_{2} \cdot \xi_{1}) \\
	&\quad 
	- \frac{1}{16} (\bar{\bm{\chi}}_{2} \left[\cancel{p}_{1}, \cancel{\xi}_{1}\right]^{B}) \langle \bar{{\psi}}_{2+1,B} {\psi}_{3} \rangle \\
	&\quad 
	- \frac{1}{2} \langle \bar{{\psi}}_{2} {\psi}_{3+1} \rangle (p_{3} \cdot \xi_{1}) 
	- \frac{1}{2}({\epsilon}_{3} \cdot \xi_{1}) \left( p_{1,k} \langle \bar{{\psi}}_{2+1} {\psi}_{3+1}^{k,B} \rangle \right) \bm{\chi}_{3,B} \\
	&\quad
	+ \frac{1}{16} \langle \bar{{\psi}}_{2} {\psi}_{3+1,A} \rangle \left(\left[\cancel{p}_{1}, \cancel{{\xi}}_{1}\right] \bm{\chi}_{3}\right)^{A} 
\fe

\subsection{4pt WT identity}
\begin{equation*}
\begin{aligned}
\label{WT: 4pt}
p_{1}^{i} \langle J_{1,i}\bar{\chi }_{2} J_{3,j} \chi _{4} \rangle &=
	-e\ \langle \bar{\chi }_{1+2} J_{3,j} \chi _{4} \rangle \ +e\ \langle \bar{\chi }_{2} J_{3,j} \chi _{1+4} \rangle\\
p_{1,i} \xi _{1,j,\zero} \langle T_{1}^{ij} {\bar{\chi}} _{2} T_{3} {\chi} _{4} \rangle  
&=
-\frac{1}{2}(\xi _{1,\zero} \cdot p_{2}) \langle {\bar{\chi}} _{2+1} T_{3} {\chi} _{4} \rangle -\frac{1}{2}(\xi _{1,\zero} \cdot p_{4}) \langle {\bar{\chi}} _{2} T_{3} {\chi} _{4+1} \rangle \\
	&
	-\frac{1}{16} \bm{\bar{\chi}}_{2,A} (\left[\cancel{p}_{1} ,\cancel{\xi }_{1,\zero}\right])^{AB} \langle {\bar{\chi}}_{2+1,B} T_{3} {\chi} _{4} \rangle\\
	&+
	\frac{1}{16} \langle {\bar{\chi}} _{2} T_{3} {\chi} _{4+1,A} \rangle (\left[\cancel{p}_{1} ,\cancel{\xi }_{1,\zero}\right])^{AB}\bm{\chi}_{4,B}\\
	&+
	(\xi _{1,\zero} \cdot \epsilon_{3})p_{3,a} \langle T_{3+1}^{a}{\bar{\chi}} _{2} {\chi} _{4} \rangle -\frac{1}{2}(\xi _{1,\zero} \cdot p_{3}) \langle T_{3+1}{\bar{\chi}} _{2} {\chi} _{4} \rangle \\
	&+
	\frac{1}{2}(\xi _{1,\zero} \cdot \epsilon_{3})(\epsilon_{3} \cdot p_{2}) \langle {\bar{\chi}} _{2+3+1} {\chi} _{4} \rangle +\frac{1}{2}(\xi _{1,\zero} \cdot \epsilon_{3})(\epsilon_{3} \cdot p_{4}) \langle {\bar{\chi}} _{2} {\chi} _{4+3+1} \rangle \\
	&-
	\frac{1}{32}(p_{1} \cdot \epsilon_{3}) \bm{\bar{\chi}}_{2,A} (\left[\cancel{\epsilon}_{3} ,\cancel{\xi }_{1,\zero}\right])^{AB} \langle {\bar{\chi}} _{2+3+1,B} {\chi} _{4,C} \rangle \bm{\chi}_{4}^{C} \\
	&+
	\frac{1}{32}(p_{1} \cdot \epsilon_{3}) \bm{\bar{\chi}}_{2,A} \langle {\bar{\chi}} _{2}^{A} {\chi} _{4+3+1}^{B} \rangle (\left[\cancel{\epsilon}_{3} ,\cancel{\xi }_{1,\zero}\right])_{BC} \bm{\chi}_{4}^{C}
\end{aligned}
\end{equation*}
\ie
2p_{1,i} \xi_{1,j} \langle T_{1}^{ij} \bar{{\psi}}_{2} T_{3} {\psi}_{4} \rangle 
	&= 
	- \langle \bar{{\psi}}_{2+1} T_{3} {\psi}_{4} \rangle (p_{2} \cdot \xi_{1}) 
	- \bar{\bm{\chi}}_{2,A} \left( p_{1,k} \langle \bar{{\psi}}_{2+1}^{k,A} T_{3} {\psi}_{4} \rangle \right) ({\epsilon}_{2} \cdot {\xi}_{1}) \\
	&\quad 
	- \frac{1}{8} \bar{\bm{\chi}}_{2,A} \left[\cancel{p}_{1}, \cancel{\xi}_{1}\right]^{AB} \langle \bar{{\psi}}_{2+1,B} T_{3} {\psi}_{4} \rangle \\
	&\quad 
	- \langle \bar{{\psi}}_{2} T_{3} {\psi}_{4+1} \rangle (p_{4} \cdot \xi_{1}) 
	- ({\epsilon}_{4} \cdot {\xi}_{1}) \left( p_{1,k} \langle \bar{{\psi}}_{2+1} T_{3} {\psi}_{4+1}^{k,B} \rangle \right) \bm{\chi}_{4,B} \\
	&\quad 
	+ \frac{1}{8} \langle \bar{{\psi}}_{2} T_{3} {\psi}_{4+1,A} \rangle \left(\left[\cancel{p}_{1}, \cancel{\xi}_{1}\right]\right)^{AB} \bm{\chi}_{4,B} \\
	&\quad 
	+ 2 \left( \langle \bar{{\psi}}_{2} T_{3+1}^{ij} {\psi}_{4} \rangle {\epsilon}_{3,i} \, p_{3,j} \right) ({\xi}_{1} \cdot {\epsilon}_{3}) 
	- (p_{3} \cdot {\xi}_{1}) \langle \bar{{\psi}}_{2} T_{3+1} {\psi}_{4} \rangle \\
	&\quad 
	- ({\xi}_{1} \cdot {\epsilon}_{3}) (p_{2} \cdot {\epsilon}_{3}) \langle \bar{{\psi}}_{2+1+3} {\psi}_{4} \rangle 
	- ({\xi}_{1} \cdot {\epsilon}_{3}) ({\epsilon}_{3} \cdot {\epsilon}_{2}) \left( p_{1,k} \bar{\bm{\chi}}_{2,A} \langle \bar{{\psi}}_{2+1+3}^{k,A} {\psi}_{4} \rangle \right) \\
	&\quad 
	- \frac{1}{16} (p_{1} \cdot {\epsilon}_{3}) \bar{\bm{\chi}}_{2,A} \left( \left[\cancel{{\epsilon}}_{3}, \cancel{{\xi}}_{1}\right] \right)^{AB} \langle \bar{{\psi}}_{2+1+3,B} {\psi}_{4,C} \rangle \bm{\chi}_{4}^{C} \\
	&\quad 
	- ({\xi}_{1} \cdot {\epsilon}_{3}) (p_{4} \cdot {\epsilon}_{3}) \langle \bar{{\psi}}_{2} {\psi}_{4+1+3} \rangle 
	- ({\xi}_{1} \cdot {\epsilon}_{3}) ({\epsilon}_{3} \cdot {\epsilon}_{4}) \left( p_{1,k} \langle \bar{{\psi}}_{2} {\psi}_{4+1+3}^{k,B} \rangle \bm{\chi}_{4,B} \right) \\
	&\quad 
	+ \frac{1}{16} (p_{1} \cdot {\epsilon}_{3}) \bar{\bm{\chi}}_{2,A} \langle \bar{{\psi}}_{2}^{A} {\psi}_{4+1+3}^{B} \rangle \left( \left[\cancel{{\epsilon}}_{3}, \cancel{{\xi}}_{1}\right] \right)_{BC} \bm{\chi}_{4}^{C} \nonumber
\fe
\ie
p_{2,i} \langle T_{1}\bar{{\psi} }_{2}^{i} T_{3} {\psi} _{4} \rangle  & 
=
-i\bar{\boldsymbol{\chi} }_{2} \left( \gamma _{i} {\epsilon}_{4,j} \langle T_{2+4}^{ij} T_{1} T_{3} \rangle \right) \boldsymbol{\chi} _{4} \\
&+\frac{1}{8} \bar{\boldsymbol{\chi} }_{2,A } \left( \left[ \cancel{p}_{1} ,\cancel{{\epsilon}}_{1} \right] \right)^{A B} \left( {\epsilon}_{1,i} \langle \bar{{\psi} }_{2+1,B}^{i} T_{3} {\psi} _{4} \rangle \right)\\
& 
+\frac{1}{8} \bar{\boldsymbol{\chi} }_{2,A } \left( \left[ \cancel{p}_{3} ,\cancel{{\epsilon}}_{3} \right] \right)^{A B} \left( {\epsilon}_{3,i} \langle \bar{{\psi} }_{2+3,B}^{i} T_{1} {\psi} _{4} \rangle \right)\\
& 
-\frac{i}{2} \left( \bar{\boldsymbol{\chi} }_{2} \cancel{{\epsilon}}_{1} \boldsymbol{\chi} _{4} \right) {\epsilon}_{1,j} {\epsilon}_{4,i} \langle T_{2+4+1}^{ij} T_{3} \rangle \\
&-
\frac{i}{2} \left( \bar{\boldsymbol{\chi} }_{2} \cancel{{\epsilon}}_{3} \boldsymbol{\chi} _{4} \right) {\epsilon}_{3,j} {\epsilon}_{4,i} \langle T_{2+4+3}^{ij} T_{1} \rangle \\
& 
-\frac{1}{16} \left( \bar{\boldsymbol{\chi} }_{2} \left[ \cancel{{\epsilon}}_{1} ,\cancel{p}_{3} \right] \left( {\epsilon}_{3,i,\zero} \langle \bar{{\psi} }_{2+1+3}^{i} {\psi} _{4} \rangle \right) \boldsymbol{\chi} _{4} \right) ({\epsilon}_{1} \cdot {\epsilon}_{3})\\
& 
-\frac{1}{32} \left( \bar{\boldsymbol{\chi} }_{2} \left[ \cancel{{\epsilon}}_{1} ,\cancel{{\epsilon}}_{3} \right] \left( {\epsilon}_{3,i,\zero} \langle \bar{{\psi} }_{2+1+3}^{i} {\psi} _{4} \rangle \right) \boldsymbol{\chi} _{4} \right) ({\epsilon}_{1} \cdot p_{3})\\
& 
-\frac{1}{16} \left( \bar{\boldsymbol{\chi} }_{2} \left[ \cancel{{\epsilon}}_{3} ,\cancel{p}_{1} \right] \left( {\epsilon}_{1,i,\zero} \langle \bar{{\psi} }_{2+1+3}^{i} {\psi} _{4} \rangle \right) \boldsymbol{\chi} _{4} \right) ({\epsilon}_{1} \cdot {\epsilon}_{3})\\
& 
-\frac{1}{32} \left( \bar{\boldsymbol{\chi} }_{2} \left[ \cancel{{\epsilon}}_{3} ,\cancel{{\epsilon}}_{1} \right] \left( {\epsilon}_{1,i,\zero} \langle \bar{{\psi} }_{2+1+3}^{i} {\psi} _{4} \rangle \right) \boldsymbol{\chi} _{4} \right) ({\epsilon}_{1} \cdot p_{3}) \\
p_{1,i} \langle \bar{{\psi} }^i_{1} {\psi}_{2} \bar{{\psi} }_{3} {\psi}_{4} \rangle
&= 
- i\bar{\boldsymbol{\chi} }_{1} \left( \gamma_i {\epsilon}_{4,j} \langle {\psi}_{2} \bar{{\psi} }_{3} T^{ij}_{4+1} \rangle \right) 
-i\bar{\boldsymbol{\chi} }_{1} \left( \gamma_i {\epsilon}_{2,j} \langle T^{ij}_{2+1} \bar{{\psi} }_{3} {\psi}_{4} \rangle \right)\\ 
&- i\bar{\boldsymbol{\chi} }_{1} \left( \gamma_i {\epsilon}_{3,j} \langle T^{ij}_{3+1} \bar{{\psi} }_{2} {\psi}_{4} \rangle \right).
\fe
\section{Majorana condition from flat space amplitude}
\label{App: Majorana Condition}
We can construct the flat space amplitude $M(\psi \psi \psi \psi)$ in polarization form by gluing the left and right 3 point vertices by the polarization states over the $S, T, U$ and then adding a contact term ansatz that includes all possible terms. We apply Ward Identities under the Majorana condition to fix the contact term and obtain the final amplitude, 
\ie
M(\bar \psi_1 \psi_2 \bar\psi_3 \psi_4) 
&=
\frac{1}{S} 
	M_{\mu_s \nu_s}(\bar\psi_1 \psi_2 h_s)  
		\eta^{(\mu_s(\mu_s'} \eta^{\nu_s)\nu_s')} 
	M_{\mu'_s \nu'_s}( h_{-s} \bar\psi_3 \psi_4)\\
&-
\frac{1}{T} 
M_{\mu_t \nu_t}(\bar \psi_3 \psi_2 h_t)
	\eta^{(\mu_t(\mu_t'} \eta^{\nu_t)\nu_t')} 
M_{\mu_t' \nu_t'}(h_{-t} \bar\psi_1 \psi_4)\\
&-
\frac{1}{U} 
M_{\mu_u \nu_u}(\bar \psi_1 \psi_3 h_u) 
	\eta^{(\mu_u(\mu_u'} \eta^{\nu_u)\nu_u')} 
M_{\mu_u' \nu_u'}(h_{-u} \bar \psi_2 \psi_4)\\
&+ M_c(\bar\psi_1 \psi_2 \bar\psi_3 \psi_4)
\fe
where we define
\ie
M_{\mu_3 \nu_3}(\bar\psi_1 \psi_2 h_3)
&=
\left[
	(\epsilon_1 \cdot \epsilon_2)(p_1-p_2)_{\mu_3}
	+ (\epsilon_{2,\mu_3})(2\epsilon_1 \cdot p_2)
	+ (\epsilon_{1,\mu_3})(-2\epsilon_2 \cdot p_1)
\right]
\bar u_1 \gamma_{\nu_3} u_2.
\fe
and we fix the contact term $M_c(\bar\psi_1 \psi_2 \bar\psi_3 \psi_4)$,
\small{
\ie
M_c(\bar\psi_1 \psi_2 \bar\psi_3 \psi_4) = 
\left( 
	\begin{aligned}
		&(\epsilon_3\cdot\epsilon_2)(\epsilon_1\cdot\epsilon_4) \bar u_3 \gamma_{\mu_t} u_2 \cdot \bar u_1 \gamma^{\mu_t} u_4\\ 
		&+ \left[ 3 (\epsilon_3\cdot\epsilon_2)(\epsilon_1\cdot\epsilon_4) 
                  -(\epsilon_4\cdot\epsilon_2)(\epsilon_1\cdot\epsilon_3)    \right] 
				\bar u_1 \gamma_{\mu_u} u_3 \cdot \bar u_2 \gamma^{\mu_u} u_4\\
        &+\frac{3}{4} (\epsilon_2 \cdot \epsilon_3)\cdot 
		\epsilon_4^{\mu_s} \bar u_1 \gamma_{\mu_s} u_2 \cdot \epsilon_1^{\nu_s} \bar u_3 \gamma_{\nu_s} u_4
        +\frac{3}{4} (\epsilon_1 \cdot \epsilon_3)\cdot 
		\epsilon_4^{\mu_s} \bar u_1 \gamma_{\mu_s} u_2 \cdot \epsilon_2^{\nu_s} \bar u_3 \gamma_{\nu_s} u_4\\
        &+\frac{3}{4} (\epsilon_3 \cdot \epsilon_4)\cdot 
		\epsilon_1^{\mu_t} \bar u_3 \gamma_{\mu_t} u_2 \cdot \epsilon_2^{\nu_t} \bar u_1 \gamma_{\nu_t} u_4
        -\frac{5}{4} (\epsilon_2 \cdot \epsilon_4)\cdot 
		\epsilon_1^{\mu_t} \bar u_3 \gamma_{\mu_t} u_2 \cdot \epsilon_3^{\nu_t} \bar u_1 \gamma_{\nu_t} u_4\\
        &-2 (\epsilon_1 \cdot \epsilon_3)\cdot 
		\epsilon_4^{\mu_t} \bar u_3 \gamma_{\mu_t} u_2 \cdot \epsilon_2^{\nu_t} \bar u_1 \gamma_{\nu_t} u_4
        +2 (\epsilon_1 \cdot \epsilon_2)\cdot 
		\epsilon_4^{\mu_t} \bar u_3 \gamma_{\mu_t} u_2 \cdot \epsilon_3^{\nu_t} \bar u_1 \gamma_{\nu_t} u_4\\
        &-\frac{1}{4} (\epsilon_1 \cdot \epsilon_4)\cdot 
		\epsilon_2^{\mu_u} \bar u_1 \gamma_{\mu_u} u_2 \cdot \epsilon_3^{\nu_u} \bar u_3 \gamma_{\nu_u} u_4
        - (\epsilon_2 \cdot \epsilon_3)\cdot 
		\epsilon_4^{\mu_u} \bar u_1 \gamma_{\mu_u} u_2 \cdot \epsilon_1^{\nu_u} \bar u_3 \gamma_{\nu_u} u_4\\
        &-\frac{1}{5} (\epsilon_1 \cdot \epsilon_2)\cdot 
		\epsilon_4^{\mu_u} \bar u_1 \gamma_{\mu_u} u_2 \cdot \epsilon_3^{\nu_u} \bar u_3 \gamma_{\nu_u} u_4
	\end{aligned}   \right).
\fe}
However, we could not find any contact term to satisfy the Ward identity without the Majorana condition. Since the WFCs must match the amplitude at the total energy pole, enforcing the Majorana condition in the amplitude spinor polarization directly leads to the boundary profile relationship expressed later in Eq. \eqref{eq:3DMajornara}.

To derive this condition, note that the Majorana condition applied to the 4D polarization spinor yields:
\begin{equation}
\bar u = u^{\text{T}} C_{-}.
\end{equation}
where we use $\text{T}$ to denote the transpose and the charge conjugation operator is defined in the Section~\ref{sec:total-energy-pole}.
This relationship is automatically satisfied if we write:
\footnote{
	This follows trivially from the identities $C_{-}^2 = 1$, $C_{-} \gamma_i^T C_{-} = \gamma_{\mu}$, and the decomposition
	\begin{equation}
	\bar u_{+} = \bar u \frac{i + \gamma_{0}}{2} = u^T_{-} C_{-}
	\end{equation}.
}
\ie
u(p) = (1 - i\slashed{\hat{p}})\bchi_\zero(p) \text{  ;  }
\bar u(p) = \bbarchi_\zero(p)(1 + i \slashed{\hat{p}}).
\fe
and require that the spinor boundary profiles are related via the Majorana condition:
\begin{equation}
\bbarchi_\zero(p) = \bchi^\text{T}_\zero(p) C_{-}. \label{eq:3DMajornara}
\end{equation}
The Majorana condition on the $u$ in the spin-$\frac{3}{2}$ polarization $\epsilon_\mu u$ is satisfied provided that $\bbarchi$ and $\bchi$, as place holder defined in \eqref{eq:boundary reference polarization constraint}, are related by the condition \eqref{eq:3DMajornara}. It is straightforward to verify that if the boundary profiles of the gravitino satisfy the Majorana condition
\begin{equation}
\bbarpsi^i_\zero(p) = \bpsi^{\text{T},i}_\zero(p) C_{-}, \label{eq:3DMajornara 3/2}
\end{equation}
then the condition on the spinor placeholder \eqref{eq:3DMajornara} and Majorana condition on the spin-$\frac{3}{2}$ polarization is ensured to be satisfied.

\section{Implications of Bulk CPT on Fermionic WFCs}
\label{app:CPT}

Here, we demonstrate that the fermion action,
\ie
S 	&= S_{\chi,bulk} + S_{\chi,b} \\ 
	&= -\int^{x_0=0}_{x_0=-\infty(1-i\epsilon)}  d^4 x \ \frac{1}{2}\ \bbarchi(\not \! \partial + i e A_\mu \gamma^\mu) \bchi - \frac{1}{2}  \bbarchi(\not\! \! \overleftarrow\partial_M -ie A_\mu \gamma^\mu) \bchi +  m \bbarchi \bchi  \\ 
&+ \int \ (- i/2)\  \bbarchi_\zero \bchi_\zero \ d^3 x 
\fe
including the boundary term, remains invariant under the standard CPT transformation:\footnote{
	Note that we define $\bar u = u^\dagger \gamma_{0,M}= i u^\dagger \gamma_0$ in this gamma matrix notation. This results in the CPT of $\bbarchi$ having an additional minus sign compared to
	 the $\eta_{\mu\nu}=diag(1,-1,-1,-1)$ notation.
}
\ie
\text{CPT}:\quad & \bchi(x^\mu) \rightarrow -\gamma_5\bchi^*(-x^\mu), \quad \bbarchi(x^\mu) \rightarrow \bbarchi^*(-x^\mu)\gamma_5,\\
i &\rightarrow -i, \quad \gamma_{M,\mu} \rightarrow \gamma_{M,\mu}^*, \quad (\p_\mu\bchi)(x^\mu) \rightarrow (\gamma_5\p_\mu\bchi^*)(-x^\mu), \quad (\p_\mu\bbarchi)(x^\mu) \rightarrow -(\gamma_5\p_\mu\bbarchi^*)(-x^\mu)
\fe
with the time boundary transformation:
\ie
\text{CPT}: -\infty \leq x_0 \leq 0 \rightarrow 0 \leq x_0 \leq \infty
\fe
This transformation reflects our time domain, as also discussed in the dS literature. \cite{Goodhew:2024eup} Additionally, the boundary integral should also transforms under the CPT transformation as follows:
\ie
\text{CPT}: \int_{x_0=\epsilon<0} d^3 x \rightarrow -\int_{x_0=-\epsilon>0} d^3 x.
\fe

Then we could translate the CPT theorem from the action to the WFCs. To achieve this, we utilize the boundary profile to expand the CPT theorem, where the field is substituted into the classical solutions. For fermionic theory, the CPT theorem can be expanded as:
\footnote{
	Notice that the $CPT(S)|_{\bchi_{cl},\bbarchi_{cl}}$ is defined by the CPT transformed action inserted by the classical solution. This will equivalently send the classical solution to its CPT image in the transformed time domain, for example,
	\ie
	\label{CPT: fermion classical solution}
	CPT(\bchi_{cl}(x^\mu)) &= 
		-\gamma_5\bchi^*_{cl}(-x^\mu)\\
	CPT(\bbarchi_{cl}(x^\mu)) &= 
		\bbarchi^*_{cl}(-x^\mu)\gamma_5
	\fe
	It is straightforward to see that all the transformations mentioned above satisfy the boundary conditions in the transformed time domain. 	
}
\ie
0 &= S|_{\bchi_{cl},\bbarchi_{cl}}- CPT(S)|_{\bchi_{cl},\bbarchi_{cl}}\\
&= \sum_{n=2} \prod^{n}_{i} \intp{i} \cdot 
-i\left(c_{n} -
c_{n,CPT} \right)^{A_1 \dots}_{A_2 \dots}
\cdot
\bbarchi_{\zero,1,A_1} \bchi_{\zero,2}^{A_2} \dots \cdot \delta^3\left(\sum_a^{n}   p_a\right). \label{eq: CPT expansion for fermionic field}
\fe
Furthermore, we aim to identify the operation on the WFCs such that $c_{n,CPT} = CPT(c_{n})$, which varies based on the distinct classical solution structures. This allows us to express the CPT implication on the WFCs as:
\ie
0 = (c_{n} - CPT(c_{n}))^{A_1 \dots}_{A_2 \dots}
\cdot
\bbarchi_{\zero,1,A_1} \bchi_{\zero,2}^{A_2} \dots.
\fe

Now, by using \eqref{3pt Fermion Expansion}, we can express the CPT-transformed boundary action with the classical solution insertion as: 
\footnote{
	In the calculations here, we utilize the identity provided by the CPT invariance of the equations of motion (EOM):
	\ie
	CPT: &(\not \! \partial +m)  K_{\bchi}(x^\mu) \left(\frac{i+\gamma_0}{2}\right)= 0 
	\rightarrow \left(\frac{i-\gamma_0}{2}\right) K^{\dagger}_{\bchi}(-x^\mu)  (-\overleftarrow{\slashed{\partial}} +m)=0
	\fe
	It follows directly that \( K^{\dagger}_{\bchi}(-x_0,p) \) satisfies the same EOM and boundary conditions at both the far past and the boundary as \( K_{\bbarchi}(x_0,p) \). This implies they are identical solutions:
	\ie
	\label{Fermion CPT: K identity}
	K^\dagger_{\bchi}(-Ex_0,p)=K_{\bbarchi}(Ex_0,p)
	\fe
	The specific form of the propagator is not required to demonstrate the identity we use. Moreover, we also use the fact that we can identify the time reversal mode as the negative energy mode: \footnote{Alternatively, we could choose not to flip the sign of the energy in the projector. However, this requires converting every WFC into the projector-sandwiched form.}
	\ie
	K_{\bchi}(-x_0,p) \bchi_{\zero}(p)&= K_{\bchi}(x_0,p)|_{E\rightarrow -E}\cdot\frac{-i\vecslashed{p}}{E-m} \cdot \bchi_{\zero}(p)\\
	\bbarchi_{\zero}(p) K_{\bbarchi}(-x_0,p)&=  \bbarchi_{\zero}(p) K_{\bchi}(x_0,p)|_{E\rightarrow -E} \cdot (\frac{i\vecslashed{p}}{E-m})
	\fe
}
\ie
&CPT(S^{(1)}_{cl,b}) 
= -\int^{x_0=\infty(1-i\epsilon)}_{x_0=0} d^4x \bbarchi^{(0),*}(-x)\gamma_5 (-ig) V^*(-x,-\p_x, -\overleftarrow{\p}_x) (-\gamma_5)\bchi^{(0),*}(-x) \\
&=\int_{d p_2}\int_{d p_3}  \left[\bbarchi_{\zero,2} \gamma_5\left(\frac{i\vecslashed{p}_2}{E_2+m}\right)\right]_{A}
\left.\left(
         c^{\dagger,B}_{3,\bbarchi_2\bchi_3,A}(p_i) 
    \right)\right|_{E\rightarrow -E, 2 \leftrightarrow 3}
    \cdot 
    \left[\frac{-i\vecslashed{p}_3}{E_3+m} 
    \gamma_5 
    \bchi_{\zero,3}\right]^{B}
\fe

If a higher spin field is involved (including the spin 3/2 field, which can be expressed as $\bpsi_\mu= \epsilon_\mu \bchi$, the product of the polarization vector and the fermion field), the CPT will also flip the sign of the polarization vector of the boundary profile. Then if we only consider there's only a pair of the fermionic fields, the COT can be more generally written as:
\begin{tcolorbox}
	\ie
	&\bbarchi_{\zero,A}(p_{\bar{\mathcal{O}}_{\chi}}) \, {c}_{c,A}^B \,\bchi_{\zero}^A(p_{\mathcal{O}_{\chi}})\\
		&+ \left[\bbarchi_{\zero}(p_{\bar{\mathcal{O}}_{\chi}})\gamma_5
		\left(\frac{i\vecslashed{p}_{\bar{\mathcal{O}}_{\chi}}}{E_{\bar{\mathcal{O}}_{\chi}}+m}\right)\right]_{A}\left.
		{c}^{\dagger,A}_{c,B}(p_i)\right|_{
			\begin{pmatrix}
			E\rightarrow -E, \\
			p_{\mathcal{O}} \leftrightarrow p_{\bar{\mathcal{O}}},\\
			\epsilon_\zero \rightarrow -\epsilon_\zero
			\end{pmatrix}
		}
		\left[ 
		\left(
			\frac{i\vecslashed{p}_{\mathcal{O}_{\chi}}}{E_{\mathcal{O}_{\chi}}+m}\right)
		\gamma_5
		\bchi_{\zero}(p_{\mathcal{O}_{\chi}})\right]^B=0.
	\fe
\end{tcolorbox}
In the above, momenta associated with fields are labeled by $p_{\mathcal{O}}$ (the fermionic one, we use $p_{\mathcal{O}_{\chi}}$), while those for their conjugates are labeled by $p_{\bar{\mathcal{O}}}$ (the conjugate fermionic one, we use $p_{\mathcal{\bar{O}_{\chi}}}$), and the subscript $c$ denotes the contact WFCs. We have verified that the CPT implication on all the contact WFCs listed in the secion \ref{sec:bootstrapping}.

\section{Polarization sums and useful Identities}
\label{app:matching identity}

According to Section \ref{sec:bootstrapping}, the limit $S\rightarrow 0$ corresponds to approaching either of the partial energy poles $E_{12s} \rightarrow 0$ and $E_{34s} \rightarrow 0$\,. In these limits, the polarization sums reduce to identical forms. Our attempt is to obtain \eqref{eqn: ORTLT to polarization sum} from \eqref{PoS} under the limit mentioned above.

We can verify this in some specific theories. For example, in the four-point function where we use the subscript $M$ to denote the exchanged field, we have
\ie
\label{OOOOTotakResT}
M_{s,\text{J}}\left( \phi _{1} \phi _{2}^{*} \phi _{3} \phi _{4}^{*}\right) &= \frac{(p_{2} -p_{1})^{\mu }(p_{4} -p_{3})_{\mu}}{S},\quad \, 
M_{s,\text{T}}\left( \phi _{1} \phi _{2} \phi _{3} \phi _{4}\right) = \frac{((p_{2} -p_{1})^{\mu }(p_{4} -p_{3})_{\mu})^2}{S},\\
M_{u,\text T}(h_1 \bar\chi_2 h_3 \chi_4)
	&= 
	\frac{ (L_u \cdot (p_2-p_4)-(\epsilon_1^T \cdot \epsilon_3^T)(E_1-E_3)(E_2-E_4)) \cdot (L_u \cdot \bar u_2 \gamma u_4-(\epsilon_1^T \cdot \epsilon_3^T)(E_1-E_3)\bar u_2 \gamma_0 u_4)}{U}, \\
M_{u,\text T}(T_1 \bar\psi_2 T_3 \psi_4)
&=
	\frac{(L_u \cdot R_u-(\epsilon_1^T \cdot \epsilon_3^T)(\epsilon_2^T \cdot \epsilon_4^T)(E_1-E_3)(E_2-E_4)) \cdot (L_u \cdot \bar u_2 \gamma u_4-(\epsilon_1^T \cdot \epsilon_3^T)(E_1-E_3)\bar u_2 \gamma_0 u_4) }{U}
\fe
in which the $L_u,R_u$ are defined in \eqref{eqn: definition of J3L} and \eqref{eqn: N of TpsiTpsi uchnl definition}.
The limit \eqref{eqn: total pole of A B limit} holds under the following useful kinematic identities,
\footnote{
	We can derive some of these identities from the fact that the 4D trace of the three-point amplitude must vanish under its total energy conservation, which corresponds to the partial energy pole of the four-point WFCs. For the $u$-channel, we focus on $E_{24u} \rightarrow 0$. Therefore, the trace of the three-point amplitude form in the four-point $u$-channel WFCs should be proportional to $E_{24u}$:
	\ie
	\label{eqn: 4Dtracetchichi}
	\eta^{\mu\nu}M(h_{-u,\mu\nu} \bar \chi_2 \chi_4)
	&=\bar u_2 (\slashed{p}^{[4]}_2 - \slashed{p}^{[4]}_4) u_4 = 0\\
	&= \bar u_2[
		-(E_2-E_4)\gamma_0
		+(p_2-p_4)^i \pi_{u,ij} \gamma^j
		+ [(p_2-p_4)^i \hat p_u^i]\vecslashed{\hat p}_u 
		]u_4\\
	&=\bar u_2 [
		(p_2-p_4)^i \pi_{u,ij} \gamma^j
		-(\frac{E_2-E_4}{E_u^2})\gamma_0 
		\left(
			E_u^2-E_{24}^2
		\right)
		]u_4,
	\fe
	Then we can rewrite the second term in the last line of \eqref{eqn: 4Dtracetchichi} in terms of the first term. It is straightforward to see that \eqref{eqn: useful identity} holds under this rewriting and other useful identities. We can apply similar calculations to $\eta_{\mu\nu} M(h_1^{TT} h_3^{TT} h_u^{\mu\nu})$ and $\eta^{\mu\nu}M(h_{-u,\mu\nu}\bar \psi^\Trans_2 \psi_4^T)$ to obtain other useful identities.}
\ie
\label{eqn: useful identity}
&\text{1.} \quad ( p_{2} -p_{1})_{\mu } \eta ^{\mu \nu }( p_{4} -p_{3})_{\nu } 
	- E_{T} \cdot \frac{( E_{2} -E_{1})( E_{4} -E_{3})}{E_{s}}\\
	&\quad =
	((p_2-p_1)_i \pi_s^{ij} (p_4-p_3)_j)- \frac{(E_2-E_1)(E_4-E_3)E_{12s} E_{34s}}{E_s^2}\\
& \text{2.} \quad E_T T^C_{OOOO} 
	= 
	((p_1-p_2)_i \pi_s^{ij} (p_1-p_2)_j)((p_4-p_3)_i \pi_s^{ij} (p_4-p_3)_j) \\
	&\quad +  E_{12s} E_{34s} \Pi^C_{1,OOOO} + E^2_{12s} E^2_{34s} \Pi^C_{2,OOOO} \\
&\text{3.} \quad \bar u_2
	\left[
			\left( (p_1-p_3)_\mu\gamma^\mu
			- E_T \frac{E_1-E_3}{E_u} \gamma_0 \right)
		\right]
	u_4 =
\bar u_2 
	\left[ 
		(p_1-p_3)_i \pi^{ij}_u \gamma_j
		-E_{13u} E_{24u} \frac{E_1-E_3}{E^2_u}\gamma_0
	\right]
u_4 \\
&	\text{4.} \quad \left[
\left(p_{3} \cdot \epsilon_{1}^{T}\right)\left(p_{2} \cdot \epsilon_{3}^{T}\right) -\left(p_{1} \cdot \epsilon_{3}^{T}\right)\left(p_{2} \cdot \epsilon_{1}^{T}\right)
\right] =
\left[
\left(p_{3} \cdot \epsilon_{1}^{T}\right)\left(p_{2} \cdot \pi_u \cdot \epsilon_{3}^{T}\right) -\left(p_{1} \cdot \epsilon_{3}^{T}\right)\left(p_{2} \cdot \pi_u \cdot \epsilon_{1}^{T}\right)
\right] \\
&	\text{5.} \quad \left[
	( p_3\cdot \epsilon_1^T) (\slashed{\epsilon_{3}}^{T})
		-(p_1\cdot \epsilon_3^T)(\slashed{\epsilon_{1}}^{T})
\right]
= 
\left[
	( p_3\cdot \epsilon_1^T) (\epsilon_{3,i}^T \pi_u^{ij} \gamma_j)
		-(p_1\cdot \epsilon_3^T)(\epsilon_{1,i}^T \pi_u^{ij} \gamma_j)
\right]\\
&\text{6.} \quad E_T T^c_{T \bar \chi T \chi} =
-\frac{1}{2}( L_u \cdot \pi_u \cdot L_u) \cdot
		[(p_2-p_4)^i \pi_{u,ij} \bbarchi_{2,\zero} (1+i\vecslashed{\hat p}_2)\gamma^j (1-i\vecslashed{\hat p}_4) \bchi_{4,\zero}] \\
		&\quad+ 
		E_{24u} E_{13u}\Pi^C_{1,T\bar \chi T \chi}
		+
		E^2_{24u} E^2_{13u} \Pi^C_{2,T\bar \chi T \chi}\\
&\text{7.} \quad (
\epsilon^\Trans_4 \cdot \epsilon^\Trans_2) E_T T^c_{T \bar \chi T \chi} 
=
-\frac{1}{4}\left(
		L_u \cdot \pi_u \cdot L_u
	\right)
	\left(
		R_u \cdot \pi_u \cdot
			\bar u_2
			\bm \gamma
			u_4
	\right) \\
		&\quad+ 
		(\epsilon^T_4 \cdot \epsilon^T_2)
		E_{24u} E_{13u} \Pi^C_{1,T\bar \chi T \chi}
		+
		(\epsilon^T_4 \cdot \epsilon^T_2)
		E^2_{24u} E^2_{13u} \Pi^C_{2,T\bar \chi T \chi}
\fe
where $L_u,R_u$ is defined in \eqref{eqn: definition of J3L}, \eqref{eqn: N of TpsiTpsi uchnl definition}, $T^c_{T \bar \chi T \chi}$ is defined in \eqref{eqn: Tc definition}, and we define other completion terms as:
\ie
\Pi^C_{1,OOOO} &= -\left(\frac{E_2-E_1}{E_s}\right)^2 (-E_s^2+E_{12}^2) - \left(\frac{E_4-E_3}{E_s}\right)^2 (-E_s^2+E_{34}^2)\\
\Pi^C_{2,OOOO} &= - \left(1+ \left(\frac{E_2-E_1}{E_s}\right)^2 \left(\frac{E_4-E_3}{E_s}\right)^2 \right) \\
T^C_{OOOO} &= \big[ -2 E_s E_{12s} E_{34s} + (E_{34s})\left(\frac{E_2-E_1}{E_s}\right)^2 (E_s^2-E_{12}^2)  \\
& + (E_{12s})\left(\frac{E_4-E_3}{E_s}\right)^2 (E_s^2-E_{34}^2) \\
& + \left(\frac{E_2-E_1}{E_s}\right)^2 \left(\frac{E_4-E_3}{E_s}\right)^2(-2 E_T E_s^2-2 E_s^3-2 E_sE_{12}E_{34}) \big]\\
\Pi^C_{1,T\bar \chi T \chi} 
	&= - \left[
		\left((\frac{E_2-E_4}{E_u^2})
		\left(
			E_u^2-E_{24}^2
		\right)\right) 
	\right] 
	\cdot \frac{1}{4} (\epsilon_1^T \cdot \epsilon_3^T)^2 \bar u_2 \gamma_0 u_4\\
\Pi^C_{2,T\bar \chi T \chi} 
	&=
	(\frac{E_1-E_3}{E_u})^2 (\frac{E_2-E_4}{E_u^2})
	\cdot \frac{1}{4} (\epsilon_1^T \cdot \epsilon_3^T)^2 \bar u_2 \gamma_0 u_4
\fe
For some identities, we only list the $s$-channel version; however, it is straightforward to extend them to other channels. On the other hand, for fermion exchange, we have 
\ie
\begin{cases}
	O^s_{A,\chi}= -i (\slashed{p}_3^{[4]}+\slashed{p}_4^{[3]})\\
	O^{s,\mu\nu}_{A,\psi}= -i\eta^{\mu\nu} (\slashed{p}_3^{[4]}+\slashed{p}_4^{[3]})
\end{cases}
\fe

For the spinor, it is clear that the amplitude factorization factor for $S\rightarrow 0$ could be obtained from the total energy pole term fixed by the partial energy pole residue:
\ie
\begin{cases}
	\label{eqn: spinor matching}
	E_{12s} &= E_{34}-E_s \rightarrow 0 :O^s_{A,\chi}= -i\slashed{p}_s^{[4]} =O^s_{L,\chi}\\
	E_{34s} &= E_{12}-E_s \rightarrow 0 :O^s_{A,\chi}= i\slashed{p}_{s,-}^{[4]} =O^s_{R,\chi}.
\end{cases}
\fe

For the gravitino, we have
\small{
\ie
\label{eqn: gravitino matching}
\begin{cases}
	\begin{aligned}
	E_{12s} = E_{34}-E_s \rightarrow 0 :
		M_{L,\mu}^s & O^{s,\mu\nu}_{A,\psi}M_{L,\nu}^s\\
	&=
		M_{L,i}^s (-i\pi_{ij,s}\slashed{p}_s^{[4]} 
		+\frac{i}{2} (1-i\slashed{\hat{p}}_s)(\slashed{\pi}_s^i \slashed{p}_s \slashed{\pi}_s^j)(\frac{1-i\gamma_0}{2})(1-i\slashed{\hat{p}}_s))M_{L,j}^s\\
	&=M^s_{L,i}O^{s,ij}_{L,\psi}M^s_{L,j}
	\end{aligned}\\
	\begin{aligned}
	E_{34s} = E_{12}-E_s \rightarrow 0 :
		M_{L,\mu}^s & O^{s,\mu\nu}_{A,\psi}M_{L,\nu}^s\\
		&=
		M_{L,i}^s (i\pi_{ij,s}\slashed{p}_{s,-}^{[4]} 
		+\frac{i}{2} (1+i\slashed{\hat{p}}_s)(\slashed{\pi}_s^i \slashed{p}_s \slashed{\pi}_s^j)(\frac{1-i\gamma_0}{2})(1+i\slashed{\hat{p}}_s))M_{L,j}^s\\
		&=
		M^s_{L,i}O^{s,ij}_{R,\psi}M^s_{R,j}.
	\end{aligned}
\end{cases}
\fe 
}
We can also use $\langle T \bar \psi T \psi \rangle$ to demonstrate that the above limit effectively works for the factor under the individual branch as $S \rightarrow 0$. The amplitude factorization for the $s$-channel exchanging gravitino $M( T \bar \psi T \psi )$ is given by
\ie
M_s(h_1 \bar\psi_2 h_3 \psi_4) 
	&=
		-i(L_s \cdot R_s-(\epsilon_1^T \cdot \epsilon_2^T)(\epsilon_3^T \cdot \epsilon_4^T)(E_1-E_2)(E_3-E_4))
		\cdot
		\bar u_2 \slashed{\epsilon}_1^T \frac{ (\slashed{p}_{3}^{[4]}+\slashed{p}_{4}^{[3]})}{S}\slashed{\epsilon}_3^T \bar u_4 
\fe
in which $L_s,R_s$ are defined in \eqref{eqn: N of TpsiTpsi s-channel definition}.
Then \eqref{eqn: gravitino matching} as a gluing factor reduction under the two branches of the partial energy pole to match $S \rightarrow 0$ will hold under the following kinematic identity:
\small{
\ie
\label{useful identity for gravitino matching}
\text{8.}& \quad (L_s \cdot R_s-(\epsilon_1^T \cdot \epsilon_2^T)(\epsilon_3^T \cdot \epsilon_4^T)(E_1-E_2)(E_3-E_4))
	-
	(\epsilon_1^T \cdot \epsilon_2^T) (\epsilon_3^T \cdot \epsilon_4^T) E_T \frac{(E_1-E_2)(E_3-E_4)}{E_s} \\
&=
(L_s \cdot \pi_s \cdot R_s) 
	-
	(\epsilon_1^T \cdot \epsilon_2^T) (\epsilon_3^T \cdot \epsilon_4^T) E_{12s} E_{34s} \frac{(E_1-E_2)(E_3-E_4)}{E_s^2}\\
\text{9.}& \quad L_{s,i_s}
	\cdot 
	\bar u_2 
	\vecslashed{\epsilon}_1^T 
	(1 -i \vecslashed{\hat p}_s)
		\cdot \frac{1+i\gamma_0}{2}
		\cdot \vecslashed{\pi}^{i_s}_s \vecslashed{\hat p}_s \vecslashed{\pi}^{j_s}_s
		\cdot \frac{1-i\gamma_0}{2}
	\cdot
	\left.\left[  
	(1 - i \vecslashed{\hat p}_s) 
	\vecslashed{\epsilon}_3^T 
	u_4
	\cdot \frac{1}{E_{34s}} 
	\cdot R_{s,j_s} 
	\right]\right|^{E_s}_{-E_s}\\
&=
	-2E_{12s}\left(1-\frac{E_1-E_2}{E_s}\right)(\bm \epsilon_1^T \cdot \bm \epsilon_2^T)(\bm \epsilon_3^T \cdot \bm \epsilon_4^T) 
	\bar u_2  
	\vecslashed{\epsilon}^\Trans_1 
	\cdot (1 -i\vecslashed{\hat p}_s)
	\cdot \frac{1-i\gamma_0}{2}
	\cdot \vecslashed{\hat p}_s
	\cdot \frac{1+i\gamma_0}{2} 
	\cdot (\frac{E_3-E_4}{E_s}+i\vecslashed{\hat p}_s)
	\vecslashed{\epsilon_3}^T u_4\\
\text{10.}& \quad \left.\left[ 
	\frac{1}{E_{12s}}  
	L_{s,i_s}
	\cdot 
   \bar u_2 
   \vecslashed{\epsilon}_1^T 
   (1 - i\vecslashed{\hat p}_s)
   \right]\right|^{E_s}_{-E_s}
	   \cdot \frac{1+i\gamma_0}{2} \cdot
	   \vecslashed{\pi}^{i_s}_s \vecslashed{\hat p}_s \vecslashed{\pi}^{j_s}_s
	   \cdot \frac{1-i\gamma_0}{2} 
	\cdot
   (1 - i\vecslashed{\hat p}_s) 
   \vecslashed{\epsilon}_3^T 
   u_4
   \cdot R_{s,j_s}\\
&=
	-2E_{34s}\left(1-\frac{E_3-E_4}{E_s}\right)(\bm \epsilon_1^T \cdot \bm \epsilon_2^T)(\bm \epsilon_3^T \cdot \bm \epsilon_4^T) 
	\bar u_2  
	\vecslashed{\epsilon}^\Trans_1 
	\cdot (\frac{E_1-E_2}{E_s} +i\vecslashed{\hat p}_s)
	\cdot \frac{1-i\gamma_0}{2}
	\cdot \vecslashed{\hat p}_s
	\cdot \frac{1+i\gamma_0}{2} 
	\cdot (1-i\vecslashed{\hat p}_s)
	\vecslashed{\epsilon_3}^T u_4
\fe}
in which we use the identity derived from the 4D $\gamma$-trace of the three-point amplitude $M(h\bar\psi\psi)$ to re-express the $\slashed{\pi}^i_s$ trace term.\footnote{
	The explicit calculation shows:
	\ie
	M(T^{TT}_1 \bar \psi^{T}_2 \psi_{s,\mu}) \gamma^\mu
		&= -\bar u_2 \vecslashed{\epsilon}^\Trans_1 \left[
			(\epsilon_1^T \cdot \epsilon_2^T)((-\slashed{P_s})-2\slashed{P_2})
			+ \vecslashed{\epsilon}_2^T (\epsilon_1^T \cdot p_2)
			- 2 \vecslashed{\epsilon}_1^T (\epsilon_2^T \cdot p_1)
		\right]\\
		&= -\bar u_2 \vecslashed{\epsilon}^\Trans_1 (\epsilon_1^T \cdot \epsilon_2^T)(\slashed{P_s}+E_{12s}\gamma_0) \\
		&= -(T^{TT}_1 \bar \psi^{T}_2 \psi_{s,0}) \gamma_0
		+ M(T^{TT}_1 \bar \psi^{T}_2 \psi_{s,i}) \pi_{s}^{ij} \gamma_j
		+ M(T^{TT}_1 \bar \psi^{T}_2 \psi_{s,i}) \hat p_s^i \hat p_s^j \gamma_j\\
		&= \bar u_2 \vecslashed{\epsilon}^\Trans_1 (\epsilon_1^T \cdot \epsilon_2^T) 
		[ -(E_1-E_2)\gamma_0 + (p_1-p_2)_i \hat p_s^i \vecslashed{\hat p}_s]
			+
			L_s
			\cdot 
			\bar u_2
			\vecslashed{\epsilon}_1^T \vecslashed{\pi}^{i_s}_s \\
	\gamma^\mu M(T^{TT}_3 \bar \psi_{-s,\mu} \psi^{T}_{4})
		&= -(\slashed{p}^{[4]}_{s,-}+E_{34s}\gamma_0) (\epsilon_3^T \cdot \epsilon_4^T) \vecslashed{\epsilon_3}^T u_4 \\
		&=  [ -(E_3-E_4)\gamma_0 + (p_3-p_4)_i \hat p_s^i \vecslashed{\hat p}_s] 
			(\epsilon_3^T \cdot \epsilon_4^T) \vecslashed{\epsilon_3}^T u_4
			+ \tilde M_{i_s}(\gamma^\Trans_3 \gamma^\Trans_4 \gamma_{-s}) 
			(\epsilon_3^T \cdot \epsilon_4^T) \slashed{\pi}_s^{i_s} \vecslashed{\epsilon_3}^T u_4.
	\fe	
}
\newpage

\bibliography{refs}
\bibliographystyle{JHEP}

\end{document}